\documentclass{aastex} 
\usepackage{emulateapj5, apjfonts, psfig, epsfig}
\begin{document}

\slugcomment{AJ, in press} 

\title{Variable Stars in the Unusual, Metal-Rich Globular Cluster 
 NGC~6388} 

\shorttitle{Variable Stars of NGC~6388}
\shortauthors{Pritzl et al.} 

\received{} 
\revised{} 
\accepted{}

\author{Barton J. Pritzl\altaffilmark{1,2} and Horace A. Smith}  
\affil{Dept.\ of Physics and Astronomy, Michigan State University, 
       East Lansing, MI 48824} 
\email{pritzl@pa.msu.edu, smith@pa.msu.edu}

\author{M\'arcio Catelan} 
\affil{Pontificia Universidad Cat\'olica de Chile, Departamento de 
       Astronom\'\i a y Astrof\'\i sica, Av. Vicu\~{n}a Mackenna 4860, 
       6904411 Macul, Santiago, Chile}
\email{mcatelan@astro.puc.cl}

\and
\author{Allen V. Sweigart}
\affil{NASA Goddard Space Flight Center, Laboratory for Astronomy and 
       Solar Physics, Code~681, Greenbelt, MD 20771}
\email{sweigart@bach.gsfc.nasa.gov}

\altaffiltext{1}{Visiting Astronomer, Cerro Tololo Inter-American
Observatory, National Optical Astronomy Observatories, which is
operated by AURA, Inc., under cooperative agreement with the
National Science Foundation.}
\altaffiltext{2}{Current address: National Optical Astronomy Observatories, 
P.O. Box 26732, Tucson, AZ 85726, email: pritzl@noao.edu}

\begin{abstract}

We have undertaken a search for variable stars in the metal-rich
globular cluster NGC~6388 using time-series $BV$ photometry.  Twenty-eight 
new variables were found in this survey, increasing the total number of 
variables found near NGC~6388 to $\sim57$.  A significant number
of the variables are RR~Lyrae ($\sim14$), most of which are probable
cluster members.  The periods of the fundamental mode RR~Lyrae are shown to 
be unusually long compared to metal-rich field stars.  The 
existence of these long period RRab stars suggests that the horizontal branch 
of NGC~6388 is unusually bright.  This implies that the metallicity-luminosity 
relationship for RR~Lyrae stars is not universal if the RR~Lyrae in 
NGC~6388 are indeed metal-rich.  We consider the alternative possibility 
that the stars in NGC~6388 may span a range in [Fe/H].  Four candidate 
Population~II Cepheids were also found.  If they are members of the cluster, 
NGC~6388 would be the most metal-rich globular cluster to contain 
Population~II Cepheids.  The mean $V$ magnitude of the RR~Lyrae is found to 
be $16.85\pm0.05$ resulting in a distance of 9.0 to 10.3~kpc, for a range of 
assumed values of $\langle M_V \rangle$ for RR~Lyrae.  We determine the 
reddening of the cluster to be $E(\bv) = 0.40\pm0.03$~mag, with differential 
reddening across the face of the cluster.  We discuss the difficulty in 
determining the Oosterhoff classification of NGC~6388 and NGC~6441 due to the 
unusual nature of their RR~Lyrae, and address evolutionary constraints on a 
recent suggestion that they are of Oosterhoff type~II.

\end{abstract} 


\section{Introduction}

NGC~6388, at ${\rm [Fe/H]}=-0.60 \pm 0.15$ (Armandroff \& 
Zinn 1988), is slightly more metal-rich than 47~Tucanae (NGC~104).  One would 
therefore expect that the color-magnitude diagram (CMD) of NGC~6388 would show 
a stubby red horizontal branch (HB) characteristic of metal-rich globular 
clusters such as 47~Tuc.  Alcaino (1981) found that NGC~6388 contains 
a strong, red HB component in addition to a small number of 
stars blueward of the red HB.\@  Hazen \& Hesser (1986) discovered a number of 
variables within the tidal radius of NGC~6388 in addition to those previously 
found (Lloyd Evans \& Menzies 1973, 1977).  They found that as many as 6 of 
these variables were RR~Lyrae (RRL) and were 
likely cluster members.   At the time, this made NGC~6388 the most metal-rich 
globular cluster known to contain RRL.\@  In a more recent study, Silbermann et 
al.\ (1994) presented ($V$, $\bv$) and ($V$, $V-R$) CMDs which confirmed Alcaino's 
previous finding that NGC~6388 contains a weak blue HB component.  An 
additional 3 RRL were found in their survey along with 4 suspected 
variables.  Unfortunately, their data were obtained under seeing conditions 
not favorable for observing variable stars in such a crowded environment. 

Rich et al.\ (1997), using {\it Hubble Space Telescope} (HST) observations, 
discovered that the CMD of NGC~6388 actually has a pronounced blue HB 
component, which stretches across the location of the instability strip.  
The presence of hot HB stars in NGC~6388 was originally suggested by Rich, 
Minniti, \& Liebert (1993) on the basis of integrated-light observations in 
the $UV$.\@
Not only does the HB have a blue component, but it also slopes upward as 
one goes blueward in a ($V$, $\bv$) CMD.\@  NGC~6388 is not, however, unique 
in these characteristics:  As we noted in Pritzl et al.\ (2001; hereafter 
Paper~I), Rich et al.\ showed that the CMD of the 
relatively metal-rich globular cluster NGC~6441 has a similar HB morphology.  
It has long been known that the HB morphology 
of globular clusters does not correlate perfectly with [Fe/H] and that
at least one parameter besides [Fe/H] is needed to
account for this (Sandage \& Wildey 1967; van den Bergh 1967). 
NGC~6441 and NGC~6388 are the only metal-rich globular clusters
known to exhibit this second parameter effect.  Very recently, the presence 
of a blue HB component was suggested also for the metal-rich globular 
cluster Terzan~5 (Cohn et al.\ 2002), a cluster which has also been found 
to contain a long-period RRL (Edmonds et al.\ 2001).

A number of possible explanations for the unusual nature of the HBs in 
NGC~6388 and NGC~6441 have been 
proposed (Sweigart \& Catelan 1998; Sweigart 1999, 2002; Paper~I; Raimondo 
et al.\ 2002).  Several of these theoretical explanations predict that the 
HBs in these clusters are unusually bright, which 
would imply unusually long periods for their RRLs.  It has already 
been shown that NGC~6441 contains a number of RRL with 
unusually long periods (Layden et al.\ 1999; Paper~I).\@  

In this paper we report new $B$ and $V$ photometry of NGC~6388 which has
led to the discovery of additional variable stars.  Preliminary results 
from these observations have already been used to argue that the 
RRLs in NGC~6388 are unusually bright for the cluster metallicity (Pritzl et 
al.\ 2000).  Here we present the results of the new study in detail 
and call attention to several unusual properties of RRLs in NGC~6388 while 
noting the similarities and differences to NGC~6441.

\section{Observations and Reductions}

NGC~6388 was observed in conjunction with NGC~6441 (see Paper~I for 
detailed discussion on the observations and reductions).  Time series 
observations of NGC~6388 were obtained at the 
0.9 m telescope at Cerro Tololo Inter-American Observatory (CTIO) using the 
Tek 2K No.\ 3 CCD detector with a field size of 13.5 arcmin per side.
Time series observations of NGC~6388 were obtained on the UT dates 
of May 26, 27, 28, and 29, and June 1, 2, 3, and 4, 1998. Exposures of 
600 seconds were obtained in both $B$ and $V$ filters.  The seeing 
ranged from 1.1 to 2.5 arcsec, with a typical seeing of 1.4 arcsec.  

Landolt (1992) and Graham (1982) standard stars were observed on June 1, 
2, and 3.  These primary standard stars 
spanned a color range from $\bv = 0.024$ to 2.326~mag, adequate to 
cover the color range of the reddened stars in NGC~6388.  Primary standards 
were observed between airmasses 1.035 and 1.392.  Only the night of June 1 
was photometric, but standards observed on the nonphotometric nights were 
incorporated in the reductions using the ``cloudy" night reduction routines 
created by Peter Stetson (private communications).  

Instrumental magnitudes $v$ and $b$ for the NGC~6388 
stars were transformed to Johnson $V$ and $B$ magnitudes using Stetson's 
{\sc trial} package.  71 local
standards within the NGC~6388 field were used to set the frame-by-frame 
zero-points for the cluster observations.  Because NGC~6388 was 
observed to higher airmass than the Landolt (1992) and Graham (1982) 
standards, the local 
standards were also used to check the adopted values of the extinction 
coefficients for the night of June 1.  The observations of the local 
standards confirmed the values of the extinction coefficients determined 
from the primary standards.  Transformation equations derived from the 
standard stars had the form:  

\begin{equation} 
v = V - 0.006 \, (\bv) + 0.159 \, (X-1.25) + C_V 
\end{equation} 

\begin{equation} 
b = B + 0.105 \, (\bv) + 0.243 \, (X-1.25) + C_B  
\end{equation} 

\noindent where $X$ is the airmass and $C_V$ and $C_B$ are the 
zero-point shifts for their respective filters.
Comparing the transformed magnitudes of the standard stars with the
values given by Graham and Landolt, we find rms residuals of 0.018 magnitudes in
$V$ and 0.006 magnitudes in $B$.\@

The photometry of stars in the NGC~6388 field could be compared with 
photometry in three earlier studies, as shown in Table~1.  First, we compared 
our photometry to the photoelectric photometry of a number of the 
standard stars in the field of NGC~6388 obtained by Alcaino (1981).  Alcaino 
reported that photometry for 11 of the 26 stars used to calibrate his NGC~6388 
data was found to be in good agreement with independent unpublished 
observations by K. C. Freeman.  We 
were able to use 14 stars from Table~I in Alcaino for comparison.  The 
difference between our data set and that of Alcaino is somewhat large, which 
may result from only being able to compare our photometry to the 
fainter Alcaino standard stars.  

A large number of comparisons could be made with the CCD photometry of 
Silbermann et al.\ (1994).  We first matched the position of the stars in 
Table~3 of Silbermann et al.\ with our data.  A number of discrepant 
values were found.  With no image to compare from Silbermann et al., it is 
difficult to ascertain whether these stars were crowded or not.  However, the 
majority of the magnitudes were in good agreement with our data.  

A comparison to the HST $B$, $V$ photometry obtained by Rich et al.\ (1997) 
of NGC~6388 was also made.  Due to the compact nature of NGC~6388 it was 
difficult finding a large number of uncrowded stars on the images obtained 
at CTIO to compare with those found using the HST data, which looked only 
at the inner regions of the cluster.  From a sample of 11 stars which were 
relatively uncrowded on our images we see that our ground-based photometry 
is brighter by about 0.025~mag.

\section{Color-Magnitude Diagram} 

\begin{figure*}[t]
  \centerline{\psfig{figure=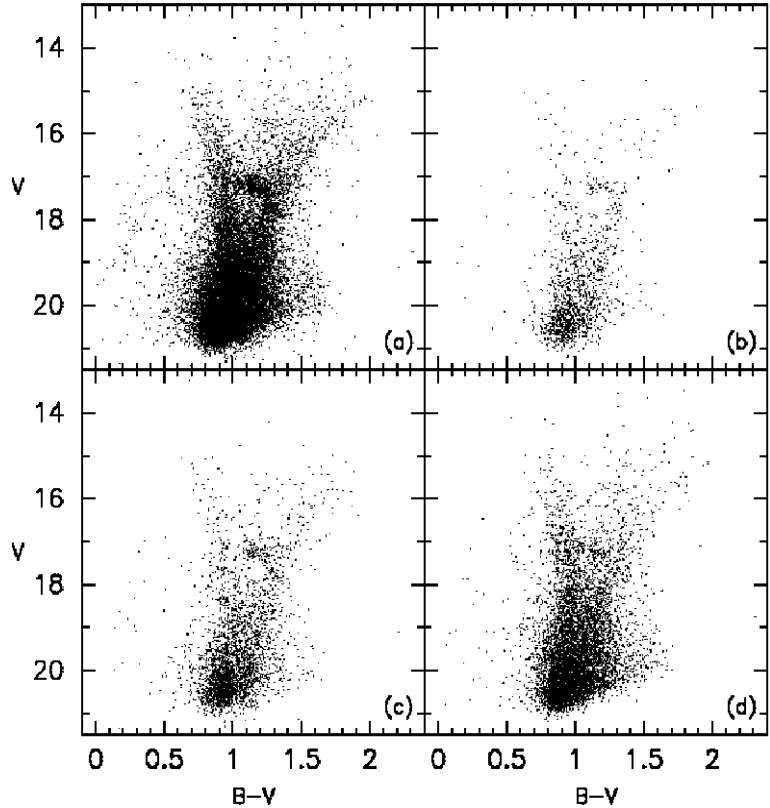,height=4.25in,width=4in}}
  \caption{Color-magnitude diagrams for the NGC~6388 stars located in the 
           complete field of view (a), out to a radius of 1.7 arcmin (b) 
           and 2.7 arcmin (c) from the cluster center, and outside a radius 
           of 6.5 arcmin from the cluster center (d).} 
  \label{Fig01} 
\end{figure*} 

Figure~1a presents a CMD consisting of 19509 stars in the field centered on 
NGC~6388.  Only stars with $\chi \le 1.5$ are shown, where $\chi$ is a 
fitting parameter in Stetson's DAOphot programs.  The red clump of the HB of 
the cluster can be seen at $V \sim 17.2$, $(\bv) \sim 1.15$.  The main sequence 
of the field can be seen extending through the cluster's HB from 
$(\bv) \sim 1.0$ to $\sim 0.7$.  Figure~1d shows the contribution of the 
field to the CMD of NGC~6388 in Figure~1a.  Stars were chosen from a radius 
greater than 6.5 arcmin from the cluster center for Figure~1d (the tidal 
radius of NGC~6388 is 6.21 arcmin; Harris 1996).

The effects of differential reddening on the red giant branch (RGB) can be seen 
in the figures.  This is especially notable in the luminosity function 
bump on the RGB at $V \sim 17.8$, $(\bv) \sim 1.3$.  A closer analysis of this 
feature could shed light on the helium content of NGC~6388 (Sweigart 1978; 
Fusi Pecci et al.\ 1990; Zoccali et al.\ 1999; Bono et al.\ 2001; Raimondo et 
al.\ 2002).  

Figure~1b shows all stars within 1.7 arcmin from the cluster center, giving 
the closest fit to the area of NGC~6388 observed by Rich et al.\ (1997).  
Because of crowding toward the center of NGC~6388, we could obtain good 
photometry for many fewer stars in the inner region than was possible with 
WFPC2.  The number of HB stars detected in the WFPC2 images is about 1350 
for NGC~6388 (Zoccali et al.\ 2000) and 1470 for NGC~6441 (Zoccali 2000, 
private communication).  In contrast, we were able to obtain good photometry 
for only $\sim70$ HB stars within the central 1.7 arcmin radius circle of 
NGC~6388, a number considerably smaller than we were able to observe close 
to the center of NGC~6441 (Paper~I).  As a result, it is difficult to use 
Figure~1b to identify any HB slope in the $V$, $\bv$ CMD.

\section{Variable Stars} 

\subsection{Discovery of New Variable Stars}

All variables were found with the same two methods as used for NGC~6441 
(see \S4.1 of Paper~I).\@  The time coverage of our observations is well suited 
for the discovery of short period variability, but not for the detection
of long period variables.  With the exception of stars falling outside of our 
field of view, all of the probable short period variable stars previously 
known were recovered during our variable star search.  The previously known 
variables that were within our field of view, but not detected in our 
survey, are likely long period variables (LPVs; Hazen \& Hesser 1986).  In 
addition, 28 probable new variable stars were detected.  In crowded regions 
close to the cluster center, the $B$ photometry proved superior to the $V$ 
photometry for identifying variable stars, presumably because of the lesser 
interference from bright red giant stars.  Finding information for the 
variable stars found in our survey is given in Table~2, where $X$,~$Y$ are 
the coordinates of the variables on the CCD [the cluster center is assumed to 
be at (1075.4,1064.4)] and $\Delta\alpha$,~$\Delta\delta$ are the differences 
in right ascension and declination from the cluster center (in arcsec).  
Finding charts for the variables can be seen in Figure~2. 

\begin{figure*}[t] 
  \centerline{\psfig{figure=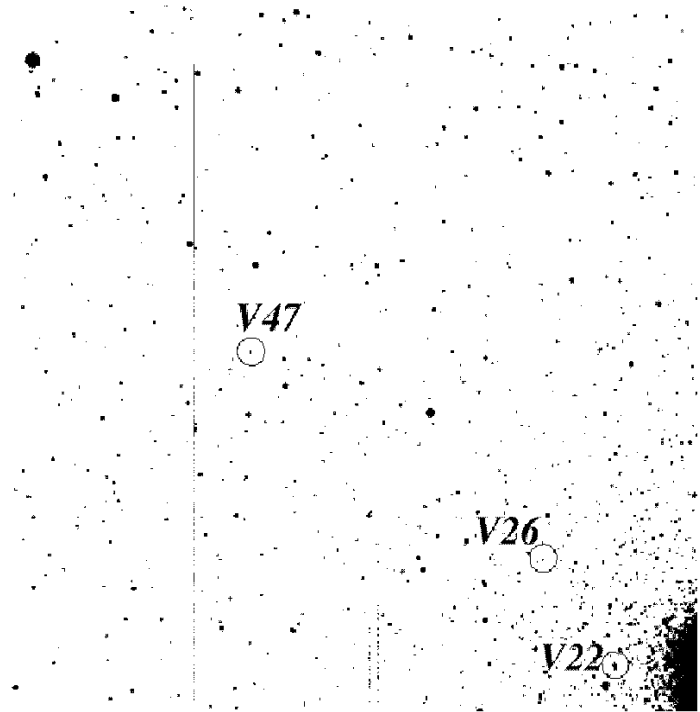,height=3.7225in,width=3.6125in}} 
  \caption{Finding charts for the NGC~6388 variable stars.  
           North is down and east is left.} 
  \label{Fig02}
\end{figure*} 

\begin{figure*}[t] 
  \figurenum{2 cont}
  \centerline{\psfig{figure=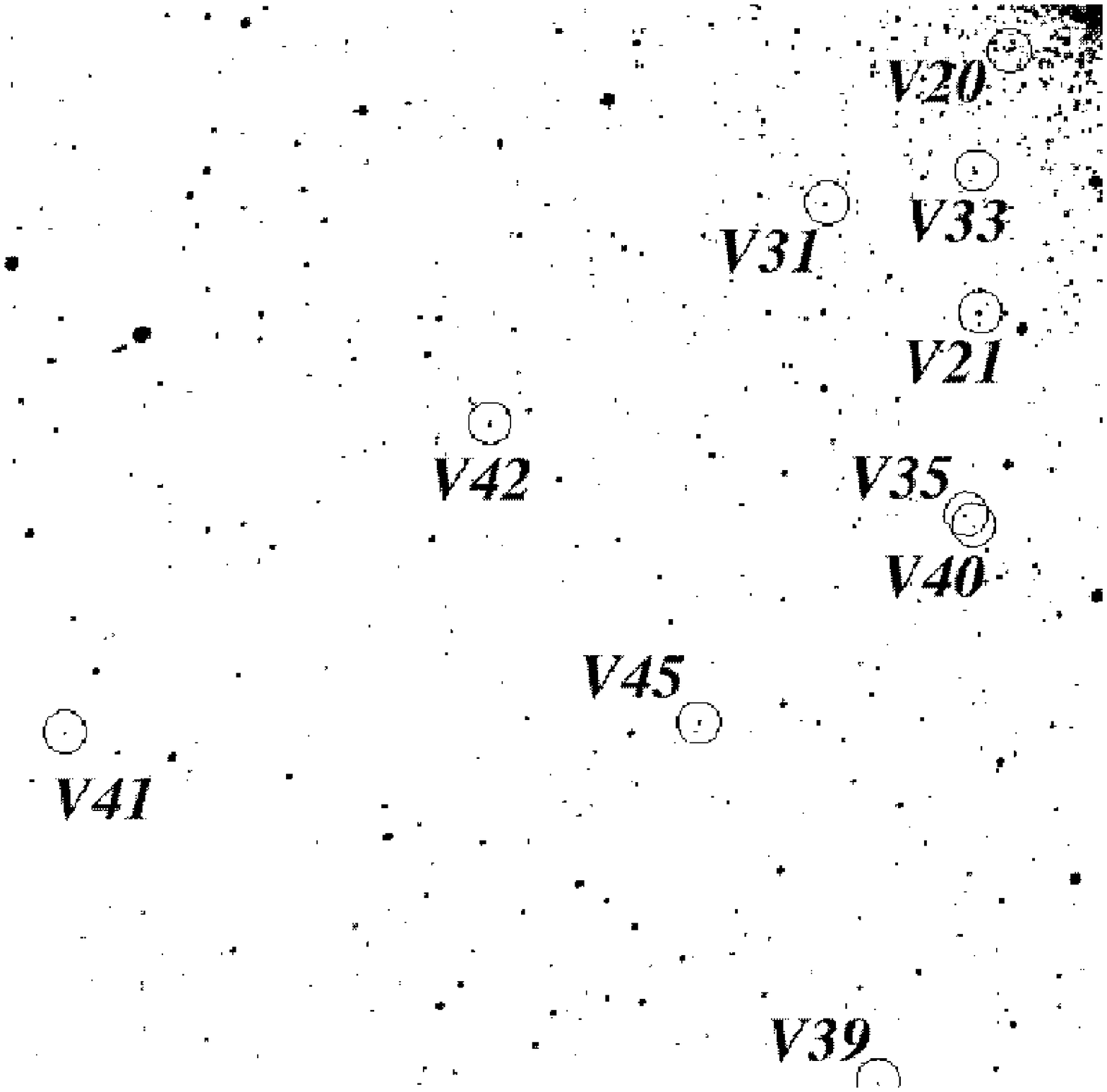,height=3.7225in,width=3.6125in}} 
  \caption{Finding charts for the NGC~6388 variable stars.  
           North is down and east is left.} 
  \label{Fig02}
\end{figure*} 

\begin{figure*}[t] 
  \figurenum{2 cont}
  \centerline{\psfig{figure=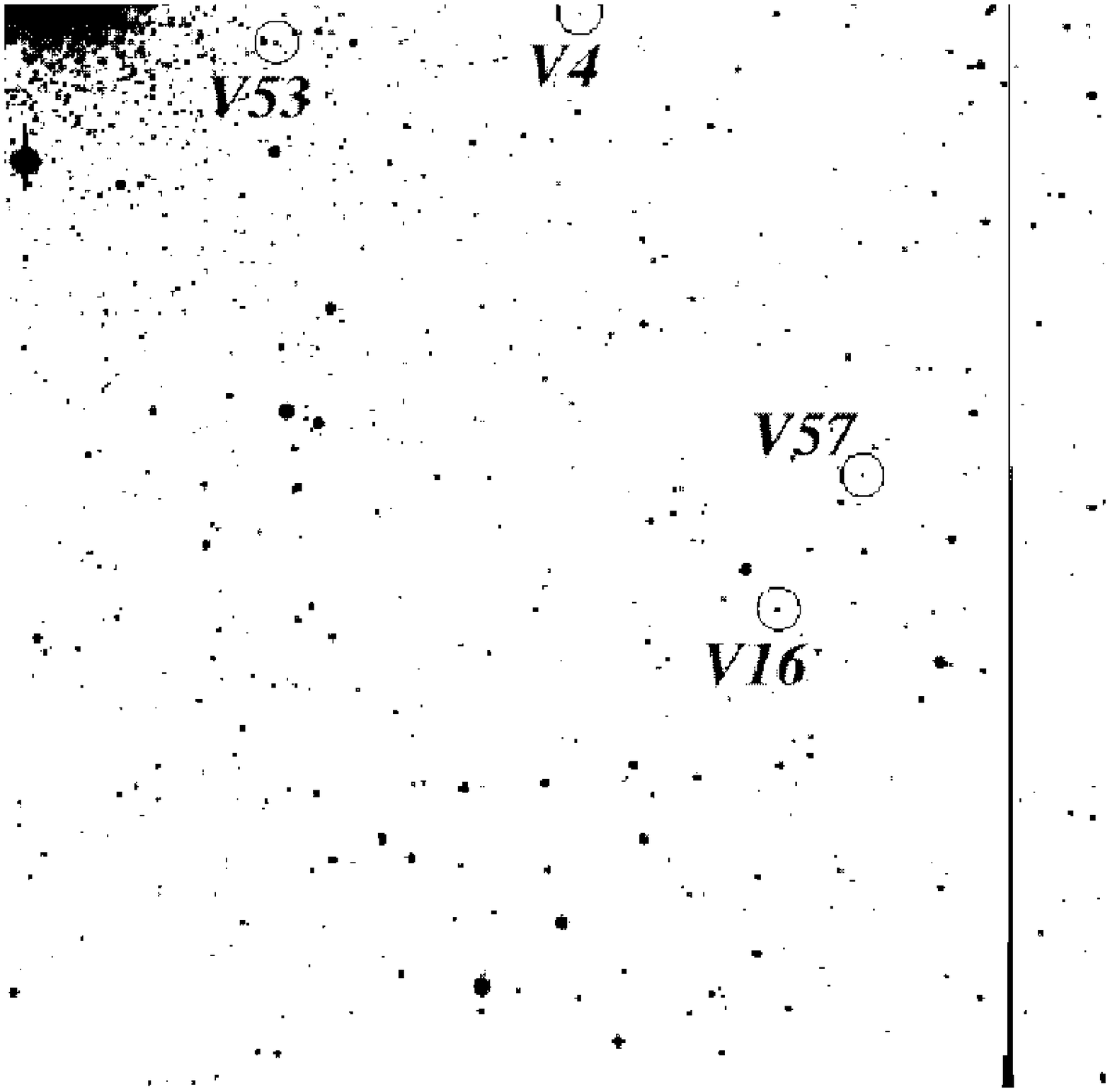,height=3.7225in,width=3.6125in}} 
  \caption{Finding charts for the NGC~6388 variable stars.  
           North is down and east is left.} 
  \label{Fig02}
\end{figure*} 

\begin{figure*}[t] 
  \figurenum{2 cont}
  \centerline{\psfig{figure=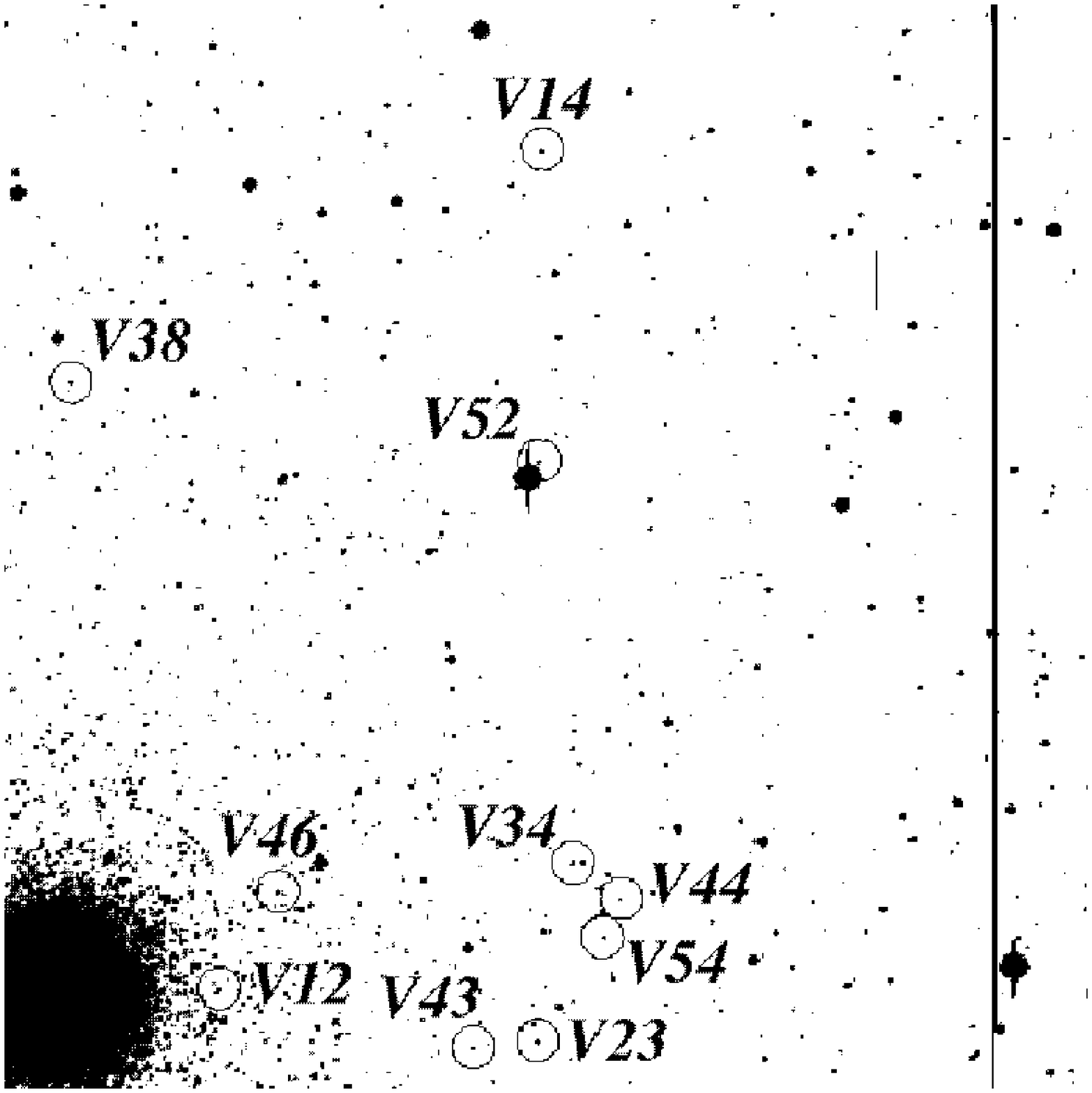,height=3.7225in,width=3.6125in}} 
  \caption{Finding charts for the NGC~6388 variable stars.  
           North is down and east is left.} 
  \label{Fig02}
\end{figure*} 

\begin{figure*}[t] 
  \figurenum{2 cont}
  \centerline{\psfig{figure=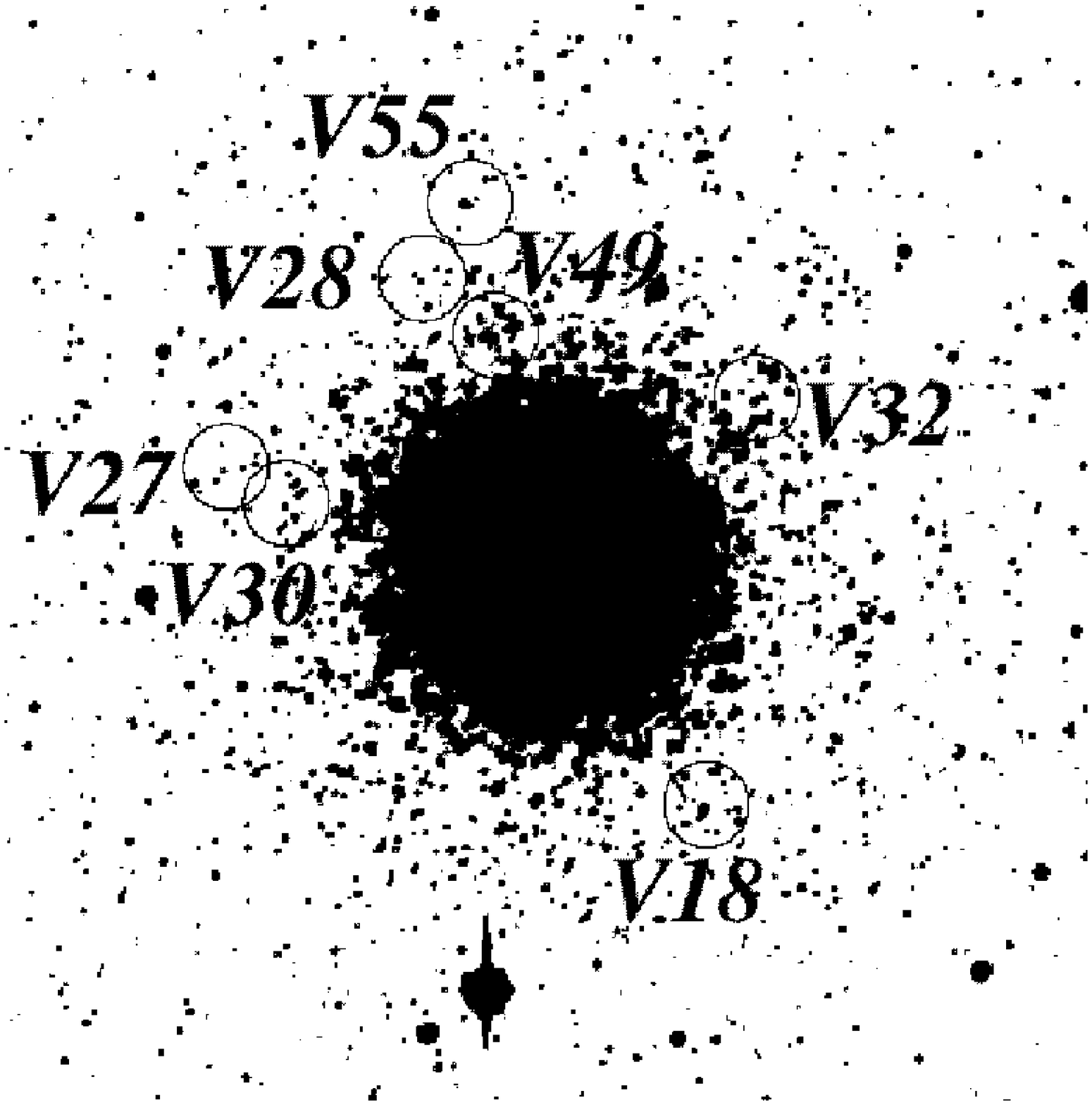,height=3.7225in,width=3.6125in}} 
  \caption{Finding charts for the NGC~6388 variable stars.  
           North is down and east is left.} 
  \label{Fig02}
\end{figure*} 

\begin{figure*}[t] 
  \figurenum{2 cont}
  \centerline{\psfig{figure=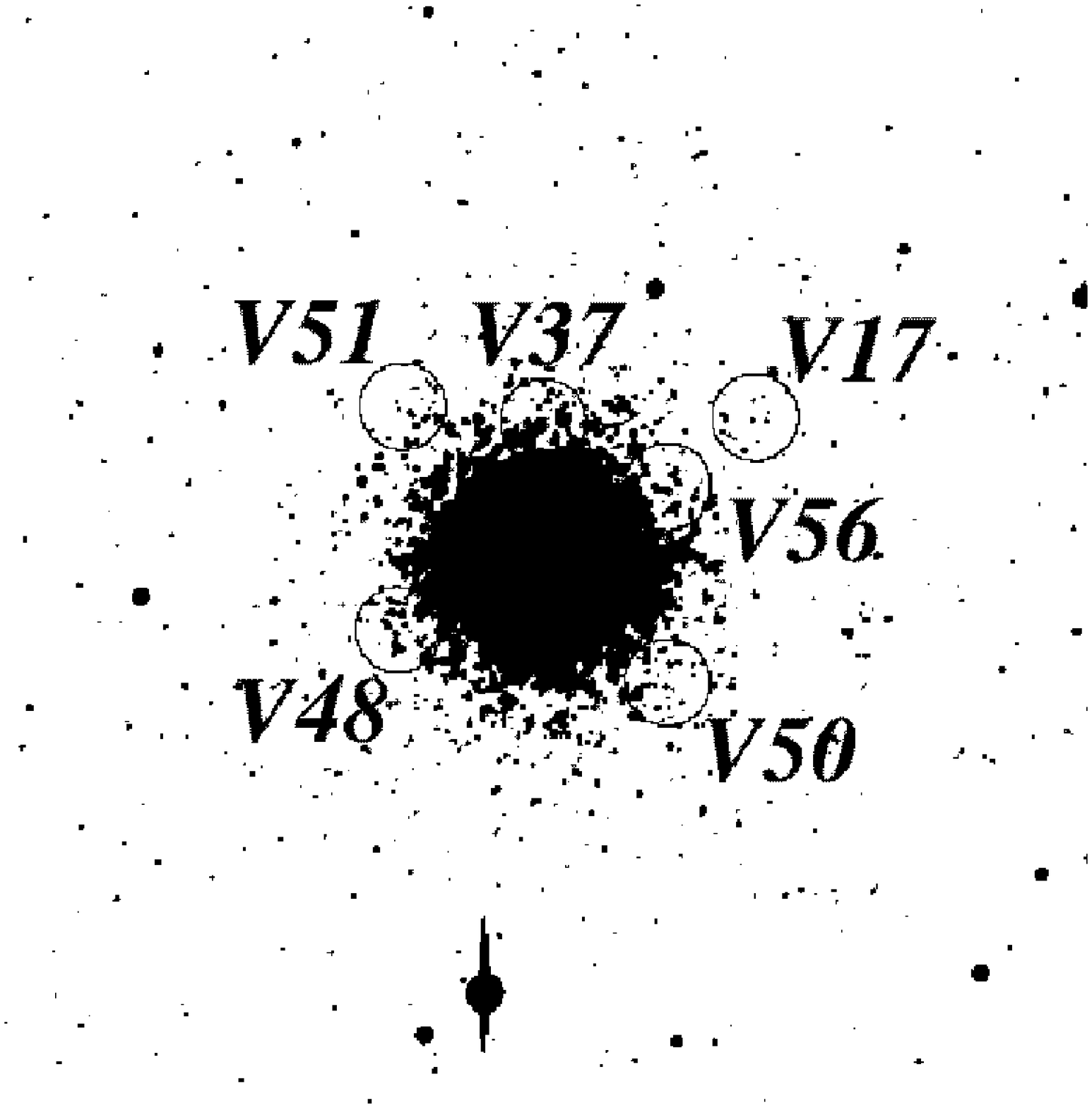,height=3.7225in,width=3.6125in}} 
  \caption{Finding charts for the NGC~6388 variable stars.  
           North is down and east is left.} 
  \label{Fig02}
\end{figure*} 

\begin{figure*}[t] 
  \figurenum{2 cont}
  \centerline{\psfig{figure=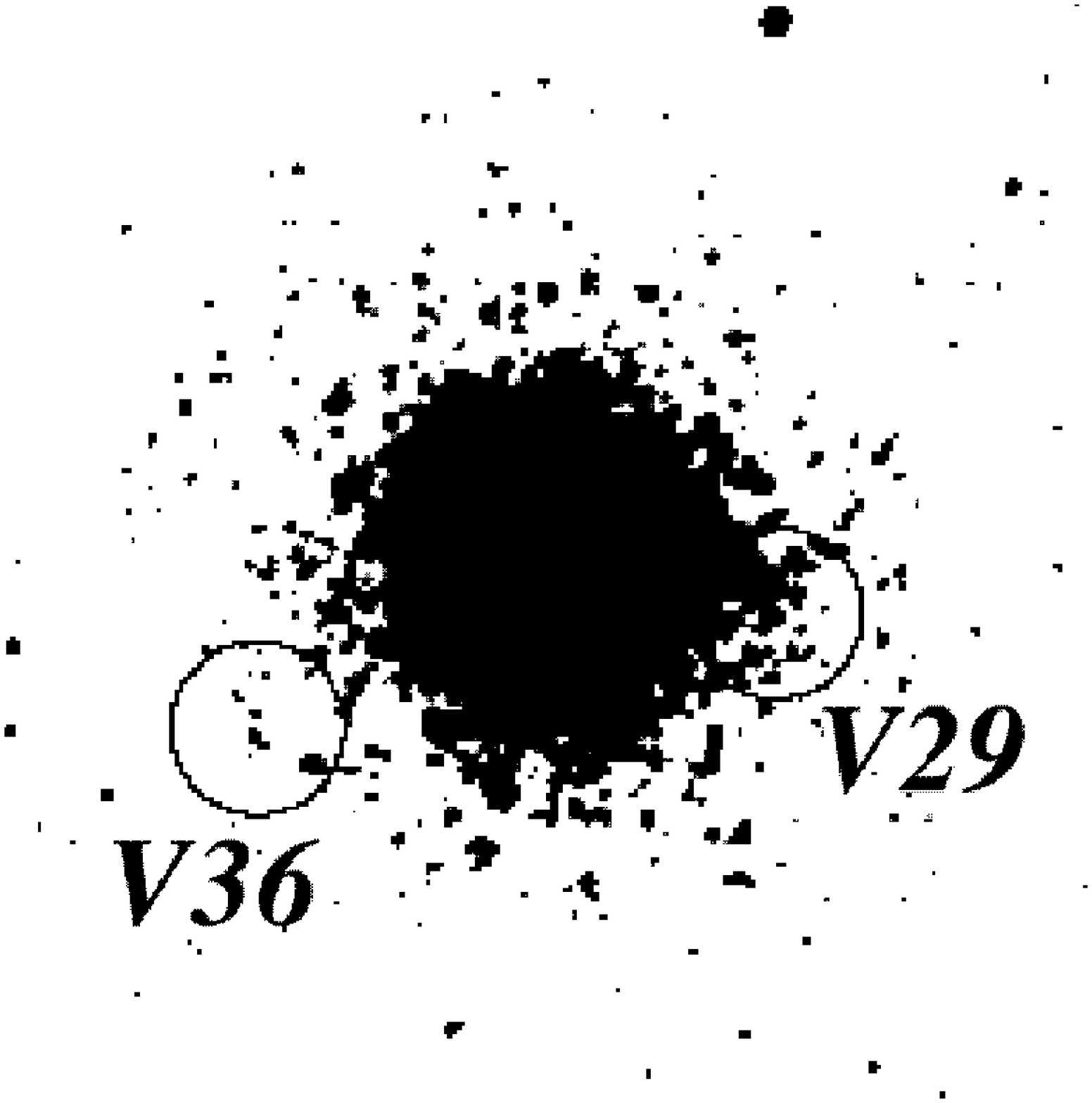,height=3.7225in,width=3.6125in}} 
  \caption{Finding charts for the NGC~6388 variable stars.  
           North is down and east is left.} 
  \label{Fig02}
\end{figure*}

\subsection{RR Lyrae stars} 

The variable stars have been studied previously by Hazen \& Hesser (1986) and 
Silbermann et al.\ (1994).  In the field of NGC~6388, the number of probable 
RRL has been increased from 10 to 14.  Figure~3 shows the position of these 
stars within the CMD of NGC~6388.  Of the previously known cluster RRL, only 
V24, which lies outside our field of view, was not rediscovered.  The variable 
stars were classified according to their period, magnitude, and shape of the 
light curve.  The mean properties of the individual RRL, and the one 
$\delta$~Scuti or SX~Phe star, found in this survey are listed in Table~3.  
Stars listed as S1, S2, and S3 in column~7 are the stars which were suspected 
to be variable as listed by Silbermann et al.  The periods determined for 
the known RRL, using the phase dispersion minimization program in IRAF, are 
found to be in good agreement with those found by Hazen \& Hesser and 
Silbermann et al.  Spline fits were used to determine the magnitude weighted 
and luminosity weighted mean magnitudes, $(\bv)_{\rm mag}$ and $\langle 
V \rangle$, respectively.  Figure~4 shows the light curves for the 
individual variable stars, where the displayed light curves are those of the 
RRL, the 
Population~II Cepheids (P2Cs), the eclipsing binaries, the $\delta$~Scuti or 
SX~Phoenicis variable, and the variables with uncertain classification, in 
that order. 

\begin{figure*}[t] 
  \centerline{\psfig{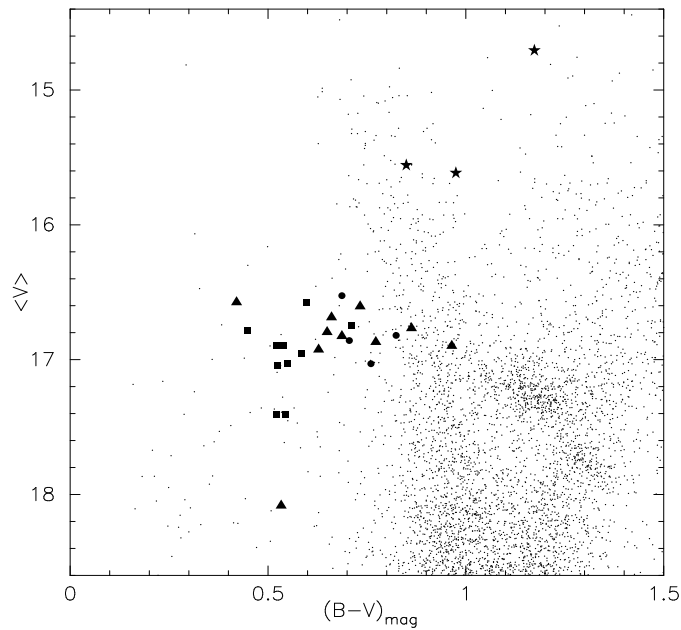}}
  \caption{Color-magnitude diagram for the fundamental mode RR~Lyrae 
           (filled circles), first overtone RR~Lyrae (filled squares), 
           suspected RR~Lyrae (filled triangles), and Population~II Cepheids 
           (five-pointed stars) in the field of NGC~6388.}
  \label{Fig03} 
\end{figure*} 

\begin{figure*}[t] 
  \centerline{\psfig{figure=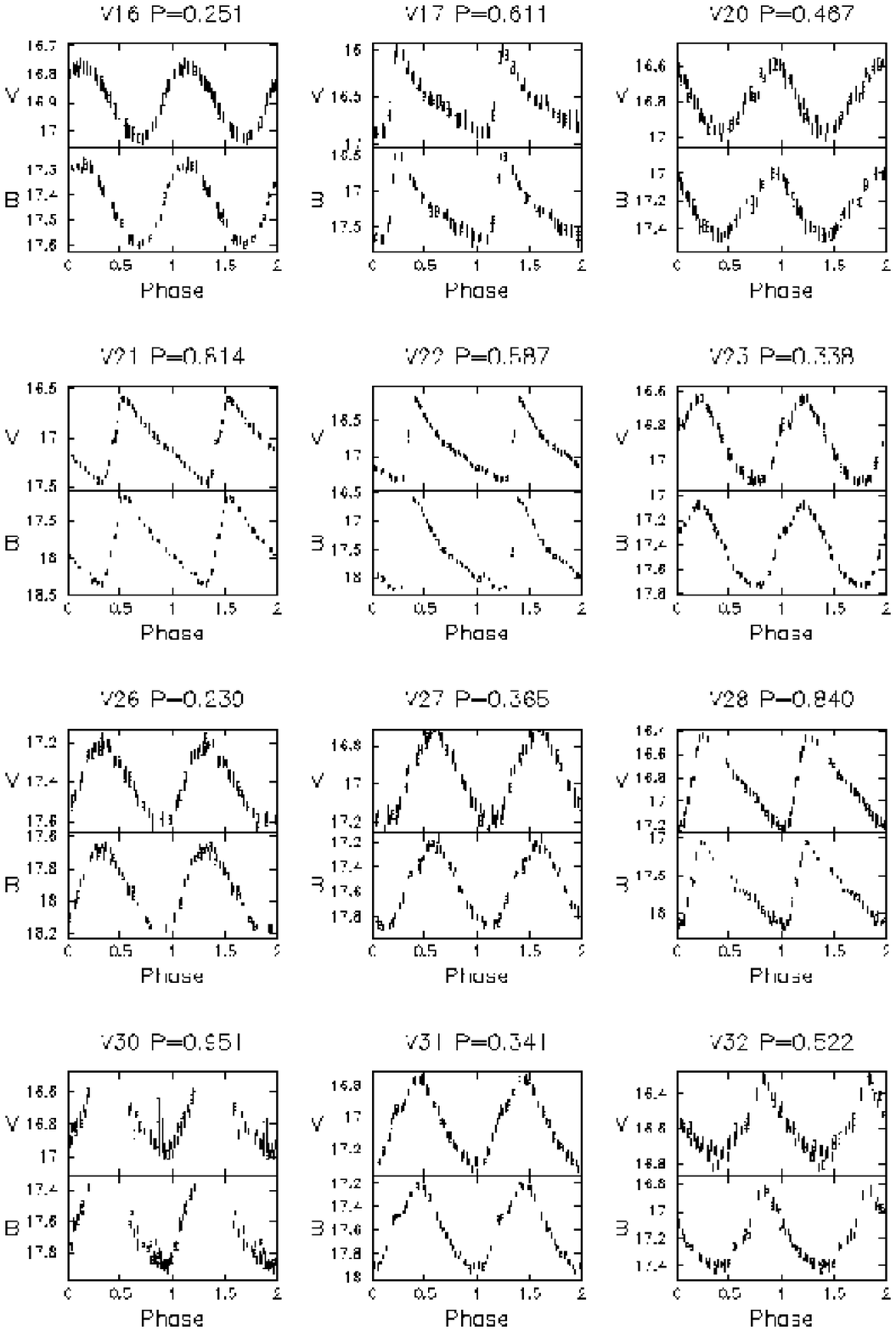,height=7.75in,width=5.5in}}  
  \caption{$V$ and $B$ light curves of NGC~6388 variable stars.} 
  \label{Fig04}
\end{figure*} 

\begin{figure*}[t] 
  \figurenum{4 cont}
  \centerline{\psfig{figure=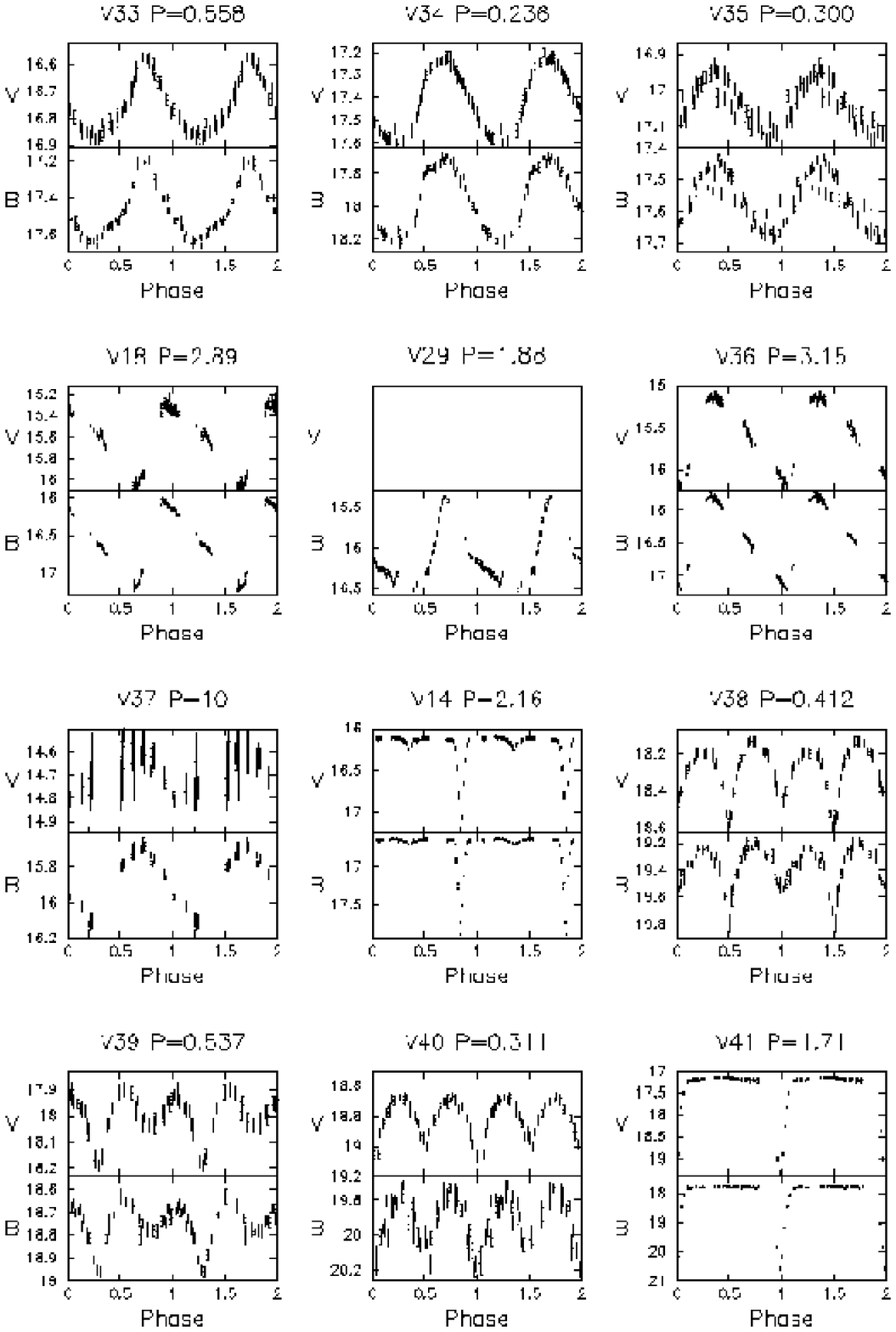,height=7.75in,width=5.5in}}  
  \caption{$V$ and $B$ light curves of NGC~6388 variable stars.} 
  \label{Fig04}
\end{figure*} 

\begin{figure*}[t] 
  \figurenum{4 cont}
  \centerline{\psfig{figure=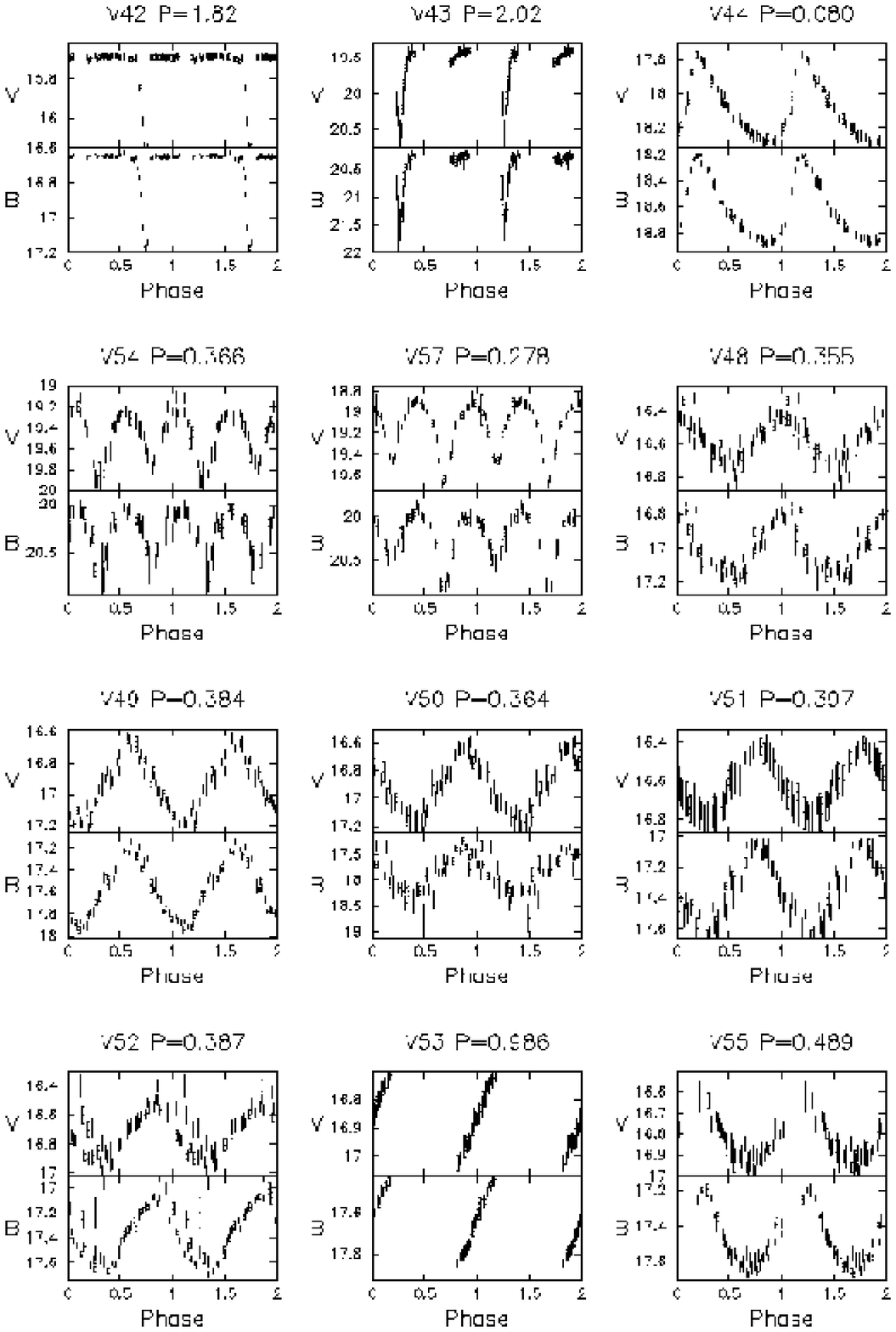,height=7.75in,width=5.5in}}  
  \caption{$V$ and $B$ light curves of NGC~6388 variable stars.} 
  \label{Fig04}
\end{figure*} 

\begin{figure*}[t] 
  \figurenum{4 cont}
  \centerline{\psfig{figure=Pritzl.fig04d.eps,height=1.5in,width=1.5in}}  
  \caption{$V$ and $B$ light curves of NGC~6388 variable stars.} 
  \label{Fig04}
\end{figure*}

The accuracy of the periods found in our survey is 
$\pm$0.001d to $\pm$0.002d, depending on the scatter and completeness of the 
light curve.  Tables~4 and 5 list the photometry for the individual 
variable stars.    

Light curves were analyzed by Fourier decomposition using the equation (order 
of fit$=8$), 

\begin{equation} 
mag = A_0 + \sum A_j \, \cos(j\omega t + \phi_j). 
\end{equation} 

\noindent
It has been shown (e.g., Clement \& Shelton 1997) that RRab and RRc stars fall 
into distinct regions in a $A_{21}$ vs.\ $\phi_{21}$ plot, where $A_{21}=A_2/A_1$ 
and $\phi_{21}=\phi_2-2\phi_1$.  As found by Simon \& Teays (1982), Figure~5 
shows for those RRL with clean light curves that the RRab fall at values 
greater than $A_{21}$ of 0.3 and the RRc fall below.  Table~6 
lists the values of the Fourier parameters for all candidate RRL, where 
$A_{j1}=A_j/A_1$ and $\phi_{j1}=\phi_j-j\phi_1$.  The errors in the phase 
differences are based on Eq.~16d in Petersen (1986).  It should be noted 
that the Fourier decomposition is dependent on the accuracy and completeness 
of the light curve.  Therefore, the RRL in NGC~6388 with high scatter or 
large gaps in their light curves have unreliable Fourier parameters, 
specifically those RRL which we were not able to definitively classify.  
We showed in Paper~I that Fourier 
decomposition parameters can be used to determine if a star is a RRc variable 
or an eclipsing binary at twice the period.  Although the method proved useful, 
there were no cases in NGC~6388 where this method was needed.

\begin{figure*}[b] 
  \centerline{\psfig{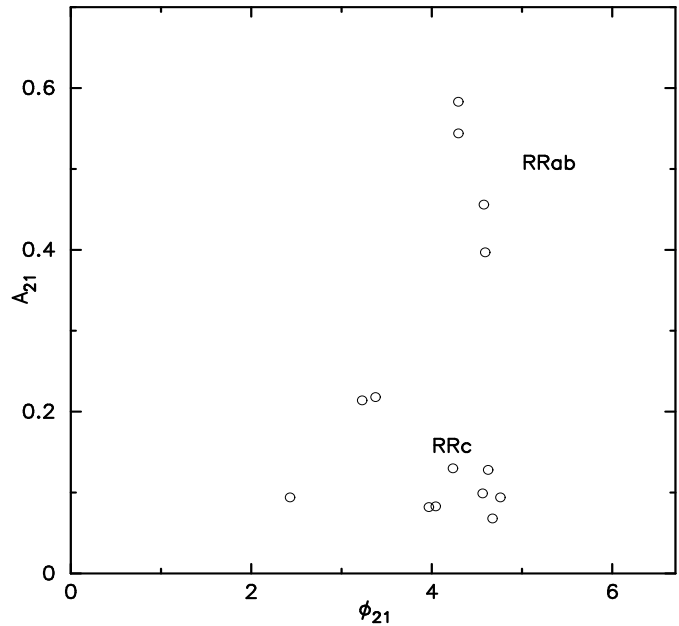}} 
  \caption{Fourier parameter plot using $A_{21}$ vs.\ $\phi_{21}$ to show 
           the distinction between RR~Lyrae types in NGC~6388.} 
  \label{Fig05}
\end{figure*}

Shown in Figure~6 are the period-amplitude diagrams for NGC~6388 given in 
$B$ and $V$.\@  There is one RRab variable, V17, whose amplitude appears to be 
low for its period.  This may be due to blending effects.  However, this 
variable does not show the shift toward redder colors seen in certain RRab 
stars in NGC~6441 which were suspected of having blended images (see 
Paper~I).  It should be noted that the Blazhko Effect can also 
reduce the amplitudes of RRab stars, but our observations do not extend over 
a long enough time interval to test for the presence of this effect.  As with 
NGC~6441, we see that NGC~6388 lacks a significant gap between the shortest
period RRab and the longest period RRc.

\subsection{Notes on Individual Variable Stars} 

\indent 
V17 - The amplitude appears to be low compared to the other fundamental mode 
RRL.\@  There is some scatter in the curve indicating a possibility of 
blending.  The ratio of the $B$ to $V$ amplitudes is not unusual, providing no 
indication that an unresolved blend with a different colored star has affected 
the photometry.  Its mean magnitude is somewhat brighter than the other 
fundamental mode RRL.\@

\indent 
V20 - There is scatter in the curve.  
It falls among the other cluster first overtone RRL in the CMD.\@

\indent 
V26 \& V34 - These two stars are similar in that they both are short period 
RRc stars and are unusually faint when compared to the other RRL.\@  Their 
amplitudes are somewhat larger than the other short period RRc stars, as well.  
Both stars would have to be dereddened by an unusually large amount to move 
them back among the other RRc stars of NGC~6388.  It may be that these two 
stars are actually nonmembers of the cluster.

\indent 
V29 - P2C variable.  Previously designated as a field RRL 
by Silbermann et 
al.\@ (1994).  No $V$ data were available due to saturation.  For this reason, 
the $B$ data were placed on the standard system using a zero-point shift of 
+3.829.  

\indent
V30 - This star fits in the CMD with the other fundamental mode RRL.\@  The best 
fit to the data is a period around 0.939d.  There is a gap in the data near 
maximum light through descending light.  This, along with scatter in the curve, 
makes finding the best period difficult.  

\indent
V32 - (S1, Silbermann et al.)  A long period RRc type variable with some 
scatter in the curve.

\indent 
V35 - One night of observations of this variable (night 4) 
falls at a different phase and 
amplitude as compared to the other nights, as shown in Figure~7.  Taking that 
night out, the data fit the period of 0.299~d.  See \S5.2 for more discussion.

\indent 
V18 and V36 - P2C variables.  The light curves in $V$ tend to have more 
scatter due 
to reaching the saturation limit of the CCD.\@  Crowding in the cluster is also 
likely to contribute to the scatter found in the light curves since both 
stars are found near the center of the cluster.  

\indent 
V37 - A probable P2C variable.  The maximum period fit to the data is 
10 days due to the length of the observing run.  The magnitude of the star gets 
fainter for the first four nights of observations, while the second four nights 
of observations show a clear maximum.  

\indent 
V44 - From the shape of the curve, location in the CMD, and the period, this 
star is either of $\delta$ Scuti or SX Phoenicis type.  From the magnitude 
of the star, it is unlikely that it is a member of NGC~6388.  

\indent 
V48 - Shows definite variability, but has a lot of scatter in the light curve 
making classification difficult and the magnitude and color unreliable.  The 
$B$ light curve does look like that of a c-type RRL.\@  The mean magnitude of 
the variable places it along the HB, although slightly bluer than the other 
RRc.  

\indent 
V49 - (S2, Silbermann et al.) A lot of scatter in the light curve makes the 
exact classification uncertain, although the $B$ curve looks somewhat like 
that of a c-type RRL.\@  Its mean magnitude and color place it among the other 
probable cluster RRc.

\indent 
V50 - The light curves show scatter, especially in $B$, making 
classification 
uncertain and the mean magnitude and color unreliable.  The $V$ curve looks 
to be RRc type.

\indent 
V51 - This star shows definite variability, but a large amount of scatter exists. 
Its magnitude and color, although somewhat unreliable, place the variable among the 
other probable NGC~6388 RRL.\@

\indent 
V52 - A variable that falls among the first overtone RRL in the CMD of NGC~6388.
An unusual light curve shape, which shows a sharp decrease in magnitude after 
maximum, makes its classification uncertain.  This variable is found next to a 
much brighter star, which may be affecting the photometry.  

\indent 
V53 - (S3, Silbermann et al.) This star shows definite variability, but we were only able to 
observe it when the star was increasing in brightness.  Therefore, the magnitude, 
color, and period given for this variable are not reliable.

\indent 
V55 - A likely c-type RRL.\@  The variable falls among the other probable RRc 
of NGC~6388.  A gap occurs during the rising light in the light curve up to 
near maximum light, making the exact classification uncertain.

\indent 
V56 - Scatter in the data makes this star's classification uncertain.  The 
mean magnitude of the variable places it along the HB of NGC~6388.

\indent 
S4 (Silbermann et al.) - We were unable to find any variability, although 
crowding may be an issue.

\subsection{Reddening} 

Reddenings for the RRab stars of NGC~6388 were determined using Blanco's (1992)  
method as outlined in Paper~I.\@  We calculated the reddenings for the 
RRab stars with good ($\bv$) light curves listed in Table~7 by using the 
averaged photometry in the phase range 0.5--0.8 and Eqs.\ 3 and 7 in Blanco 
(1992).    

The mean reddening value for the 4 RRab stars 
which are believed to be probable members of NGC~6388 is $E(\bv)=0.40\pm0.03$.  
The range in reddening values agrees with previous determinations that 
NGC~6388 is subject to differential reddening, although the range is not as 
large 
as that of NGC~6441.  The smaller range in differential reddening may be due 
to the smaller sample of RRab stars, but it is most likely due to NGC~6388 
being farther away from the Galactic plane.

Silbermann et al.\ (1994) determined, in a similar fashion, 
that the reddenings for V17 and V29 are 0.48 and 0.47, 
respectively, with uncertainties of $\pm$0.04.  One reason for the 
discrepancies between the reddenings determined by this survey and that 
of Silbermann et al.\ is their use of $\Delta{S}=8$, corresponding to a 
metallicity lower than the one adopted in this paper ($\Delta{S}=3.22$).  
Another explanation for the differences in reddenings may be the 
high scatter in the light curves found by Silbermann et al.
Alcaino (1981) derived the reddening of NGC~6388 to be $E(\bv)=0.41$ 
from the color of the giant branch as compared to that in 47~Tuc.
Zinn (1980) and Reed, Hesser, \& Shawl (1988) obtained $E(\bv)=0.35$
and 0.39, respectively, from their analysis of the integrated cluster light.  
The Schlegel, Finkbeiner, \& Davis (1998) reddening value is 0.415 for 
NGC~6388.

We find the mean magnitude of the RRL, leaving out those RRL of uncertain 
classification, to be $\langle V_{\rm RR} \rangle = 16.85 \pm 0.05$ and 
$16.93 \pm 0.06$ without and with the two faint RRc stars V26 and V34, 
respectively.  (For the rest of the paper, we include V26 and V34 among the 
NGC~6388 RRL except where noted.)  If we assume $A_V = 3.2 E(\bv) = 1.28$, and 
if $M_V$ for the RRL is between $+0.5$ to $+0.8$, then the distance modulus for 
NGC~6388 is in the range of 14.77 to 15.07~mag for the case where we do not 
include V26 and V34.  The resulting distance to the cluster would be from 9.0 
to 10.3~kpc.  For comparison, Harris (1996) lists a distance of 11.5~kpc to 
NGC~6388 with a reddening of $E(\bv)=0.40$, $V_{\rm HB}=17.25$, and 
$M_V({\rm HB})=0.71$.  The fainter $\langle V_{\rm RR} \rangle$ value, which 
includes V26 and V34, would only increase the distance by approximately 
0.3~kpc.

There is some question as to whether or not there is a metallicity 
spread among the stars in NGC~6388 (see \S5.6).  It is therefore interesting 
to see how the reddening derived from the RRab stars depends on the adopted 
metallicity.  If the metal abundances of the RRL were much 
lower than that which we have adopted, they would have less line blanketing 
and the reddenings which we would derive would be greater.  For example, if 
we had adopted ${\rm [Fe/H]} = -2.0$, our derived reddening value for 
NGC~6388 would increase by 0.05~mag, giving a reddening significantly larger 
than found by most other methods.

It should be emphasized that the type of RRL in NGC~6388 may be different in 
nature from those which Blanco used in establishing his relationship between 
metallicity, period, and intrinsic color.  As was done for NGC~6441, we have 
assumed that Blanco's (1992) formula applies to the RRab stars in NGC~6388.  
This may not be the case if, as is presented in this paper, the RRab stars 
in NGC~6388 are unusually bright for their metallicity, although the reddening 
derived in this study matches well with those from other studies.

\subsection{Cepheids}
Silbermann et al.\ (1994) found a bright variable, V29, which they believed to 
be a foreground RRL.\@  We find this star together with V18 from Hazen \& 
Hesser (1986), and two additional variables to be P2Cs.  The mean 
properties of these variables are listed in Table~8.  V18 was listed  
by Hazen \& Hesser as a star with a period $< 2$ days.  
Of the 4 Cepheids found, 3 have periods of less than 10 days, making them 
members of the subset of P2Cs  
known as BL Herculis stars.  As was noted in \S4.3, V37 has 
a period of around 10 days, classifying it as a W Virginis-type 
P2C.\@  Although V29 was saturated in $V$, using light curve fitting 
programs created by Andrew Layden 
(Layden \& Sarajedini 2000, and references therein) we were able to find 
$\langle B \rangle = 16.035$.  From the zero-point shift (\S4.3), V29 seems to 
be oddly brighter than V18 and V36.  The reason for this difference is 
uncertain, although crowding may be an effect.  The properties of the Cepheids 
are discussed further in \S5.3.

\subsection{Eclipsing Binaries and LPVs} 

We were able to find a number of eclipsing
binary stars within our field of view, which are not likely to be members 
of NGC~6388.  Only V14 listed by Hazen \& Hesser (1986) was recovered.  
Table~9 lists 
photometric data for the binary stars.  Due to our sampling it was 
somewhat difficult to determine accurate periods for detached binaries.  

The time coverage of our observations was not suitable for the detection of 
LPVs.  Only two of the previously suspected variables 
in NGC~6388 were determined to be LPVs by this survey, V4 and V12.  Three 
additional LPVs were found, V45, V46, and V47.  The LPVs 
found by this study and their locations can be found in Table~2.

\begin{figure*}[t] 
  \centerline{\psfig{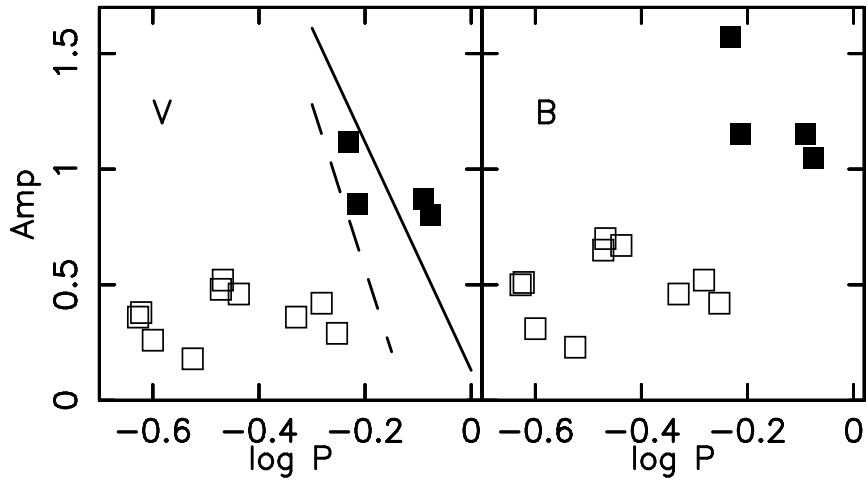}} 
  \caption{Period-amplitude diagram for NGC~6388 in $V$ and 
           $B$ showing fundamental mode RR~Lyrae (filled squares) and first 
           overtone RR~Lyrae (open squares).  The dashed and solid lines 
           represents the Oosterhoff types~I and II lines, respectively, as 
           given by Clement (2000; private communication).} 
  \label{Fig06}
\end{figure*}

\section{Discussion}

\subsection{General Properties of NGC~6388 RR Lyrae} 

We compare the average properties of the RRL in NGC~6388 and NGC~6441 to those 
of RRL in Oosterhoff types~I and II clusters, M3 and M15, in Table~10.  The 
values for M3 and M15 are taken from Table 3.2 of Smith (1995).  For 
comparison, Clement et al.\ (2001) found the mean periods for the RRL in 
Galactic globular clusters to be 0.559~d (RRab) and 0.326~d (RRc and RRd) for 
Oosterhoff type~I clusters and 0.659~d (RRab) and 0.368~d (RRc and RRd) for 
type~II clusters.  The RRab in 
NGC~6388 have unusually long periods for a metal-rich cluster, as was shown 
in Pritzl et al.\ (2000).  From what is known of the periods of metal-rich 
field RRL, one would expect the mean period of the RRab stars in NGC~6388 to 
be even shorter than those from Oosterhoff type~I clusters.  In fact, the mean 
period of the RRab in NGC~6388 is longer than in typical Oosterhoff type~II 
clusters, as was also the case for NGC~6441.  It is also interesting to note 
the unusually high $N_{\rm c}/N_{\rm RR}$ ratio (where $N_{\rm c}$ is the 
number of RRc stars and $N_{\rm RR}$ is the total number of RRL in the system) 
for NGC~6388.  Even if, as discussed previously, the RRc stars V26 and V34 are 
non-members, this ratio would remain high:  $N_{\rm c}/N_{\rm RR} = 0.57$. 

In Figure~8 of Paper~I, we showed how the trend of decreasing period with 
increasing metallicity for the Oosterhoff types~I and II globular clusters is 
broken by the mean periods of the RRab stars in NGC~6388 and NGC~6441.  It 
appears, as was concluded by Pritzl et al.\ (2000), that NGC~6388 does not 
fall into either Oosterhoff group according to its metallicity.  However, it 
has been suggested that the Oosterhoff dichotomy may be due to evolutionary 
effects.  
Lee, Demarque, \& Zinn (1990) postulated that the RRL in Oosterhoff type~II 
clusters have evolved away from a position on the zero-age HB (ZAHB) on 
the blue side of the instability strip.  This leads to the RRL in 
Oosterhoff type~II clusters having longer periods and higher luminosities than 
the RRL in Oosterhoff type~I clusters, whose ZAHBs are thought to be more 
heavily populated in general.  
This idea has been used to argue that the location of RRab stars in the 
period-amplitude diagram may be more a function of Oosterhoff type than 
metallicity (Clement \& Shelton 1999; 
Lee \& Carney 1999; Clement \& Rowe 2000).  While most of the RRab in NGC~6441 
fall along the Oosterhoff type~II line as defined by Clement (2000; private 
communications; see Fig.~7 in Paper~I), the RRab in NGC~6388 appears to be
scattered in the 
period-amplitude diagram (see Figure~6).  Two of the RRab fall at longer 
periods than given by the Oosterhoff type~II line, while two others fall 
in-between or near the Oosterhoff type~I line (see \S5.7 for further 
discussion).

\begin{figure*} 
  \centerline{\psfig{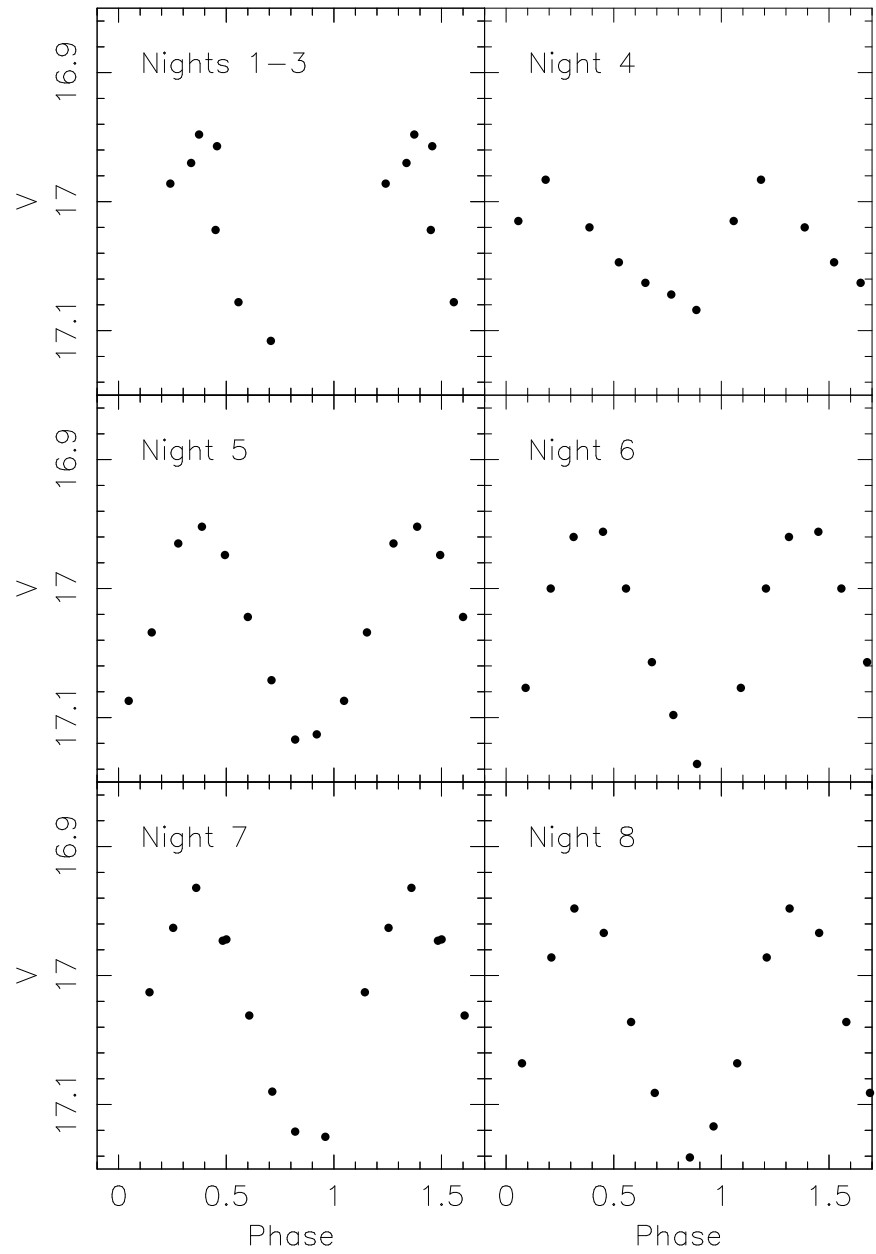}} 
  \caption{Nightly light curves for V35.  Night~4 appears to have a different 
           amplitude and phase than the other nights.}
  \label{Fig07}
\end{figure*}

The derived properties of the RRc stars can also give some indication to the 
Oosterhoff type of the cluster (e.g., Table~4 of Clement \& Rowe 2000).  Using 
the Fourier parameters in Table~6 along with Eqs.\ 2-5 in Clement \& Rowe, 
which derive from Simon \& Clement (1993a, 1993b), we calculated the masses, 
luminosities, temperatures, and absolute magnitudes listed in Table~11 for the 
RRc in NGC~6388.  Only those RRc's with errors in the $\phi_{31}$ measurement 
less than 0.50 and periods less than 0.44d were included.  The mean values for 
the mass, $\log\,(L/L_{\odot})$, $T_{\rm eff}$, and $M_V$ are $0.48\,M_{\odot}$, 
1.62, 7495~K, and 0.82.  As with the derived properties of the RRc variables 
in NGC~6441, the masses for the RRc stars in NGC~6388 are low (see Paper~I for 
further discussion).  Placing NGC~6388 in Table~4 of Clement \& Rowe according 
to the mean $\log\,(L/L_{\odot})$ and $T_{\rm eff}$ values, we 
see that NGC~6388 falls among the Oosterhoff type~I clusters.  This contradicts 
what was indicated by the location of the RRab stars in the period-amplitude 
diagram.  These results further illustrate the difficulty in classifying 
NGC~6388, along with NGC~6441.  

We also calculated the properties for the RRab stars using the 
Jurcsik-Kov\'{a}cs 
method (Jurcsik \& Kov\'{a}cs 1996; Kov\'{a}cs \& Jurcsik 1997).  After 
correcting the Fourier parameters in Table~6 to work in the sine-based 
Jurcsik \& Kov\'{a}cs equations, we derived the parameters shown in Table~12 
using Eqs.\ 1, 2, 5, 11, 17, and 22 from Jurcsik (1998), correcting the values 
of $\log\,(L/L_{\odot})$ and $\log\,T_{\rm eff}$ by +0.1 and +0.016 
(Jurcsik \& Kov\'{a}cs 1999).  It should be noted that Eqs. 17 and 22 were 
not meant to calculate the $\log\,(L/L_{\odot})$ and mass values for 
individual stars, but we list the values for illustrative purposes.  The mean 
values for the mass, $\log\,(L/L_{\odot})$, $\log\,T_{\rm eff}$, $M_V$, and 
[Fe/H] are 0.56~$M_{\odot}$, 1.69, 3.82, 0.66, and -1.21, respectively.  When 
compared to the data in Figure~1 of Jurcsik 
\& Kov\'{a}cs (1999), the mean value of $\log\,(L/L_{\odot})$ for the RRab in 
NGC~6388 is about 0.1 brighter than their data at a metallicity of 
${\rm [Fe/H]} = -0.60$ which agrees with the idea that the RRL in NGC~6388 
are unusually bright for the metallicity of the cluster.  Meanwhile, the mean 
$\log\,T_{\rm eff}$ values are about 0.02 lower and the mean mass is consistent 
with the data given in Jurcsik \& Kov\'{a}cs (1999).  The derived mean value 
for $M_V$, which does not include the $\log\,(L/L_{\odot})$ correction of 
$+0.1$ from Jurcsik \& Kovacs (1999), is about 0.49 mag brighter than given 
by Eq.\ 5 in Kov\'{a}cs \& Jurcsik (1996) using ${\rm [Fe/H]} = -0.60$. 

Similar to what we found for the RRab in NGC~6441 (Paper~I), the metallicity 
derived from the RRab in NGC~6388 is unusually low, ${\rm [Fe/H]}=-1.2$ 
($-1.4$ on the Zinn \& West 1984 scale).  It is uncertain how well the 
Jurcsik-Kov\'{a}cs method applies to RRab of unusually long periods.  
Moreover, it is hard to reconcile the 
RGB morphology of NGC~6388 with that of more metal-poor globular clusters 
(Raimondo et al.\ 2002).  While a lower metallicity may explain the blue 
extension to the HB, it would not explain the red clump.  We discuss the 
possibility of a metallicity spread in \S5.6. 

In Figure~8 we present an updated histogram for the period distribution of 
the RRL in NGC~6388 according to their period.  As seen in Pritzl et al.\ 
(2000), this provides an additional way to demonstrate how the properties of 
the RRL contradict the metallicity of the parent cluster.  The number 
of RRc in NGC~6388 is high even when compared to Oosterhoff type~II clusters 
rather than being the relatively small number expected for the more 
metal-rich Oosterhoff type~I clusters. 

\begin{figure*} 
  \centerline{\psfig{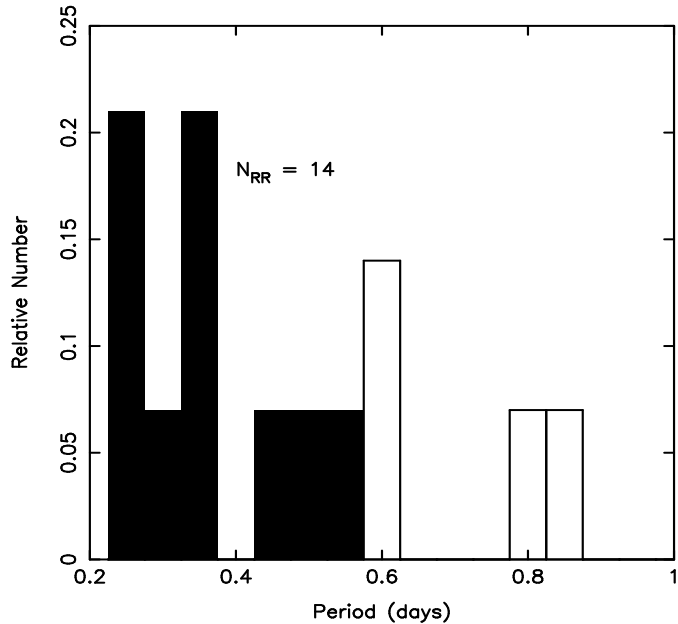}} 
  \caption{Period distribution histogram for the RR~Lyrae in NGC~6388.  The 
           dark area is occupied by c-type RR~Lyrae.  The light area is 
           occupied by ab-type RR~Lyrae.  Only RR~Lyrae with certain 
           classification are included.}
  \label{Fig08}
\end{figure*} 

Another way in which NGC~6388 is distinguished from the typical Oosterhoff 
groups is the high ratio of long period ($P>0.8$d) RRab.  Although there 
is only a small number, 50\% (2 of 4) of the RRab have longer periods.  The 
large proportion of long period RRab, a property shared by NGC~6441 (Pritzl et 
al.\ 2000, 2001), is not seen in other globular clusters, with the exception 
of NGC 5897 (Clement \& Rowe 2001).

\subsection{RRc Variables} 

The existence of RRL is unusual in globular clusters as 
metal-rich as NGC~6388.  The presence of RRc is especially unusual due to 
the typically red HB morphology associated with metal-rich globular clusters.  
This provides a unique opportunity to investigate the 
properties of these stars in an environment not seen in other globular 
clusters.  Although the RRc stars in NGC~6388 span a wide range of periods, 
the mean period is 0.36d which is comparable to the value for a typical 
Oosterhoff type~II cluster (see Table~10).

As was noted with NGC~6441, the RRc stars in NGC~6388 have some distinguishing 
features.  One such feature is the bump in the light curve seen during 
the rise to maximum brightness.  While this bump occurs at a phase of 
about 0.2 before maximum for the RRc with periods between 0.33~d and 0.37~d, 
the RRc with periods greater than 0.4~d have the bump occurring nearer phase 0.3 
before maximum.  No bumps are seen in the light curves for the RRc with 
periods less than 0.31d.

Layden et al.\ (1999) originally pointed out that the RRc stars in NGC~6441 
may exhibit longer than usual rise times.  Similar to NGC~6441 (Layden et al.\ 
1999; Paper~I), the range of rise times for the RRc stars in 
NGC~6388 is about 0.45 to 0.50 in phase with the longer period stars having 
longer rise times.  Compared to the range listed in Layden et al.\ for RRc 
stars from the {\it General Catalog of Variable Stars} (Kholopov 1985), 
0.32 to 0.46, the rise times seen in the RRc stars of NGC~6388 are somewhat 
longer than usual.

The RRL in NGC~6388 show an unusually short gap between the longest period 
RRc star and the shortest period RRab star, similar to NGC~6441 (Paper~I).\@  
Van Albada \& Baker (1973) found that a gap of about 0.12 should be expected 
between the logarithms of the periods of the shortest period RRab star and the 
longest period RRc star, assuming that the RRab and RRc stars have the 
same mass and luminosity, and that there exists a single transition line 
in effective temperature between the first overtone and fundamental mode 
pulsation domains in the HR diagram.  Many globular clusters show such a gap, 
indicating that these assumptions may hold for most globular clusters.  The 
absence of such a gap in NGC~6388 and NGC~6441 implies that one or more of 
the assumptions is incorrect in their case.  

The RRc star V35 deserves a special mention.  As noted in \S4.3 and 
seen in Figure~7, the phase and amplitude of the star are different on 
the night of 29 May than on the other nights.  
There are no nearby stars affecting the photometry and there is no
indication that night has any problems since the photometry for other 
variables shows no anomalies.  It may be the case that the star is going 
through a mode switch, mode instability, or exhibiting non-radial pulsations 
(e.g., M55:  Olech et al.\ 1999).  More photometry of V35 may help clear up 
this issue.

An interesting feature seen in NGC~6388, but not in NGC~6441, is the occurrence 
of ``short" period c-type RRL.\@  As seen in the period-amplitude diagram 
for NGC~6388 (Figure~6), the c-type RRL, with the exception of V35, seem to 
fall into three distinct groups:  The ``longer" period RRc centered at 
$\log P = -0.288$, the ``intermediate" period RRc centered at $\log P = 
-0.459$, and the ``shorter" period RRc centered at $\log P = -0.617$.  There 
does not appear to be any distinction between the shorter period RRc found 
in NGC~6388 and the more ``intermediate" period RRc, according to their Fourier 
parameters, but there is a clear distinction when the Fourier parameters of 
the longer period RRc are compared to those of the ``intermediate" period 
RRc (see \S5.5).  
The light curves of the shorter period RRc seem to show more scatter 
during maximum light as compared to the light curves of the other RRc members 
of NGC~6388, and are slightly more asymmetric, although the photometry 
obtained in this survey is not accurate 
enough to make a conclusive argument for this.  It is of interest to note that 
two of the three shorter period RRc stars are fainter than the other probable 
RRc of NGC~6388, as shown in Figure~3 and discussed in \S4.3.  It cannot be 
determined if this effect 
is due to the differential reddening in NGC~6388, but it can be said that these 
two RRc stars, which also happen to be the shortest period RRc, do not fall in 
the same part of the NGC~6388 field.  

It has been argued by some authors that short period RRc stars with 
$P<0.35$d and amplitudes less than 0.3~mag may in 
fact be pulsating in the second overtone mode (e.g., Clement, Dickens, \& 
Bingham 1979; 
Walker 1994; Walker \& Nemec 1996).   The MACHO collaboration 
found a maximum in the Large Magellanic Cloud RRL period distribution at 
0.28 days (Alcock et al.\ 
1996), arguing that this corresponded to the second overtone, RRe, stars.  
Alcock et al.\ found that the RRL located about this range showed skewed 
light curves, as was modeled by Stellingwerf et al.\ (1987).  Clement \& Rowe 
(2000), using the RRc stars in $\omega$ Centauri (NGC~5139), argue that a 
number of the 
shorter period stars were pulsating in the second overtone mode.  They showed 
that these RRc stars also have low amplitudes ($A_V < 0.3$~mag).  Overall, 
the short period RRc stars in NGC~6388 do not have amplitudes as low as these 
stars.  However, if we disregard V26 and V34, which may be nonmembers (\S4.3), 
the remaining two RRc stars, V16 and V35, do have low amplitudes, making them 
similar in both period and amplitude to the stars in $\omega$ Cen which 
Clement \& Rowe identified as second overtone pulsators.  However, a case has 
also been made that short period RRL of this type are not second overtone  
pulsators.  Kov\'{a}cs (1998) argued that these variables are RRc variables 
at the short period end of the instability strip.  It is beyond the scope of 
this paper to argue for or against the classification of the shorter period 
RRc-type stars as second overtone pulsators.  In any case, further observations 
of these variables would help to improve pulsation models and help 
explain why such a large range in RRc periods exists (0.24 - 0.56 days) 
in NGC~6388.

\subsection{Cepheids} 

The occurrence of P2Cs in globular clusters is not uncommon.  Yet, if 
the probable Cepheids found in the field of NGC~6388 are indeed members 
of this cluster, NGC~6388 would be the most metal-rich globular cluster 
known to contain Cepheids.  A review by Harris (1985) listed the globular 
clusters containing Cepheid (16 GCs) or RV Tauri variables (5 GCs).  
Harris confirmed that the globular clusters which contained P2Cs also 
have blue horizontal branches (Wallerstein 1970).  It was further noted 
by Harris that BL Her, P2Cs with periods $< 10$ days, may be most 
frequent in clusters which have extended blue tails on the horizontal 
branch.  Smith \& Wehlau (1985) found that all of the globular clusters 
known to contain Cepheids have $B/(B+R) > 0.50$, where $B$ is the number 
of stars blueward of the RRL gap and $R$ is the number of stars redward 
of the RRL gap.  The reverse is not true, however.  All clusters with 
$B/(B+R) > 0.50$ do not contain Cepheids.  Smith \& Wehlau 
also noted that W Vir stars, P2Cs with periods $> 10$ days, tend 
to be in the most metal-rich of the globular clusters which have blue 
HBs.  Finally, it was also shown that the clusters with Cepheids 
are also the brighter, more massive, clusters, especially those 
clusters which have two or more Cepheids.  In a summary of the variable 
stars in Galactic globular clusters, Clement et al.\ (2001) use the 
HB ratio $(B-R)/(B+V+R)$, where $B$ and $R$ follow the above definitions and 
$V$ is the number of RRL, finding that most globular 
clusters with P2Cs have $(B-R)/(B+V+R)>0$, with two exceptions: NGC~2808 
and Palomar~3.  Pal~3 provides the most serious 
challenge to explain since it lacks any 
blue extension to the HB (Borissova, Ivanov, \& Catelan 2000).

NGC~6388 does exhibit some of the features listed above.  It does have 
a blue component to its HB.\@  Although the $B/(B+R)$ fraction was not 
determined for NGC~6388 in this study due to the high contamination from 
field stars, according to the number counts in Zoccali et al.\ (2000) 
$B/(B+R) = 0.15$.  Even though this disagrees with the idea that P2Cs are 
only found in globular cluster with $B/(B+R) > 0.50$, the relatively large 
number of BL~Her stars found in 
NGC~6388 agrees with Harris' (1985) idea that they are more 
frequent in clusters with extended blue tails.  The one candidate W Vir 
star in NGC~6388 follows the idea that these stars tend to be in the most 
metal-rich of the globular clusters with well-developed blue HBs.  NGC~6388 is 
also one of the 
brightest globular clusters known in the Galaxy, confirming the tendency 
of clusters containing Cepheids to be brighter than those that do not.  

There has been some debate on the P2C period-luminosity relations and 
on the question of whether or not P2Cs pulsate in 
both the fundamental and first overtone modes or solely in the fundamental 
mode (see Nemec, Nemec, \& Lutz 1994; McNamara 1995; and references therein).  
For this study, we would like to have an idea of the absolute magnitudes of 
the Cepheids in NGC~6388 in order to estimate the distance to the cluster 
and the absolute magnitude of the RRL.\@  Using Eqs.\ 7, 8, 11, and 12 from 
McNamara (1995), we calculated the absolute magnitudes listed in Table~13.  
It should be noted that the absolute magnitudes calculated for the P2Cs 
in Nemec et al., which were also used by McNamara, are based on the 
distance moduli and reddenings of the systems for which they are associated.  
Although there is some differential reddening across the face of NGC~6388, 
we use the mean reddening of the cluster, $E(\bv)=0.40$, in determining 
the extinctions, $A_V=3.2E(\bv)=1.28$ and $A_B=4.1E(\bv)=1.64$, to deredden 
the Cepheid magnitudes.  We want to restate that the mean magnitudes of these 
stars are somewhat uncertain due to the scatter and gaps in the light 
curves (see Figure~4).  Examining the resulting distances in Table~13, V29 
and V37 have distances much less than V18 and V36.  We noted previously 
(\S4.5) that V29 seemed unusually bright for its period.  The precise 
$\langle B \rangle$ magnitude is uncertain since the star is saturated in 
$V$.\@  For V37, the exact period is uncertain due to an incomplete light curve, 
and thus its magnitude may also be uncertain.  The scatter in the light curve 
also makes the magnitude uncertain.  It also could be that V29 and V37 are not 
members of NGC~6388, but this seems unlikely due to their proximity to the 
cluster center (see Figure~2).  The distance we find using the visual 
estimates of V18 and V36 is 10.6~kpc, which matches well with the distance 
estimates made from the RRL.  

If we adopt the distance of 10.6~kpc and $A_V=1.28$, we find 
$M_{V,{\rm RR}}=+0.44$ for $V_{\rm RR}=16.85$~mag.  This value is brighter than 
one would expect for the metallicity of NGC~6388.  Using the $M_{V,{\rm RR}}$ 
relations from Nemec et al.\ (1994) and ${\rm [Fe/H]}=-0.60$, one finds the 
absolute magnitudes of the RRL in NGC~6388 to be approximately +1.00.  This is 
much fainter than would would expect given that the P2C period-luminosity 
relation derives from their data.  However, given 
${\rm [Fe/H]}=-0.60$ and using Eq.\ 7 in Lee et al.\ (1990), 
we find $M_{V,{\rm RR}}=+0.72$ on their scale.  

An interesting question to ask is:  Why does NGC~6388 contain Cepheid stars, 
but NGC~6441 does not?  It is possible that our survey was 
incomplete in finding any Cepheids in NGC~6441 (Paper~I), but this does not 
seem to be 
the case since no Cepheids were found in the survey of Layden et al.\ 
(1999), either.  Assuming that our survey 
was complete, and no Cepheids occur in NGC~6441, the answer to this question 
may give hints as to the origin of the Cepheids.  Both clusters 
are among the brightest known and both have similar HB morphologies.  Along 
with having similar metallicities, it would seem that if one 
of this pair of clusters contained Cepheids, the other would have them too.  
It was suggested by Smith \& Wehlau (1985), from 
the models of Mengel (1973) and Gingold (1976), that P2Cs may 
evolve from horizontal branch stars which already have low envelope masses.  
Sweigart \& Gross (1976) predicted that 
clusters with blue horizontal branches \textit{and} higher metal abundances 
would produce horizontal branch stars with especially low envelope masses.  
It can be seen in the CMDs for NGC~6388 and NGC~6441, by 
Rich et al.\ (1997), that the blue ``tail" in NGC~6388 appears to be more 
populated than in NGC~6441.  This may explain the difference between NGC~6388 
and NGC~6441, although it may just be a selection effect since we are talking 
about such small numbers.

\subsection{Period-Amplitude Diagram} 

We revisit the period-amplitude diagram in Figure~9 and compare the RRab in 
NGC~6388 to the RRab in other globular clusters (M15: Silbermann \& Smith 1995, 
Bingham et al.\ 1984; M68: Walker 1994; M3: Carretta et al.\ 1998; 47~Tuc: 
Carney, Storm, \& Williams 1993) and metal-rich field RRab.  As discussed in 
Pritzl et al.\ 
(2000), the RRab in NGC~6388 fall at unusually long periods compared to field 
RRab of similar metallicities.  Similar to NGC~6441 (Paper~I), NGC~6388 breaks 
the trend of increasing period with decreasing metallicity for a given amplitude.  
Although no direct measurements of the metallicity for the RRL in NGC~6388 are 
available (see \S5.6), it is interesting to see that V9 of 47~Tuc 
(${\rm [Fe/H]}=-0.76$) falls in the same general location as the RRab in 
NGC~6388 and NGC~6441, though shifted away from the mean locus occupied by 
the RRL in these two globulars by $\Delta\log\,P \approx +0.1$ at fixed 
amplitude.

\begin{figure*} 
  \centerline{\psfig{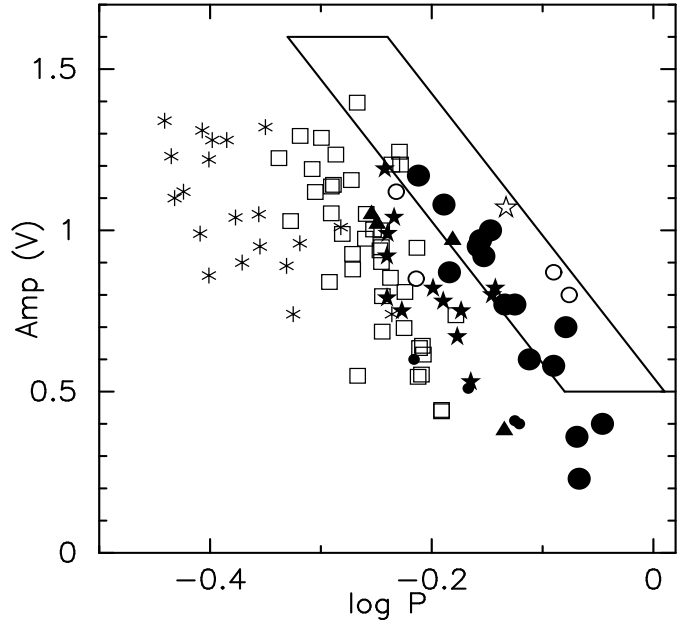}} 
  \caption{Period-amplitude diagram for the ab-type RR~Lyrae of NGC~6388 
           (open circles) as compared to field RR~Lyrae of 
           ${\rm [Fe/H]} \ge -0.8$ (asterisks), V9 in 47~Tucanae (open star), 
           M3 (open boxes), M15 (filled stars), and M68 (filled triangles).  
           The filled circles represent the RRab stars in NGC~6441 from 
           Pritzl et al.\ (2001).  The boxed area, taken from Figure~9 of 
           Layden et al.\ (1999), denotes the region predicted by Sweigart \& 
           Catelan (1998) where the RR~Lyrae should be located according to a 
           helium-mixing scenario.} 
  \label{Fig09}
\end{figure*} 

The RRab in NGC~6388 even fall at longer periods when compared to the RRab 
in the Oosterhoff type~II cluster M15.  This seems to contradict our previous 
finding that the RRab in NGC~6388 are scattered about a singular locus for all 
Oosterhoff type~II globular clusters as given by Clement 
(2000, private communication) (see \S5.1).  An important tool in creating these 
lines was the compatibility condition (Jurcsik \& Kov\'{a}cs 1996) which 
helped in determining which light curves are ``normal" or not.  An analysis 
done in Paper~I (cf.\ \S5.3) of the sample of RRab in M15 showed that a few 
of these stars which satisfy the compatibility condition fall between the 
Oosterhoff 
lines given by Clement.  It is interesting to note that only V22 of NGC~6388 
is close to complying with the compatibility condition (column~9 of Table~6).  
Although our time-coverage is not long enough to observe the Blazhko Effect, it 
would seem unlikely that all of the RRab in NGC~6388 would show this effect 
given that only about 20\% of the RRab stars are thought to show the Blazhko 
Effect among the field and globular cluster populations alike.  Also, the 
number fraction possibly decreases with increasing metallicity (see 
Table~5.3 in Smith 1995 and references therein). 

Another way to examine how the RRab in NGC~6388 compare to those in M15 is 
to analyze the period shift of NGC~6388 from M3.  Due to the small number of 
RRab stars in NGC~6388, it is difficult to make a general statement on the 
period shift.  Two of the RRab, V17 and V22, appear to have period shifts 
close to those for the RRab in NGC~6441 ($\sim 0.08$ in $\log\,P$; see \S5.3 
in Paper~I), but the other 
two, V21 and V28, have period shifts more on the order of 0.15 implying they 
are somewhat brighter than the other RRab in NGC~6388 and NGC~6441. 
As can be seen from Figure~9, this property is shared by 
V9 in 47~Tuc. However,  
our photometry does not indicate that V21 and V28 are unusually bright compared 
to the other RRab.

\subsection{Comparisons to Long Period RR Lyrae in $\omega$ Centauri} 

It has been shown that $\omega$ Cen is a globular cluster 
containing mostly RRL similar to those in Oosterhoff type~II clusters 
along with a smaller population similar to RRL in Oosterhoff type~I 
clusters (Butler, Dickens, \& Epps 1978; Caputo 1981; Clement \& Rowe 2000).  
It is interesting to note that $\omega$ Cen also contains a 
number of longer period RRL similar to what is seen in NGC~6388 and which are 
not typically found in either Oosterhoff type~I or Oosterhoff type~II 
clusters.  These longer period ($P>0.8$~d) stars are similar in both period 
and amplitude to the long period RRab in NGC~6388 and NGC~6441, although 
long period RRab form a much smaller fraction of the total RRab population 
in $\omega$~Cen than in those two clusters. 

In addition to the longer period RRab stars, $\omega$ Cen contains 
a number of longer period RRc, similar to V20, V32, and V33 in 
NGC~6388.  In Paper~I we showed how the longer period RRc formed 
a distinct group at lower $\phi_{21}$ in a $\phi_{21}$ vs.\ 
$A_{21}$ plot for $\omega$ Cen (see \S5.4, and Figure~11 in Paper~I).\@  
This trend was also seen in NGC~6441.  An examination of the $\phi_{21}$ 
and $A_{21}$ values for the longer period RRc in NGC~6388 (Table~6) 
shows that the three RRc which stand apart from the rest of the RRc 
at shorter $\phi_{21}$ values in Figure~5 are indeed the three 
longer period stars.

In Figure~10, we plot a sample of $\omega$ Cen RRc stars 
according to their metallicity.  The periods were taken from 
Kaluzny et al.\ (1997).  When there was more than one entry for 
a single star, the values were averaged.  The [Fe/H] values come 
from Rey et al.\ (2000).  The RRc classifications were taken from 
Butler et al.\ (1978).  There appears to be a slight trend of 
increasing metallicity with decreasing period and increasing amplitude.  
The RRc of NGC~6388 in the period range 
of 0.3d to 0.4d fall among the more metal-rich RRc of $\omega$ 
Cen, while the longer period ones ($P>0.45$d) fall among some of the RRc 
in $\omega$ Cen which have a more ``intermediate" metallicity.  
Two other longer period RRc are included in this plot, V70 in 
M3 (Kaluzny et al.\ 1998; Carretta et al.\ 1998) and V76 in M5 
(Kaluzny et al.\ 2000).  Although the period of V76 is somewhat 
shorter than the periods of the other long period stars, it is unusually long 
when compared to the other RRc stars in M5.  We emphasize that such 
long period RRc variables are extremely rare in either Oosterhoff type~I or 
Oosterhoff type~II globular clusters, no additional examples being known to 
exist besides V70 in M3 and V76 in M5.  

\begin{figure*} 
  \centerline{\psfig{figure=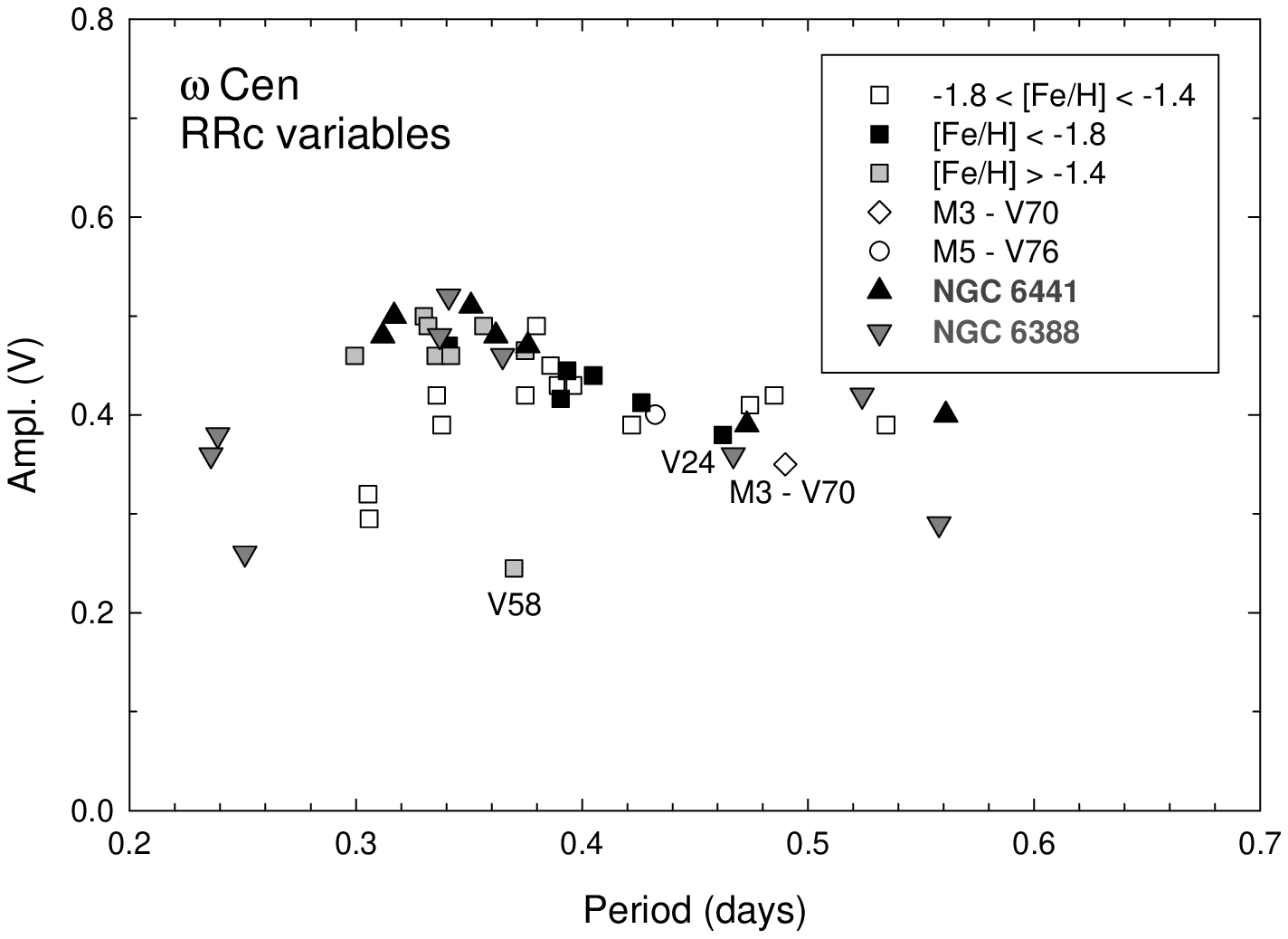,height=3.25in,width=4.25in}} 
  \caption{Period-amplitude diagram for the c-type RR~Lyrae in $\omega$ Cen 
           in comparison to those found in NGC~6388 and NGC~6441.  V35 has 
           been left out of this diagram due to the problems discussed in 
           \S5.2.} 
  \label{Fig10}
\end{figure*} 

The unusual nature of NGC~6388 and NGC~6441 is shown again by the 
occurrence of these longer period RRc stars.  More importantly, as was 
discussed above, it may be that these stars hint at a more 
metal-poor component in NGC~6388 (and NGC~6441), with properties similar 
to the corresponding ones that give rise to the $\omega$ Cen long-period 
RRc's.  In this paper we have assumed 
that the RRL in NGC~6388 have the same metallicity as the 
overall cluster value since no direct metallicity measurements 
for the RRL currently exist.

\subsection{A Metallicity Spread in NGC~6388?} 

The possibility of a spread in metallicity in NGC~6388 and NGC~6441 
was first proposed by Piotto et al.\ (1997).  More recently Sweigart 
(2002) explored the metallicity-spread scenario through stellar 
evolution calculations.  The models showed that the upward slope 
in the HB seen in NGC~6388 and NGC~6441 can be produced by a spread 
in metallicity assuming that all of the stars are coeval and that 
the mass loss efficiency, as measured by the Reimers (1975) mass loss 
parameter, is independent of [Fe/H].  (Please see Paper~I for an 
in-depth discussion on the metallicity spread issue in this cluster 
and NGC~6441.)  

In an attempt to produce synthetic CMDs for $\omega$ Cen, Ree et al\ 
(2002) also attempted to model the CMDs of NGC~6388 and NGC~6441.  
Their population models showed that two distinct populations could be 
contained within the broad RGBs of the two clusters.  In addition, the HB 
could be explained by adding an older, metal-poor HB and a younger, metal-rich 
HB, with the age and metallicity spread being about 1.2~Gyr and 0.15~dex, 
respectively.  
Ree et al.\ went on further to suggest that if there is indeed a metallicity 
spread in NGC~6388 and NGC~6441, these two clusters may in fact be 
relics of disrupted dwarf galaxies as has been suggested for $\omega$ 
Cen and M54.  The one conundrum with this hypothesis is that, unlike 
$\omega$ Cen, where you have a mostly metal-poor population 
with some metallicity spread towards higher metallicities, for NGC~6388 and 
NGC~6441 one would have a mostly metal-rich population with a small metal-poor 
population.  Their synthetic CMDs also failed to reproduce the sloping nature 
of the HB.\@   In order to have self-enrichment up to such a high 
metallicity, a progenitor galaxy with which NGC~6388 and NGC~6441 may have 
been associated would have to have been much more massive than the 
Sagittarius dwarf spheroidal, or than the conjectured dwarf spheroidal 
progenitor of $\omega$ Cen.

Recently, Raimondo et al.\ (2002) argued that all metal-rich globular clusters 
should have tilted red HBs.  In discussing the specific cases of NGC~6388 and 
NGC~6441, they felt that the slopes in the red HBs were real and not artifacts 
of differential reddening.  Raimondo et al.\ also used the point where the 
HB magnitudes increase to make the blue tail as defined in Brocato et al.\ 
(1998) to match up more metal-poor globular clusters with well defined 
blue tails to NGC~6388.  Matching the CMDs in this way allowed them to 
argue that the blue component of the HB of 
NGC~6388 cannot derive from metal-poor 
progenitors since there is not a corresponding large number of RGB stars 
blueward of the given RGB of the cluster.  This 
provides an important constraint and challenge to the possibility of the 
blue HB in NGC~6388 and NGC~6441 being metal-poor.  However, it should be 
noted that if NGC~6388 and NGC~6441 have metallicity spreads rather than 
two distinct populations, the metal-poor RGB stars would scatter 
over a larger area blueward of the metal-rich RGB.\@  This would serve to 
make the metal-poor giant stars less conspicuous than in the case where 
you have two distinct metallicities.  In any case, more studies into this 
hypothesis should be done and clearly, metallicity determinations need to be 
made of the RRL in NGC~6388 and NGC~6441 to resolve this issue.  It should be 
noted that Raimondo et al.\ did not address the presence of long-period RRL in 
NGC~6388 and NGC~6441, and did not provide a physical explanation for the 
presence of blue HB stars in these clusters.

\subsection{Are NGC~6388 and NGC~6441 Oosterhoff Type~II Globular Clusters?  
Evolutionary Constraints} 

Our preceding discussion has indicated that there are problems with the 
unambiguous classification of NGC~6388 and NGC~6441 into either Oosterhoff 
group (see also Pritzl et al.\ 2000, 2001). In a recent analysis of this 
problem, Clement et al.\ (2001) suggested that the period-amplitude relation 
of the RRab variables in NGC~6441 is most consistent with that cluster being 
classified as an Oosterhoff type~II system (see also Walker 2000). 
As we have seen, the RRab 
variables in NGC~6388 scatter about the Oosterhoff type~II region with some 
falling near the Oosterhoff type~I regions of the period-amplitude diagram.  
Several authors (Lee et al.\ 1990; Clement \& Shelton 1999; Clement et al.\ 
2001) have proposed that evolution from ZAHB positions blueward of the 
instability strip may play an important role in determining the pulsational 
properties of RRL in Oosterhoff type~II clusters. We thus need to consider 
whether the RRL in NGC~6388 and NGC~6441 could have evolved in this fashion 
from blue HB progenitors. To do this, we must address possible problems with 
the general evolutionary scenario in which RRL in Oosterhoff 
type~II globular clusters are all evolving to the red from ZAHB positions on 
the blue HB. \@ 

This evolutionary scenario for Oosterhoff type~II RRL raises 
the following key questions: (1) Do blue HB stars spend enough time within 
the instability strip as they evolve redward to the asymptotic giant branch 
to produce the observed numbers of RRL in Oosterhoff type~II globular 
clusters?\@  (2) Does the 
predicted period-effective temperature relation for a given cluster depend 
on the stellar distribution along the blue HB?\@  (3) Is the predicted 
period-effective temperature relation independent of metallicity? 

Renzini \& Fusi Pecci (1988) and Rood \& Crocker (1989) have argued that blue 
HB stars do not spend sufficient time in the instability strip to account for 
the observed number of RRL in the Oosterhoff type~II clusters.  We 
decided to reanalyze these results by investigating the predicted loci and 
evolutionary timescales of blue HB models as they evolve across the 
instability strip in both the Hertzsprung-Russell (HR) and period-temperature 
diagrams.  Given the range in metallicities among Oosterhoff type~II clusters, 
and also the possibility of a metallicity spread in NGC~6388 and NGC~6441, we 
considered four metallicity values, namely: $Z = 0.0005$, 0.001, 0.002 and 
0.006 which correspond to ${\rm [Fe/H]}$ values of $-1.93$, $-1.63$, $-1.33$, 
and $-0.85$ for ${\rm[}\alpha{\rm/Fe]}=0.48$ and $-1.57$, $-1.27$, $-0.97$, 
and $-0.50$ for ${\rm[}\alpha{\rm/Fe]}=0.00$.  The ${\rm[}\alpha{\rm/Fe]}=0.48$ 
case is appropriate for the $Z = 0.0005$, 0.001, and 0.002 values.  
For $Z=0.006$, the alpha elements may be somewhat less enhanced than at 
${\rm [Fe/H]} < -1$.  In this case, a more appropriate value for [Fe/H] may be 
somewhere between the two values given.  (Although the lowest value of $Z$ 
does not correspond to the lower metallicities of globular clusters such as 
M15, we will see below that our conclusions would be essentially unchanged 
had we extended the range of metallicities considered to even lower $Z$ 
values.)  A helium abundance $Y_{\rm MS} = 0.23$ was assumed in all cases. 
The new HB sequences computed for this analysis are based on the same 
assumptions as the models of Sweigart \& Catelan (1998). In addition, we 
assumed an instability strip width $\Delta \log\, T_{\rm eff} \sim 0.085$ 
(Smith 1995), and the fundamental pulsation equation from van Albada \& Baker 
(1971). 

At any given metallicity, there is a critical mass $M_{\rm HB,ev}$ above which 
stars will evolve to the blue while crossing the instability strip, but below 
which stars will cross the instability strip only during their final redward 
evolution back to the asymptotic giant branch.  Therefore, $M_{\rm HB,ev}$ 
defines the reddest possible HB morphology that a globular cluster could have 
and still produce exclusively redward-evolving RRL.\@  In practice, 
one should expect the actual HB morphology of any globular cluster in which 
all RRL are indeed evolving to the red to be much bluer than 
defined by $M_{\rm HB,ev}$, due to the presence of lower-mass stars---in other 
words, ``real clusters'' do not have delta functions for their HB mass 
distributions.  They presumably cannot have higher HB masses than 
$M_{\rm HB,ev}$ or else they would not be Oosterhoff type~II in the 
evolutionary sense described above.  Table~14 gives the variation of 
$M_{\rm HB,ev}$ with $Z$, as well as some key number fractions in the 
resulting HR diagrams. 

\begin{figure*}
  \centerline{\psfig{figure=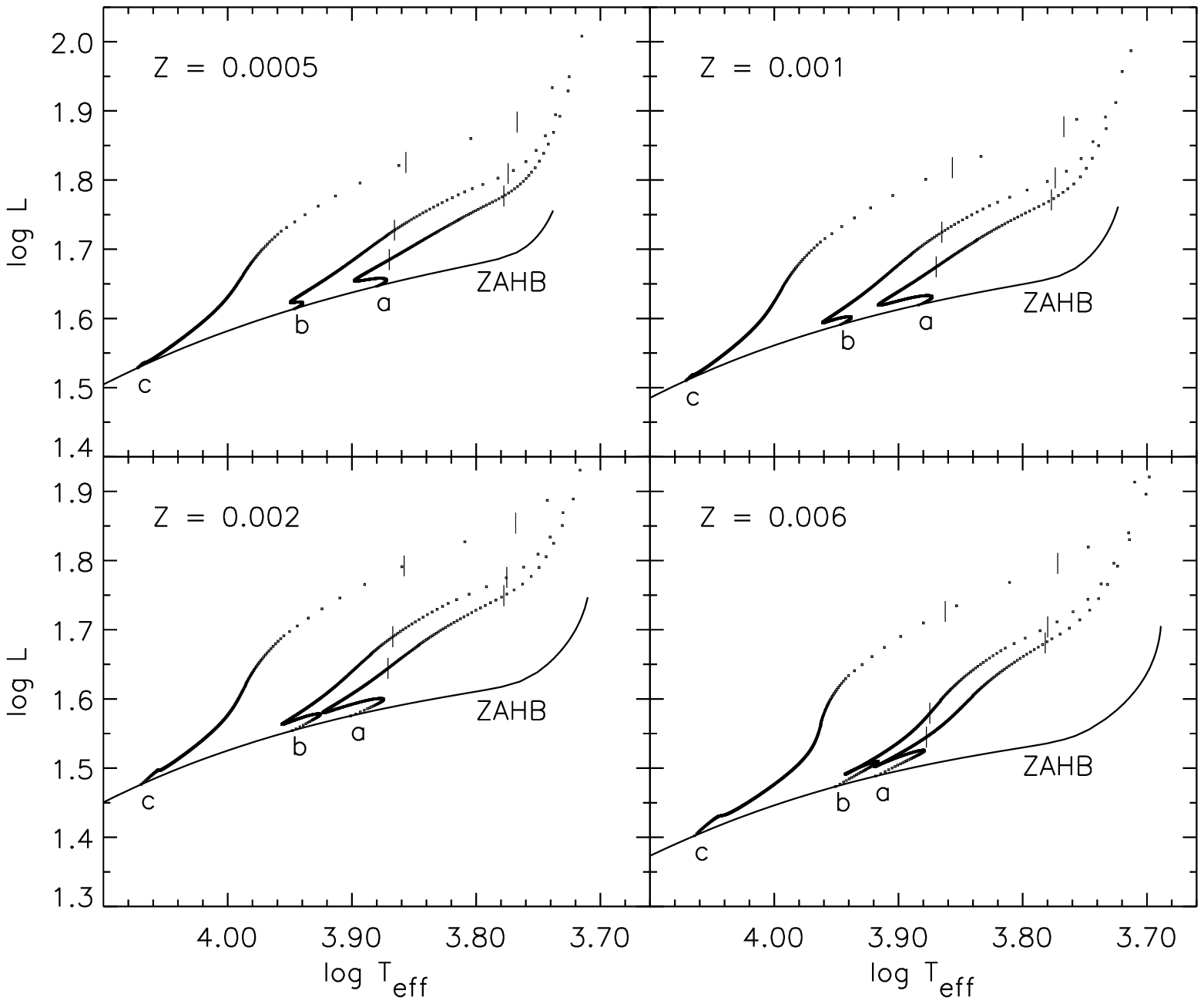,height=6in,width=7.0in}} 
  \caption{Evolutionary tracks for blue HB stars with heavy-element 
           abundances $Z$ of 0.0005, 0.001, 0.002 and 0.006. In each panel,
           label ``a" identifies the HB track corresponding to the 
           $M_{\rm HB,ev}$ value shown in Table~14. Label ``b" indicates the 
           HB track corresponding to a ZAHB position with $(\bv)_0 = 0$, 
           which is thecolor of the peak of the blue HB distribution in 
           NGC~6441; and label ``c" corresponds to a ZAHB position with 
           $(\bv)_0 = -0.1$, which is the color of the peak of the blue HB 
           distribution in NGC~6388.  Each track is plotted as a series of 
           points separated by a time interval of 200,000 yr in order to 
           indicate the rate of evolution.  The two thin vertical lines along 
           each track mark the blue and red edges of the instability strip. 
           The solid curves give the ZAHB locus for each metallicity.  Note 
           the sharp drop in the time spent within the instability strip 
           between tracks ``a" and ``c."} 
  \label{Fig11}
\end{figure*} 

Figure~11 shows the actual HB tracks for specific ZAHB locations, where the 
tracks for $M_{\rm HB,ev}$ are labeled ``a''.  Tracks ``b" and ``c" start 
their evolution at a bluer position on the ZAHB, corresponding to 
$(\bv)_0 = 0$ and $-0.1$; these are estimated to represent the peak of the 
blue HB distributions in NGC~6441 and NGC~6388, respectively---cf.\ Fig.~7 
in Piotto et al.\ (1999).  To illustrate the rate of evolution, each HB track 
is plotted as a sequence of points separated by a time interval of 200,000~yr.  
The thin vertical lines along each track denote the blue and red edges of 
the instability strip.  It is obvious from Figure~11 that the bluest HB stars 
evolve rapidly through the instability strip when compared to those for 
$M_{\rm HB,ev}$.  This is even more clearly illustrated in Figure~12 where 
we plot the ratio of the time spent in the instability strip to the time 
blueward of the instability strip for all sequences with $M \le M_{\rm HB,ev}$ 
against the time-averaged value of $\log\,T_{\rm eff}$ over the part of the 
HB track blueward of the instability strip.  This choice for abscissa gives 
a better indication of where the stars are coming from on the blue HB than 
the ZAHB value of $\log\,T_{\rm eff}$.  Figure~12 shows how the fraction of 
the HB phase which a blue HB star spends in the instability strip decreases 
rapidly as $\log\,T_{\rm eff}$ increases.  Overall the variation of $V/B$ with 
$\log\,T_{\rm eff}$ is quite similar for all four metallicity values, 
although the maximum value of $V/B$ does depend somewhat on the metallicity.  
We conclude that, on the basis of their evolutionary rates, only stars within 
a relatively small range in $\log\,T_{\rm eff}$ just blueward of the 
instability strip have a significant chance of producing redward-evolving 
RRL, irrespective of the cluster metallicity.  In terms of Figure~11, this 
corresponds primarily to the region in-between ``a" and ``b," corresponding 
mostly to the color range $0 \lesssim (\bv)_0 \lesssim 0.2$. 

\begin{figure*}[t]
  \centerline{\psfig{figure=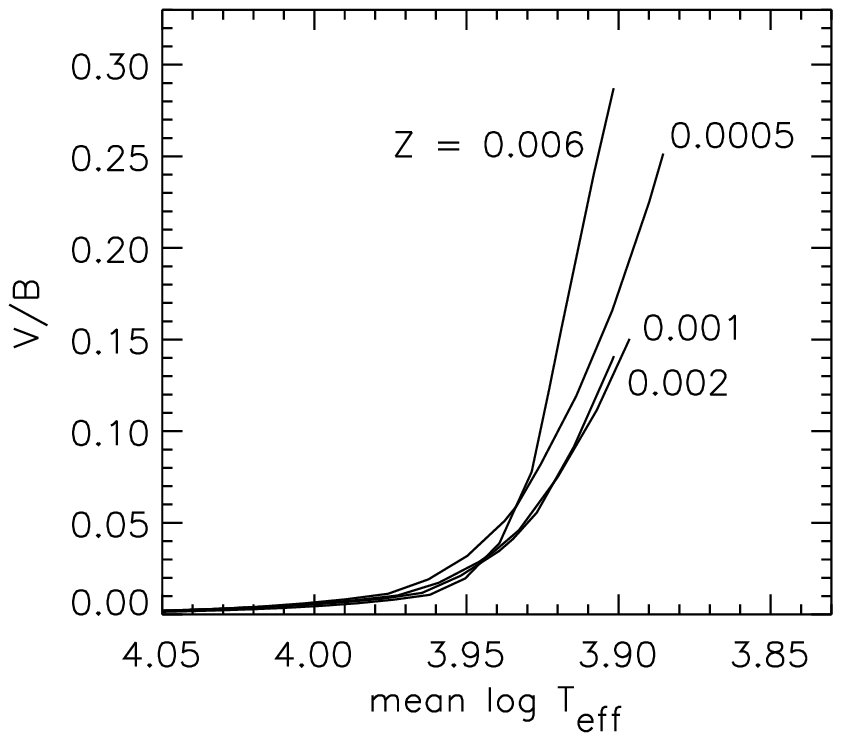}} 
  \caption{Number ratio $V/B$ of variables to blue HB stars predicted by HB 
           evolutionary tracks with heavy-element abundances $Z$ of 0.0005, 
           0.001, 0.002 and 0.006. The ordinate gives the ratio of the time 
           spent within the instability strip to the time spent blueward of
           the instability strip for blue HB tracks that only enter the 
           instability strip during the final evolution back to the 
           asymptotic giant branch.  The abscissa gives the time-averaged 
           value of $\log\,T_{\rm eff}$ over the part of the HB track 
           blueward of the instability strip. The low temperature end of each 
           curve is set by the mass $M_{\rm HB,ev}$ (see text). Note the sharp 
           decline in the fraction of the HB lifetime spent within the 
           instability strip as the mean HB effective temperature increases.} 
  \label{Fig12}
\end{figure*} 

The direct answer to the first question we posed at the beginning of this 
section is that only a relatively small portion of those HB stars blueward 
of the instability strip would spend enough time in the instability strip 
to be detected as RRL.  Therefore, in order to produce a rich 
Oosterhoff type~II RRL population, one would need to have a very strong 
population of HB stars just blueward of the instability strip.  This agrees 
with the conclusions of Renzini \& Fusi Pecci (1988) and Rood \& Crocker 
(1989), who point out that the redward evolving blue HB scenario 
might work for globular clusters which have relatively few RRL 
such as M92, but not for Oosterhoff type~II clusters with more substantially 
populated RRL instability strips, such as M15.  Based on our ground-based 
data, we find that NGC~6388 has a $V/B$ ratio of approximately 0.20 to 0.33 
depending on whether or not we include those RRL of uncertain classification.  
The highest ratio that can be derived from the HB simulations discussed below 
is approximately 0.08 to 0.085.  Therefore, it appears that not all of the RRL 
in NGC~6388 could have derived from redward evolving blue HB stars. 

In Figure~13 we have plotted the $\log\,P$-$\log\,T_{\rm eff}$ version of 
Figure~11.  The dots are again separated by 200,000 yr, and the pulsation 
periods were calculated from van~Albada \& Baker (1971).  This figure 
illustrates how the stars evolving from the blue HB occupy different loci in 
the $\log\,P$-$\log\,T_{\rm eff}$ diagram.  
It is important to bear in mind, in 
determining the period-temperature distribution for a given cluster, the 
strong dependence of the $V/B$ ratio on the HB $T_{\rm eff}$.  For example, 
there would have to be about 4 times as many ``b" stars as ``a" stars to 
produce the same number of RRL.\@  For the ``c" stars, the ratio is about 
50-100 times.  We also found that the locus of the corresponding tracks 
changes very little with metallicity for $Z=0.0005$, 0.001, and 0.002.  For 
these metallicities the decrease in mass with increasing $Z$ compensates for 
the decrease in luminosity so that there is little change in the period.  
Thus, for Oosterhoff type~II metallicities, the $\log\,P$-$\log\,T_{\rm eff}$ 
relation seems more a function of where the tracks come from on the blue 
ZAHB rather than their metallicity.  On the other hand, there is a significant 
difference between $Z=0.002$ and 0.006.  The smaller masses of the $Z=0.006$ 
sequences cannot compensate for their lower luminosities.  This result should 
be kept in mind when interpreting the Oosterhoff classification of bright, 
metal-rich RRL, such as V9 in 47~Tuc ($[{\rm Fe/H}] \simeq -0.7$, 
$[\alpha/{\rm Fe}] \simeq +0.2$; Carney 1996), V1 in Terzan~5 (metallicity 
around solar; Edmonds et al.\ 2001), the field star AN~Ser 
($[{\rm Fe/H}] \simeq -0.04$; Sweigart \& Catelan 1998 and references 
therein)---and, of course, the RRL in NGC~6388 and NGC~6441. 

Figure~14 shows a set of synthetic HB simulations for the $Z = 0.0005$ case, 
which is a metallicity representative of typical Oosterhoff type~II 
globular clusters, particularly when $\alpha$-enhancement is taken into 
account. These simulations were computed using {\sc sintdelphi}, described 
in Catelan, Ferraro, \& Rood (2001b). In all cases, we have assumed a mass 
dispersion 
$\sigma_M = 0.03\,M_{\sun}$, but truncated the (Gaussian) mass deviate at 
$M_{\rm HB,ev}$ in order to ensure that any RRL present is evolving to the 
red. In order to minimize the statistical fluctuations and better illustrate 
the main trends, 1500 HB stars were used in each simulation.  The upper left 
panel of Figure~14 shows the case in which the peak of the HB mass 
distribution is very close to ``a," corresponding to an input value 
$\langle M_{\rm HB} \rangle = 0.66\,M_{\sun}$. The subsequent panels show the 
variation in HB morphology and in the expected RRL properties as one 
decreases $\langle M_{\rm HB} \rangle$ in steps of $0.015\,M_{\sun}$.  
Figure~15 is the same as Figure~14, the only difference being that the CMD 
was zoomed on the region around the instability strip, in order to better 
illustrate the expected RRL luminosity distributions.  Similarly, Figure~16 
shows the corresponding location of the RRL in the 
$\log\,P-\log\,T_{\rm eff}$ plane. In the latter, both RRab and RRc variables 
are shown, the periods for the latter having been ``fundamentalized" 
as outlined in Catelan (1993).  The 
lower envelope to the distribution of variables in the upper left panel is 
overplotted on all panels; this line corresponds closely to line ``a" in the 
upper left panel of Figure~13. The mean shift in periods at fixed 
$T_{\rm eff}$ away from this line, as well as the corresponding standard 
deviation, is also indicated in the plots.

\begin{figure*}[t]
  \centerline{\psfig{figure=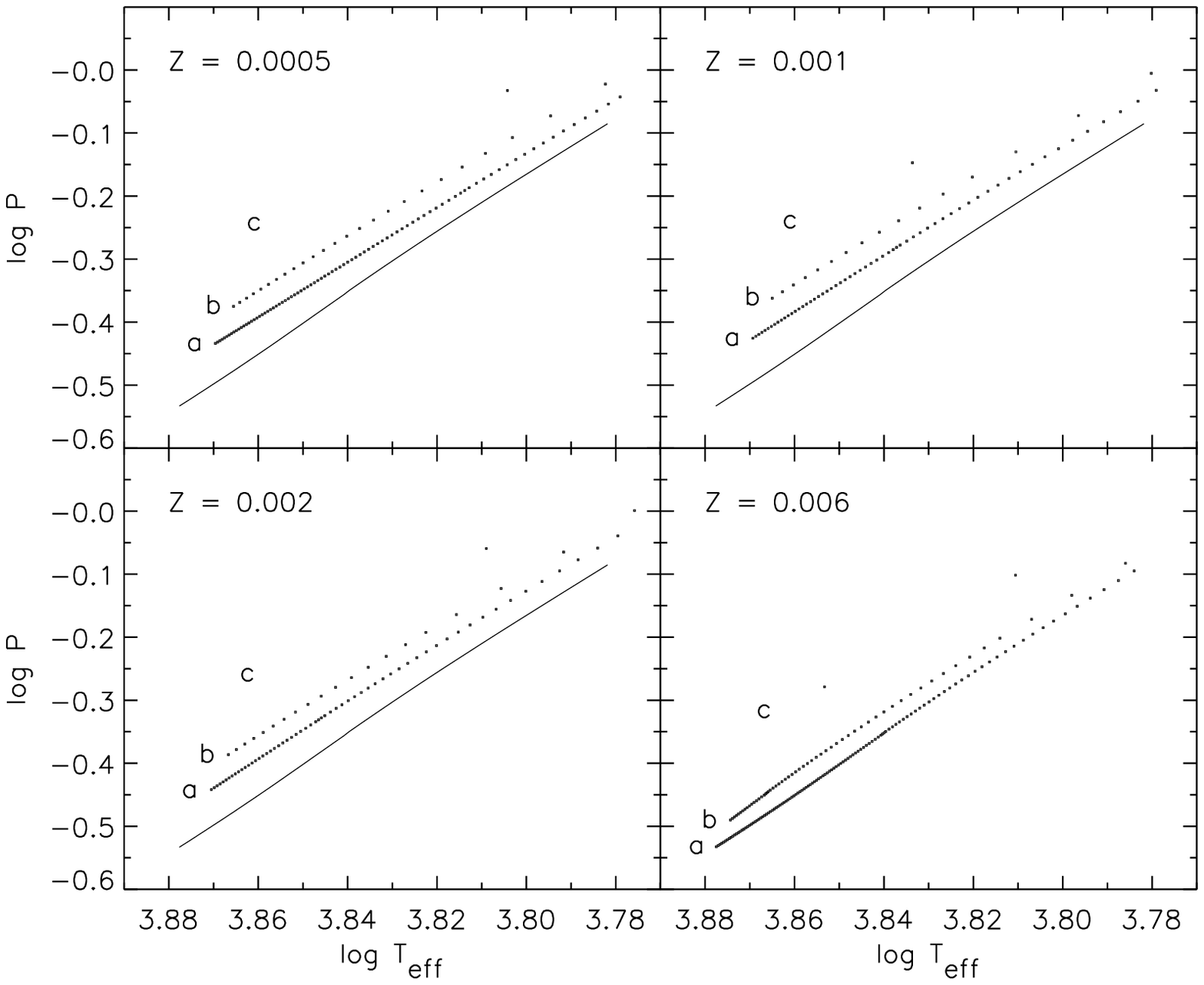}}
  \caption{Location of the HB evolutionary tracks from Figure~11 in the 
           period-temperature plane. Periods were computed from the pulsation 
           equation of van Albada \& Baker (1971). Only the part of the 
           tracks within the instability strip are plotted. Each track is 
           labeled at the blue edge of the instability strip. As in Figure~11 
           the tracks are plotted as a series of points separated by a time 
           interval of 200,000 yr. The ``c" tracks evolve so rapidly that 
           they contain only 1 or 2 points within the instability strip.  A 
           finer time resolution would show that the ``c" tracks run 
           approximately parallel to the ``a " and ``b" tracks. For 
           comparison the period-temperature locus of the ``a" track with 
           $Z = 0.006$ is plotted as a thin line in the three lower 
           metallicity panels. Note that corresponding tracks in the panels 
           for $Z \le 0.002$ have very similar loci. In contrast, the tracks 
           for $Z = 0.006$ are shifted towards shorter periods.} 
  \label{Fig13}
\end{figure*}

\begin{figure*}[t]
  \centerline{\psfig{figure=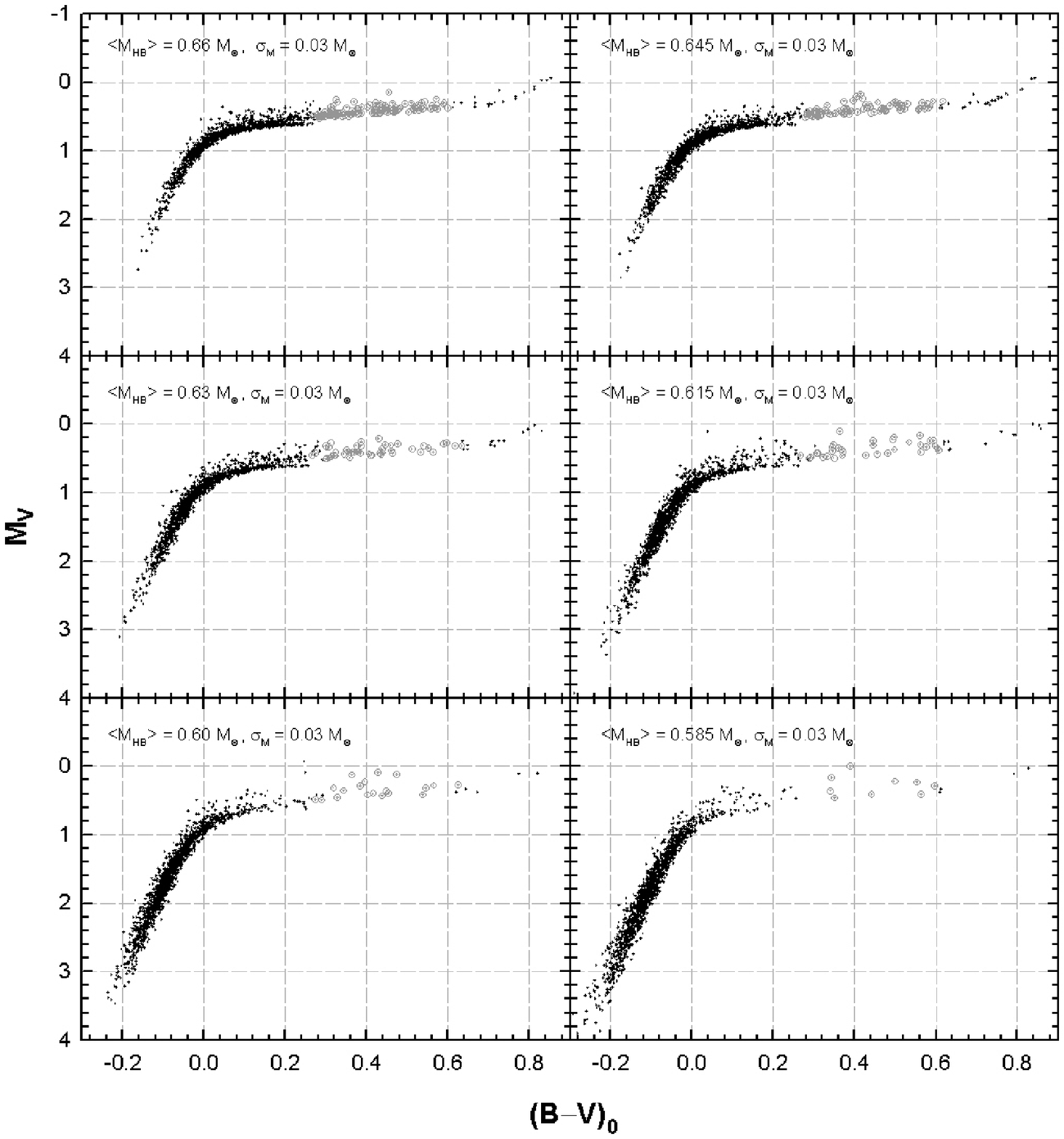}}
  \caption{Synthetic HB simulations of Oosterhoff type~II globular clusters 
           with all RR~Lyrae evolved away from a position on the blue ZAHB.\@  
           RR~Lyrae are indicated by encircled dots. From upper left to 
           lower right, progressively bluer HB types are shown. The input 
           values of the mean ZAHB mass and corresponding dispersion are 
           indicated in each panel. Note that, even in the upper left panel, 
           the HB morphology does not get redder than $(B-R)/(B+V+R) = +0.89$, 
           which is much bluer than Oosterhoff type~II globulars such as M15 
           and M68.} 
  \label{Fig14}
\end{figure*}  

Figures~15 and 16 show that, with a distribution in $M_{\rm HB}$ peaked 
around ``a," a natural concentration of the RRL towards the locus 
occupied by ``a" occurs. However, even for this higher value of 
$\langle M_{\rm HB} \rangle$, the implied scatter in $\log\,P$ at fixed 
$T_{\rm eff}$ is not negligible, and may easily reach 
$\Delta\log\,P \approx 0.05$. Moreover, there is a tendency for the mean 
locus to shift towards longer periods with decreasing 
$\langle M_{\rm HB} \rangle$, implying that, apart from statistical 
fluctuations, globular clusters are expected to produce a longer period 
shift at fixed $T_{\rm eff}$ as the HB morphology gets bluer (see also Lee 
1990).  Note also that the scatter in $\log\,P$ at a fixed $T_{\rm eff}$ also 
shows a tendency to increase as the HB gets bluer. Such intrinsic scatter 
often corresponds to a significant fraction of the separation between 
Clement \& Shelton's (1999) Oosterhoff type~I and type~II lines 
($\Delta\log\,P \sim 0.1$). The dispersion of the RRL in the 
$\log\,P-\log\,T_{\rm eff}$ plane questions the definition of a single line 
where the Oosterhoff type~II  RRL should fall, even in the context of the 
redward evolving blue HB scenario; the scatter would naturally be even larger, 
had the models not been artificially truncated at $M_{\rm HB,ev}$.  In this 
sense, it is worth noting that, in Fig.~2 of Clement \& Shelton (1999), 
RRL's deviate from the Oosterhoff type~II line at fixed amplitude by 
$\sim 0.02$, $\sim 0.035$, and $\sim 0.02$ in $\log\,P$ for M9, M68, and M92, 
respectively.  Scatter away from the Oosterhoff type~II line is also 
apparent in Fig.~10 of Clement \& Rowe (2000). 

\begin{figure*}[t]
  \centerline{\psfig{figure=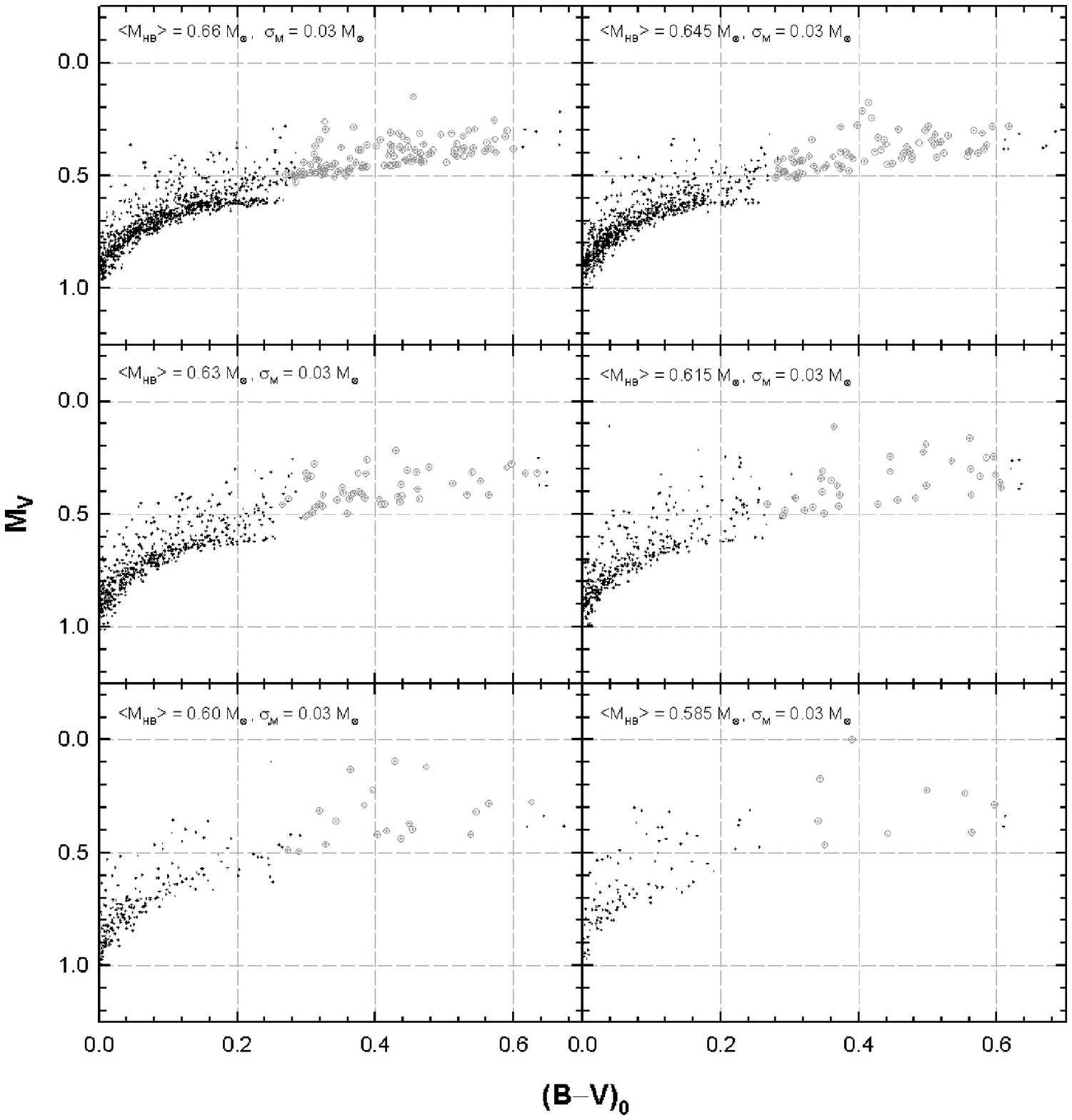}}
  \caption{As in Figure~14, but zooming on the distributions around the 
           RR~Lyrae level. Note the increasing concentration of the RR~Lyrae 
           distribution towards its lower envelope (which is well approximated 
           by line ``a" in the upper left panel of Figure~11) as the HB gets 
           redder; the presence of intrinsic scatter in all the simulations; 
           and the decrease in the number of variables as the HB gets bluer.} 
  \label{Fig15}
\end{figure*} 

\begin{figure*}[t]
  \centerline{\psfig{figure=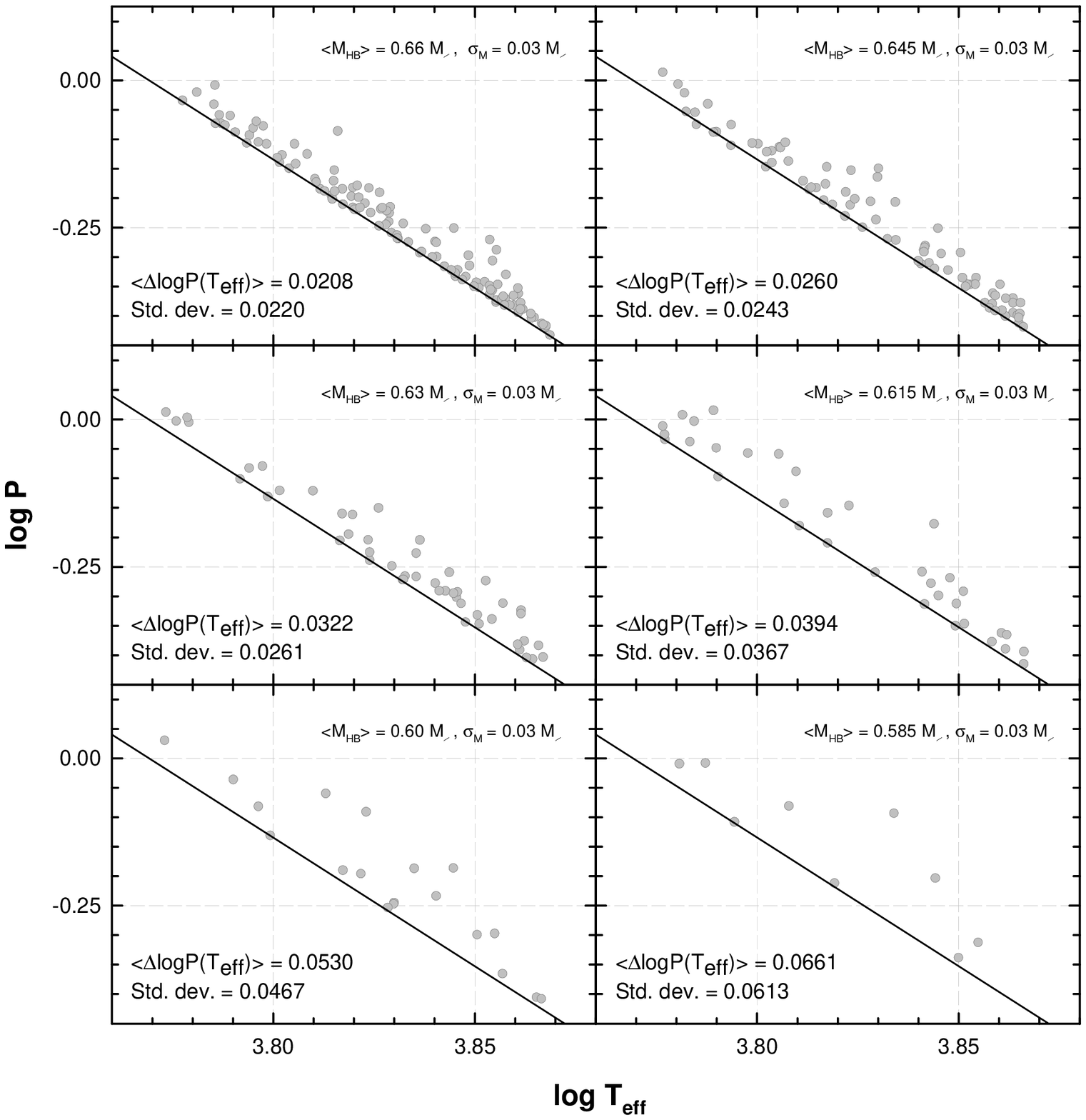}}  
  \caption{As in Figure~15, but showing the $\log\,P-\log\,T_{\rm eff}$ 
           distribution. A line has been drawn on all panels that corresponds 
           to the lower envelope of the higher-$\langle M_{\rm HB}\rangle$ 
           distribution (which is well approximated by line ``a" in the upper 
           left panel of Figure~13). The mean shift in periods (at fixed 
           $T_{\rm eff}$) for each of the simulations is indicated, as is the 
           corresponding standard deviation.  Note the shift towards longer 
           mean periods and the increase in the intrinsic scatter as the HB 
           gets bluer.} 
  \label{Fig16}
\end{figure*} 

It is important to note that the predicted period shift between 
the ZAHB and ``a" is only a relatively small fraction of the actual period 
shift between Clement \& Shelton's (1999) lines for the two Oosterhoff groups. 
It is only when the mean Oosterhoff type~II distribution clusters around ``b," 
not ``a," that we obtain a period shift of approximately the correct size. A 
smaller (observed) period shift would be possible if, as predicted by 
synthetic HB simulations, stars in Oosterhoff type~I globulars encompassed a 
relatively wide range in absolute magnitudes and periods at fixed color and 
temperature, respectively (see, e.g., Fig.~1a in Lee et al.\ 1990, and 
Figs.~2-3 in Catelan 1993). This, in turn, would be consistent with the 
results of Sandage (1990), who finds intrinsic scatter in the CMD and in the 
$\log\,P-\log\,T_{\rm eff}$ diagrams of both Oosterhoff type~II and type~I 
globulars, such scatter being systematically more pronounced for the latter. 
In particular, in the case of a cluster like M3, with a continuously 
populated ZAHB across the instability strip from the red HB to the blue HB 
``tail" (see, e.g., Fig.~1 in Moehler 2001), a sharp dichotomy between ``ZAHB" 
and ``evolved" stars is certainly 
not predicted by the models. These results indicate that intrinsic scatter 
must be adequately taken into account when attempting to classify individual 
RRL in terms of Oosterhoff type (Clement \& Shelton 1999; Clement \& 
Rowe 2000). 

We can now adequately address the second and third questions we proposed at 
the beginning of this section.  As we have discussed above, in order to produce 
a significant number of RRL according to the redward evolving blue HB scenario, 
a globular cluster would need to have a very strongly populated blue HB just 
to the blue of the instability strip.  Therefore, most of the RRL would fall 
along the same locus in the $\log\,P - \log\,T_{\rm eff}$ diagram, regardless 
of metallicity.  However, it is possible to produce discrepant points in the 
$\log\,P - \log\,T_{\rm eff}$ diagram from those RRL which originate from HB 
stars bluer than those found close to the instability strip.  The tradeoff is 
that one would not find as many of these stars due to their rapid evolution 
through the instability strip, unless of course the ZAHB mass distribution 
compensates for the variation in evolutionary timescales as a function of 
$M_{\rm HB}$.  

Regarding NGC~6388 and NGC~6441, Table~14 clearly implies that, if these 
clusters are Oosterhoff type~II in the evolutionary sense discussed 
above---that is, with their RRL evolving to the red---and if the 
RRL have the cluster metallicity $Z \approx 0.006$, then the HB morphology of 
the population to which these stars are associated must necessarily be 
characterized by: 

\begin{eqnarray} 
\nonumber 
(B-R)/(B+V+R) \gg 0.74 \\ 
B/(B+V+R) \gg 0.76. 
\end{eqnarray} 

\noindent Increasing the helium abundance with increasing $Z$ would lead to 
even more stringent requirements on the number ratios. 

Obviously, the HB morphology of either NGC~6388 or NGC~6441 is not at all 
this blue. On the contrary, both these clusters have predominantly red HBs.  
Therefore, in order for NGC~6388 and NGC~6441 to be Oosterhoff type~II 
clusters with all their RRL evolved away from the blue ZAHB, we 
must resort to some different explanation. We have the following alternatives: 

i) Assume that all cluster stars have the same metallicity. Then the 
distribution in some quantity along the ZAHB---such as the total mass, 
helium core mass, helium abundance---must be intrinsically bimodal, the red 
HB component being decoupled from the RRL plus blue HB component. Even though 
in this case the constraints on the number ratio between blue and red HB 
stars are lifted, those on the number ratio between blue HB and RRL stars 
remain very tight, with $V/B \ll 0.3$ (cf.\ Table~14). These restrictions are 
even more severe if the cluster variables have a somewhat lower metallicity 
than $Z = 0.006$ (cf.\ \S5.1), as can be seen from Table~14; 

ii) We can also relax the requirement that all cluster stars have the same 
metallicity. This means that we can invoke a lower metallicity for the RRL 
plus blue HB component. Note, however, that, unlike in the ``metallicity 
range'' scenario of Sweigart (2002), the ZAHB {\em must} necessarily be 
depopulated in the RRL region, or the result will not be an Oosterhoff 
type~II cluster with all RRL evolved from the blue. Therefore, a 
metal-poor component giving rise to RRL must necessarily be detached 
from the red HB component, with no ZAHB stars populating the region between 
the blue HB and the red HB.\@ In other words, if the range in temperatures 
on the NGC~6441 and NGC~6388 HBs is due to a range in metallicity, and if the 
clusters are Oosterhoff type~II (with all variables evolved), it thus follows 
that their metallicity distributions must be bimodal.  Note that this would be 
an even more dramatic bimodality than is thought to exist in $\omega$~Cen 
(e.g., Pancino et al.\ 2000).  Such a bimodality, if indeed present, would 
help reconcile the predicted presence of short-period (near-ZAHB) RRL 
in Sweigart's (2002) scenario with the lack of such short-period RRL in 
NGC~6388 and NGC~6441. 

It should be noted, in addition, that in either scenario the (sloped) red HB 
component must necessarily {\em not} produce RRL in any significant 
numbers, or else the ``evolutionary Oosterhoff type~II hypothesis'' would 
break down; this, in turn, also limits the extent to which some of the 
physical parameters, such as the helium abundance, can vary along the red HB 
component (as thought to be necessary to produce the sloped HB that 
characterizes both NGC~6388 and NGC~6441; Sweigart \& Catelan 1998). This is 
because HB stellar evolution naturally predicts that stars commencing their 
HB evolution just redward of the red edge of the instability strip will 
eventually evolve along ``blue loops'' and cross the instability strip, thus 
becoming blueward-evolving RRL. In particular, red HB stars with a 
fairly high helium abundance, which evolve along very extended blue loops 
during a significant fraction of their HB lifetime (compare, for instance, 
the $Y = 0.23$ HB tracks in Fig.~1 of Sweigart 1999 with the $Y = 0.43$ HB 
tracks presented in the same figure, as well as with the helium mixing tracks 
in Fig.~3 of that paper), would lead to a population of RRL clearly 
violating the assumption that RRL in Oosterhoff type~II globular 
clusters are all evolving to the red. 

Given the evolutionary issues discussed above, it is conceivable that the 
particular selection criteria used in Clement \& Shelton (1999) in 
defining their ``normal" stars is too strict in the sense of leading to the 
rejection of 
many of those stars which are naturally expected to scatter around their 
Oosterhoff types~I and II lines.  If so, this might alleviate the 
requirement that RRL in Oosterhoff type~II clusters be evolved away 
from the blue ZAHB---and, conversely, that RRL in Oosterhoff type~I 
clusters remain on the ZAHB during their entire lifetimes.  In this sense, it 
is worth noting that the $Z = 0.0005$ case shown in Table~14 and Figures~11 
and 13-16 should provide a fairly realistic description of metal-poor 
Oosterhoff type~II clusters with $\alpha$-elements enhanced by a factor of 
about three and with ${\rm [Fe/H]} \simeq -2$---such as M68. However, M68 has 
an HB type $(B-R)/(B+V+R) = 0.44$ (Walker 1994), whereas 
$(B-R)/(B+V+R)_{\rm min} = 0.7$ for the $Z = 0.0005$ case (Table~14).  Hence 
it is not 
possible to successfully model an Oosterhoff type~II cluster such as M68 if 
one requires that all its RRL have evolved from blue ZAHB positions.  Again, 
this is in agreement with similar arguments put forth by Renzini \& Fusi Pecci 
(1988) and Rood \& Crocker (1989). 

We conclude that, if the Clement et al.\ (2001) suggestion that NGC~6388 and 
NGC~6441 are both Oosterhoff type~II is confirmed, this will most likely 
require a revision of the evolutionary interpretation of the Oosterhoff 
dichotomy, suggesting that there may be different reasons why RRL 
in Oosterhoff type~II globular clusters are brighter than those in Oosterhoff 
type~I globular clusters---evolution away from the blue ZAHB being one such 
reason, but perhaps not the only one, even for metal-poor clusters such as M68 
and M15. If, on the other hand, it should turn out that RRL in all 
Oosterhoff type~II globulars are indeed evolved, then extant evolutionary 
calculations seriously underestimate the duration of the late stages of HB 
evolution. The latter is, in fact, a possibility that will have to be 
seriously considered in future investigations. As well known, there is  
considerable evidence pointing to redward evolution for RRL in 
Oosterhoff type~II globular clusters beyond that considered in the 
present analysis---including the RRc-to-RRab number 
ratios and mean (secular) period change rates (Smith 1995 and references 
therein). Uncertainties currently affecting HB models, 
and which may be relevant when considering this issue, include those 
directly related to HB evolution and those affecting it indirectly, by 
means of the properties of RGB models by the time they reach the tip of 
the RGB.\@  Among the former may be listed the 
$^{12}{\rm C}(\alpha,\gamma)^{16}{\rm O}$ reaction rates, since this  
reaction dominates over the triple-$\alpha$ reaction at the late stages 
of HB evolution, and the treatment of 
``breathing pulses" which occur when He is close to being exhausted in 
the cores of HB stars (see \S2.2.4 in Catelan et al. 2001a for a related 
discussion and references). Among the latter, most important are those that 
affect the determination of the size of the He-core mass at the He-flash, 
$M_{\rm c}$. This is because HB stars which start their evolution with 
efficient H-burning shells spend most of their time not on the ZAHB proper, 
but evolving on ``blueward loops" in the HR diagram (see, e.g., Figure~11).  
The size of the blue loops is determined, among other things, by the balance 
between energy generation in the He-core and in the H-burning shell, thus being 
strongly dependent on the value of $M_{\rm c}$ (Iben \& Rood 1970; Sweigart \& 
Gross 1976).  

If blue HB stars evolved more directly to 
the red, instead of spending so much time evolving to the blue along these 
``blueward loops," HB tracks such as those depicted in Figure~11 would spend 
much more time in the instability strip, so that HB simulations like those 
in Figures~14-16 would give rise to larger numbers of RRL.\@  In this sense, 
one might suspect that, by decreasing the metallicity further down with 
respect to the values that we have considered, we might also be able to 
suppress the blueward loops (e.g., Sweigart 1987).  However, current models do 
not support that conclusion.  Prominent blueward loops are also seen in models 
with even lower metallicities, covering the range in [Fe/H] that is perhaps 
more appropriate to Oosterhoff type~II clusters such as M15.  In particular, 
blue loops similar to the ones that we find to affect the maximum possible 
number of redward-evolving RRL have also been found in the following studies:  
Sweigart (2002):  $Z = 0.0003$ (his Fig.~1); Yi, Lee, \& Demarque (1993): 
$Z = 0.0001$ and $Z=0.0004$ (their Figs.~1b and 1c); Caloi, D'Antona, \& 
Mazzitelli (1997):  $Z=0.0001$ and $Z=0.0002$ (their Fig.~2); Dorman (1992):  
${\rm [Fe/H]}=-2.26$ and $-2.03$, oxygen-enhanced (his Figs.~2-4).  This 
also implies that our arguments, dependent as they are on the occurrence 
of such blue loops, remain valid when computations are extended toward 
lower metallicities, and do not depend on whose sets of evolutionary tracks 
one uses.

Metallicity effects not being a tenable explanation, analysis of  a possible 
increase in $M_{\rm c}$ thus becomes the next logical step in the search 
for a way to reduce the size of the blue loops and thus produce larger 
numbers of redward-evolving RRL in Oosterhoff type~II clusters.  In fact, 
an increase in $M_{\rm c}$ was earlier suggested by Castellani \& 
Tornamb\`e (1981) and Catelan (1992) as a possible way to help account 
for the evolutionary properties of RRL in Oosterhoff type~II 
globulars. The value of $M_{\rm c}$ is uncertain due to a variety of factors, 
as extensively discussed by Catelan, de Freitas Pacheco, \& Horvath (1996) 
and Salaris, Cassisi, \& Weiss (2002). Particularly noteworthy is the 
current lack of reliable 
electron conductive opacities for the physical conditions 
characterizing the interiors of RGB stars. Outside the canonical framework, 
the interplay between stellar rotation and the value of $M_{\rm c}$ has 
still not been firmly established; it is certainly a possibility that 
rapidly rotating RGB cores will be able to attain higher masses by the 
time of the He-flash than non-rotating RGB cores (Mengel \& Gross 1976). 

Due to these uncertainties, it remains unclear whether the failure of the 
current generations of HB models to account for the properties of 
Oosterhoff type~II globulars is fatal to the evolutionary scenario, or 
whether future improvements in the input physics and in producing 
realistic models of HB and RGB stars will prove that the evolutionary 
scenario is indeed the correct explanation.  What {\em is} clear, at present, 
is that the evolutionary interpretation of the Oosterhoff dichotomy is 
far from being conclusively settled---and even more so in the cases of 
NGC~6388 and NGC~6441, whose detailed properties differ in several 
important respects from those of typical Oosterhoff type~II globulars.

\section{Summary and Conclusions}

The unusual horizontal branch morphology in NGC~6388 is confirmed 
in this ground-based study.  Typical of a metal-rich globular cluster, 
NGC~6388 contains a strong red component to the horizontal 
branch.  A blue component can also be seen extending through the 
instability strip, sloping upward in $V$ with decreasing ($\bv$).  
This second-parameter effect cannot be explained by age or increased 
mass loss along the red giant branch (Sweigart \& Catelan 1998).  
It may also be that NGC~6388 has a spread in metallicity, so that the 
blue component would then be due to a lower metallicity.

The number of known RR~Lyrae has been increased to 14.  The 
periods of the RRab stars are found to be unusually long for the 
metallicity of NGC~6388.  In fact, as seen in a period-amplitude 
diagram comparing NGC~6388 to other globular clusters, the 
periods for the RRab stars are as long as, if not longer than, those 
found in Oosterhoff type~II clusters.  A small number of long period 
RRc stars were found in NGC~6388, resulting in a smaller than 
expected gap between the shortest period RRab and longest period 
RRc stars.  Long period RRc stars are uncommon in globular 
clusters with the exceptions of $\omega$ Cen and NGC~6441. 

The reddening was determined to be $E(\bv)=0.40\pm0.03$~mag with 
some differential reddening for NGC~6388.  The mean $V$ magnitude 
for the horizontal branch from the RR~Lyrae, excluding the possible field 
RR~Lyrae V26 and V34, was found 
to be $16.85\pm0.05$~mag leading to a range in distance of 9.0 
to 10.3~kpc, depending on the adopted HB absolute magnitude.

Four candidate Population~II Cepheids were found in the field of NGC~6388.  
Although their existence agrees with the extended blue tail in NGC~6388, 
their likely membership makes NGC~6388 the most metal-rich 
globular cluster known to contain Population~II Cepheids. However, 
such stars may still be fairly metal-poor, in the case there is an 
internal metallicity spread in the cluster.  The distance of 
NGC~6388 derived from the Cepheids is 10.6~kpc. 

NGC~6388 does not appear to fit in to the typical Oosterhoff 
classification scheme.  The long periods of the RRab in NGC~6388 
contradict the trend of increasing period with decreasing 
metallicity, for a given amplitude.  If there is no spread in 
metallicity in NGC~6388, this implies that the 
metallicity-luminosity relationship for RR~Lyrae is not universal.
We provide a detailed discussion, based on theoretical HB models 
and simulations, of the recent suggestion that NGC~6441---and, by 
analogy, NGC~6388---can be classified as Oosterhoff type II, also 
confirming previous difficulties in accounting for the Oosterhoff 
dichotomy in terms of an evolutionary scenario.

\acknowledgments

This work has been supported by the National Science Foundation under 
grants AST 9528080 and AST 9986943.
A. V. S. gratefully acknowledges support from NASA Astrophysics
Theory Program proposal NRA-99-01-ATP-039.

We would like to thank the referee for his/her comments and suggestions which 
helped to clarify the paper.  B. P. would like to thank Peter Stetson for the 
use of his reduction programs and his assistance in getting them to run 
properly.  Thank you to Nancy Silbermann for sharing her knowledge of the 
reduction programs.  Thank you to Suzanne Hawley and Tim Beers for their 
insightful comments.  Thank you to Brian Sharpee for the use of his spline 
programs.  We would also like to thank Christine Clement for the use of her 
Fourier deconvolution program, for providing analytical formulae for her 
Oosterhoff lines in the period-amplitude diagram, and for useful 
discussions.  We also thank Manuela Zoccali for 
providing number counts along the HBs of NGC~6388 and NGC~6441.

\begin{deluxetable}{ccc} 
\tablewidth{0pc}
\footnotesize
\tablecaption{Mean Differences in Photometry\label{tbl-1}}
\tablehead{
\colhead{Reference} & \colhead{$\Delta\,V$} & \colhead{$\Delta\,B$}
          }
\startdata
Alcaino & 0.03$\pm$0.02 & 0.02$\pm$0.03 \\
Silbermann et al. & 0.007$\pm$0.008 & -0.004$\pm$0.008 \\ 
HST & 0.035$\pm$0.007 & 0.025$\pm$0.010 \\ 
\enddata
\tablecomments{difference = reference magnitude - magnitude in present study} 
\end{deluxetable}

\begin{deluxetable}{cccccc} 
\tablewidth{0pc}
\footnotesize
\tablecaption{Locations of Variable Stars\label{tbl-2}}
\tablehead{
\colhead{ID} & \colhead{X} & \colhead{Y}
& \colhead{$\Delta\alpha$ (2000)} & \colhead{$\Delta\delta$ (2000)} 
& \colhead{Classification}
          }
\startdata
V4  & 1560.0 &  999.9 & -192.2 &   26.0 & LPV \\
V12 & 1228.2 & 1067.0 &  -60.7 &   -0.8 & LPV \\
V14 & 1532.5 & 1858.0 & -184.0 & -310.9 & Binary \\
V16 & 1743.4 &  447.1 & -263.2 &  243.2 & RRL \\
V17 & 1170.8 & 1129.4 &  -38.1 &  -25.4 & RRL \\
V18 & 1147.5 &  949.0 &  -28.3 &   45.4 & P2C \\
V20 &  938.0 &  965.0 &   54.9 &   38.8 & RRL \\
V21 &  911.2 &  722.6 &   66.3 &  133.9 & RRL \\
V22 &  901.0 & 1076.3 &   69.2 &   -4.9 & RRL \\
V23 & 1528.6 & 1016.9 & -179.8 &   19.3 & RRL \\
V26 &  793.3 & 1228.3 &  111.5 &  -64.7 & RRL \\
V27 &  923.4 & 1106.7 &   60.2 &  -16.8 & RRL \\
V28 & 1014.8 & 1194.5 &   23.7 &  -51.1 & RRL \\
V29 & 1124.4 & 1052.1 &  -19.4 &    4.9 & P2C \\ 
V30 &  951.3 & 1089.8 &   49.2 &  -10.1 & RRL \\
V31 &  768.4 &  823.1 &  122.7 &   94.3 & RRL \\
V32 & 1170.9 & 1138.8 &  -38.1 &  -29.1 & RRL \\
V33 &  907.3 &  854.0 &   67.4 &   82.4 & RRL \\
V34 & 1562.0 & 1184.7 & -193.6 &  -46.6 & RRL \\
V35 &  898.0 &  535.9 &   72.1 &  207.2 & RRL \\
V36 & 1004.7 & 1025.3 &   28.2 &   15.3 & P2C \\
V37 & 1072.7 & 1127.3 &    0.9 &  -24.7 & P2C \\
V38 & 1088.9 & 1638.4 &   -7.2 & -225.3 & Binary \\
V39 &  815.2 &    9.6 &  106.7 &  413.7 & Binary \\
V40 &  904.7 &  525.6 &   69.5 &  211.3 & Binary \\
V41 &   62.1 &  332.7 &  404.6 &  285.8 & Binary \\
V42 &  456.0 &  620.1 &  247.3 &  173.6 & Binary \\
V43 & 1467.1 & 1010.7 & -155.3 &   21.6 & Binary \\
V44 & 1607.4 & 1151.7 & -211.5 &  -33.6 & $\delta$~Scuti or SX~Phe \\
V45 &  649.7 &  341.7 &  171.3 &  283.1 & LPV \\
V46 & 1283.6 & 1158.4 &  -82.9 &  -36.6 & LPV \\
V47 &  359.0 & 1523.1 &  283.0 & -181.0 & LPV \\
V48 & 1004.9 & 1030.6 &   28.1 &   13.2 & RRL? \\
V49 & 1048.9 & 1168.0 &   10.2 &  -40.7 & RRL? \\
V50 & 1130.0 & 1006.1 &  -21.5 &   23.0 & RRL? \\
V51 & 1006.9 & 1134.3 &   27.0 &  -27.5 & RRL? \\
V52 & 1530.4 & 1564.3 & -182.2 & -195.6 & RRL? \\
V53 & 1278.0 &  973.1 &  -80.1 &   36.1 & RRL? \\
V54 & 1589.2 & 1115.1 & -204.1 &  -19.2 & Binary \\
V55 & 1037.2 & 1229.8 &   14.7 &  -65.0 & RRL? \\
V56 & 1130.4 & 1097.4 &  -22.1 &  -13.8 & RRL? \\
V57 & 1821.4 &  571.7 & -294.6 &  194.4 & Binary \\
\enddata
\end{deluxetable}

\begin{deluxetable}{ccccccc}
\tablewidth{0pc}
\tablecolumns{7} 
\tablecaption{Mean Properties of Pulsating Variable Stars with $P<1$~d 
\label{tbl-3}}
\tablehead{
\colhead{ID} & \colhead{Period} & \colhead{$\langle V \rangle$}
& \colhead{$(\bv)_{\rm mag}$} & \colhead{$A_V$} 
& \colhead{$A_B$} & \colhead{Comments}  
          }
\startdata
V16 & 0.251 & 16.895 & 0.538 & 0.26 & 0.31 & c \\ 
V17 & 0.611 & 16.525 & 0.687 & 0.85 & 1.15 & ab \\ 
V20 & 0.467 & 16.787 & 0.447 & 0.36 & 0.46 & c \\ 
V21 & 0.814 & 17.030 & 0.760 & 0.87 & 1.15 & ab \\ 
V22 & 0.587 & 16.858 & 0.706 & 1.12 & 1.57 & ab \\
V23 & 0.338 & 16.893 & 0.520 & 0.48 & 0.65 & c \\ 
V26 & 0.239 & 17.403 & 0.522 & 0.38 & 0.51 & c; Field? \\ 
V27 & 0.365 & 16.953 & 0.585 & 0.46 & 0.67 & c \\ 
V28 & 0.840 & 16.820 & 0.824 & 0.80 & 1.05 & ab \\ 
V30 & 0.951 & 16.766 & 0.863 & \nodata & \nodata & ab? \\ 
V31 & 0.341 & 17.031 & 0.550 & 0.52 & 0.70 & c\\ 
V32 & 0.522 & 16.578 & 0.598 & 0.42 & 0.52 & c, S1 \\ 
V33 & 0.558 & 16.747 & 0.712 & 0.29 & 0.42 & c \\ 
V34 & 0.236 & 17.409 & 0.544 & 0.36 & 0.50 & c; Field? \\ 
V35 & 0.300 & 17.041 & 0.525 & 0.18 & 0.23 & c \\
V44 & 0.080 & 18.082 & 0.533 & 0.50 & 0.67 & $\delta$~Scuti or SX~Phe \\ 
V48 & 0.355 & 16.574 & 0.421 & \nodata & 0.34 & c? \\ 
V49 & 0.384 & 16.925 & 0.628 & 0.52 & 0.65 & c?, S2 \\ 
V50 & 0.364 & 16.898 & 0.965 & 0.54 & \nodata & c? \\ 
V51 & 0.397 & 16.604 & 0.733 & 0.34 & 0.48 & c? \\ 
V52 & 0.387 & 16.686 & 0.661 & 0.23 & 0.55 & c? \\ 
V53 & 0.986 & 16.868 & 0.773 & \nodata & \nodata & ab?, S3 \\ 
V55 & 0.489 & 16.795 & 0.649 & \nodata & \nodata & c? \\ 
V56 & 0.552 & 16.825 & 0.687 & \nodata & \nodata & c? \\
\enddata 
\end{deluxetable}

\begin{deluxetable}{cccccc}
\tablewidth{0pt}
\footnotesize
\tablecaption{Photometry of the Variable Stars (V)\label{tbl-4}}
\tablehead{
\colhead{} & \multicolumn{2}{c}{V16} & & \multicolumn{2}{c}{V17} \\ 
\cline{2-3} \cline{5-6} \\ 
\colhead{HJD-2450000} & \colhead{$V$} & \colhead{$\sigma_{V}$} & & 
 \colhead{$V$} & \colhead{$\sigma_{V}$} 
          }
\startdata
966.266  & 16.787  &  0.018 &&  16.365  &  0.064 \\
959.172  & 16.970  &  0.017 &&  16.724  &  0.035 \\
959.237  & 16.791  &  0.019 &&  16.782  &  0.025 \\
960.112  & 17.001  &  0.018 &&  16.277  &  0.018 \\
960.212  & 16.837  &  0.018 &&  16.532  &  0.033 \\
961.301  & 16.829  &  0.018 &&  16.219  &  0.028 \\
961.335  & 16.937  &  0.018 &&  16.272  &  0.022 \\
961.367  & 17.008  &  0.019 &&  16.406  &  0.025 \\
962.065  & 16.850  &  0.017 &&  16.565  &  0.036 \\
962.117  & 16.988  &  0.032 &&  16.544  &  0.053 \\
962.155  & 17.029  &  0.017 &&  16.629  &  0.074 \\
962.216  & 16.845  &  0.021 &&  16.747  &  0.071 \\
962.257  & 16.760  &  0.018 &&  16.789  &  0.059 \\
962.294  & 16.797  &  0.019 &&  16.760  &  0.113 \\
962.330  & 16.883  &  0.022 &&  16.879  &  0.060 \\
965.076  & 16.844  &  0.018 &&  16.473  &  0.024 \\
965.114  & 16.955  &  0.018 &&  16.532  &  0.048 \\
965.146  & 17.014  &  0.021 &&  16.558  &  0.065 \\
965.183  & 17.016  &  0.018 &&  16.595  &  0.047 \\
965.216  & 16.926  &  0.016 &&  16.660  &  0.036 \\
965.248  & 16.801  &  0.016 &&  16.716  &  0.039 \\
965.280  & 16.775  &  0.018 &&  16.733  &  0.053 \\
965.313  & 16.820  &  0.016 &&  16.753  &  0.058 \\
965.346  & 16.901  &  0.019 &&  16.783  &  0.063 \\
966.062  & 16.804  &  0.018 &&  16.881  &  0.058 \\
966.094  & 16.878  &  0.017 &&  16.600  &  0.045 \\
966.135  & 16.982  &  0.017 &&  16.011  &  0.055 \\
966.167  & 17.026  &  0.018 &&  16.056  &  0.060 \\
966.203  & 16.984  &  0.016 &&  16.185  &  0.060 \\
966.233  & 16.867  &  0.016 &&  16.291  &  0.065 \\
966.327  & 16.839  &  0.019 &&  16.516  &  0.075 \\
967.050  & 16.776  &  0.020 &&  16.674  &  0.043 \\
967.082  & 16.843  &  0.018 &&  16.715  &  0.065 \\
967.114  & 16.921  &  0.018 &&  16.756  &  0.060 \\
967.146  & 16.996  &  0.017 &&  16.731  &  0.089 \\
967.188  & 17.018  &  0.016 &&  16.820  &  0.055 \\
967.243  & 16.848  &  0.018 &&  16.886  &  0.055 \\
967.276  & 16.774  &  0.016 &&  16.862  &  0.069 \\
967.308  & 16.788  &  0.016 &&  16.696  &  0.051 \\
967.345  & 16.869  &  0.018 &&  16.084  &  0.070 \\
968.056  & 16.786  &  0.018 &&  16.295  &  0.026 \\
968.089  & 16.840  &  0.017 &&  16.412  &  0.032 \\
968.122  & 16.921  &  0.017 &&  16.498  &  0.032 \\
968.163  & 17.009  &  0.018 &&  16.532  &  0.024 \\
968.195  & 17.014  &  0.017 &&  16.598  &  0.031 \\
968.236  & 16.889  &  0.017 &&  16.614  &  0.029 \\
968.274  & 16.779  &  0.017 &&  16.662  &  0.031 \\
968.307  & 16.775  &  0.017 &&  16.702  &  0.021 \\
\enddata
\tablecomments{The complete version of this table is in the electronic 
edition of the Journal.  The printed edition contains only a sample.}
\end{deluxetable}

\begin{deluxetable}{cccccc}
\tablewidth{0pt}
\footnotesize
\tablecaption{Photometry of the Variable Stars (B)\label{tbl-5}}
\tablehead{
\colhead{} & \multicolumn{2}{c}{V16} & & \multicolumn{2}{c}{V17} \\ 
\cline{2-3} \cline{5-6} \\ 
\colhead{HJD-2450000} & \colhead{$B$} & \colhead{$\sigma_{B}$} & & 
 \colhead{$B$} & \colhead{$\sigma_{B}$} 
          }
\startdata
966.274 &  17.287  &  0.012 &&  17.103  &  0.072 \\
959.183 &  17.491  &  0.011 &&  17.474  &  0.040 \\
959.246 &  17.299  &  0.014 &&  17.541  &  0.031 \\
960.120 &  17.584  &  0.013 &&  16.897  &  0.017 \\
961.309 &  17.377  &  0.008 &&  16.709  &  0.059 \\
961.343 &  17.509  &  0.010 &&  16.847  &  0.031 \\
962.057 &  17.386  &  0.014 &&  17.249  &  0.031 \\
962.097 &  17.473  &  0.012 &&  17.274  &  0.039 \\
962.147 &  17.598  &  0.011 &&  17.326  &  0.076 \\
962.188 &  17.520  &  0.007 &&  17.481  &  0.078 \\
962.205 &  17.437  &  0.006 &&  17.525  &  0.073 \\
962.228 &  17.327  &  0.007 &&  17.535  &  0.066 \\
962.265 &  17.260  &  0.011 &&  17.581  &  0.073 \\
962.302 &  17.343  &  0.017 &&  17.675  &  0.092 \\
962.337 &  17.450  &  0.020 &&  17.649  &  0.057 \\
965.068 &  17.365  &  0.013 &&  17.072  &  0.018 \\
965.106 &  17.487  &  0.014 &&  17.201  &  0.033 \\
965.138 &  17.582  &  0.014 &&  17.283  &  0.068 \\
965.191 &  17.564  &  0.008 &&  17.338  &  0.037 \\
965.223 &  17.432  &  0.007 &&  17.413  &  0.039 \\
965.255 &  17.288  &  0.007 &&  17.544  &  0.061 \\
965.288 &  17.296  &  0.008 &&  17.523  &  0.055 \\
965.320 &  17.354  &  0.010 &&  17.550  &  0.055 \\
965.354 &  17.471  &  0.017 &&  17.626  &  0.060 \\
966.054 &  17.296  &  0.017 &&  17.679  &  0.058 \\
966.086 &  17.394  &  0.011 &&  17.404  &  0.051 \\
966.127 &  17.513  &  0.011 &&  16.512  &  0.048 \\
966.159 &  17.582  &  0.011 &&  16.517  &  0.061 \\
966.195 &  17.563  &  0.007 &&  16.698  &  0.055 \\
966.241 &  17.355  &  0.007 &&  16.924  &  0.060 \\
966.334 &  17.388  &  0.013 &&  17.277  &  0.068 \\
967.058 &  17.281  &  0.018 &&  17.467  &  0.059 \\
967.090 &  17.381  &  0.018 &&  17.542  &  0.073 \\
967.121 &  17.483  &  0.011 &&  17.564  &  0.068 \\
967.154 &  17.571  &  0.008 &&  17.553  &  0.070 \\
967.195 &  17.578  &  0.007 &&  17.623  &  0.073 \\
967.235 &  17.402  &  0.008 &&  17.634  &  0.051 \\
967.268 &  17.298  &  0.007 &&  17.664  &  0.063 \\
967.300 &  17.281  &  0.011 &&  17.507  &  0.050 \\
967.337 &  17.377  &  0.012 &&  16.814  &  0.054 \\
968.046 &  17.276  &  0.017 &&  16.804  &  0.030 \\
968.081 &  17.345  &  0.011 &&  16.966  &  0.024 \\
968.115 &  17.443  &  0.008 &&  17.107  &  0.025 \\
968.155 &  17.576  &  0.011 &&  17.217  &  0.019 \\
968.187 &  17.590  &  0.010 &&  17.263  &  0.021 \\
968.249 &  17.364  &  0.007 &&  17.357  &  0.025 \\
968.282 &  17.281  &  0.007 &&  17.399  &  0.020 \\
968.315 &  17.306  &  0.007 &&  17.470  &  0.024 \\
\enddata 
\tablecomments{The complete version of this table is in the electronic 
edition of the Journal.  The printed edition contains only a sample.}
\end{deluxetable}

\begin{deluxetable}{cccccccccc}
\tablewidth{0pt}
\footnotesize
\tablecaption{Fourier Coefficients for the RR~Lyrae\label{tbl-6}}
\tablehead{
\colhead{ID} & \colhead{$A_{\rm 1}$} & \colhead{$A_{\rm 21}$} & 
\colhead{$A_{\rm 31}$} & \colhead{$A_{\rm 41}$} &\colhead{$\phi_{\rm 21}$} & 
\colhead{$\phi_{\rm 31}$} & \colhead{$\phi_{\rm 41}$} & \colhead{$D_m$} 
& \colhead{Classification} 
          }
\startdata
V16 & 0.128 & 0.094 & 0.031 & 0.012 & 4.760 & 3.216$\pm$0.374 & 1.909 & \nodata & RRc \\
V17 & 0.295 & 0.583 & 0.360 & 0.196 & 4.294 & 2.222$\pm$0.068 & 0.477 & 5.35 & RRab \\
V20 & 0.180 & 0.094 & 0.092 & 0.050 & 2.430 & 5.393$\pm$0.347 & 3.854 & \nodata & RRc \\
V21 & 0.332 & 0.456 & 0.191 & 0.098 & 4.577 & 2.879$\pm$0.092 & 1.068 & 9.01 & RRab \\
V22 & 0.391 & 0.544 & 0.337 & 0.203 & 4.295 & 2.196$\pm$0.035 & 0.524 & 3.10 & RRab \\
V23 & 0.236 & 0.082 & 0.035 & 0.071 & 3.968 & 4.390$\pm$0.323 & 2.982 & \nodata & RRc \\
V26 & 0.197 & 0.128 & 0.030 & 0.030 & 4.624 & 2.669$\pm$0.552 & 2.585 & \nodata & RRc \\
V27 & 0.234 & 0.068 & 0.042 & 0.048 & 4.673 & 5.496$\pm$0.506 & 3.332 & \nodata & RRc \\
V28 & 0.331 & 0.397 & 0.133 & 0.081 & 4.592 & 2.973$\pm$0.117 & 1.429 & 13.17 & RRab \\
V31 & 0.257 & 0.083 & 0.055 & 0.055 & 4.044 & 5.082$\pm$0.182 & 3.207 & \nodata & RRc \\
V32 & 0.187 & 0.218 & 0.142 & 0.111 & 3.377 & 0.206$\pm$0.260 & 4.277 & \nodata & RRc \\
V33 & 0.132 & 0.214 & 0.130 & 0.027 & 3.229 & 0.674$\pm$0.188 & 5.040 & \nodata & RRc \\
V34 & 0.187 & 0.130 & 0.085 & 0.056 & 4.234 & 3.017$\pm$0.173 & 2.268 & \nodata & RRc \\
V35 & 0.093 & 0.037 & 0.020 & 0.026 & 4.340 & 5.924$\pm$0.859 & 1.672 & \nodata & RRc \\
V48 & 0.152 & 0.070 & 0.157 & 0.067 & 4.334 & 4.345$\pm$0.594 & 3.766 & \nodata & RRc? \\
V49 & 0.234 & 0.054 & 0.097 & 0.109 & 3.291 & 6.002$\pm$0.422 & 3.480 & \nodata & RRc? \\
V50 & 0.242 & 0.036 & 0.116 & 0.126 & 4.355 & 5.997$\pm$0.392 & 4.553 & \nodata & RRc? \\
V51 & 0.174 & 0.077 & 0.106 & 0.053 & 2.772 & 5.139$\pm$0.291 & 4.591 & \nodata & RRc? \\
V52 & 0.144 & 0.111 & 0.205 & 0.286 & 4.188 & 3.479$\pm$1.458 & 3.541 & \nodata & RRc? \\
V53 & 5.838 & 0.621 & 0.264 & 0.402 & 0.179 & 3.366$\pm$1.186 & 3.655 & \nodata & RRab? \\
V55 & 0.098 & 0.577 & 0.762 & 0.758 & 1.511 & 4.910$\pm$0.192 & 2.133 & \nodata & RRc? \\
V56 & 0.171 & 0.054 & 0.236 & 0.095 & 5.967 & 1.148$\pm$0.352 & 0.170 & \nodata & RRc? \\
\enddata
\end{deluxetable}

\begin{deluxetable}{cccc}
\tablewidth{0pc}
\footnotesize
\tablecaption{Reddening Determinations\label{tbl-7}}
\tablehead{
\colhead{ID} & \colhead{$E(\bv)$} 
          }
\startdata
V17 & 0.365 \\ 
V21 & 0.388 \\ 
V22 & 0.405 \\ 
V28 & 0.431 \\ 
\enddata
\end{deluxetable}

\begin{deluxetable}{cccccccc} 
\tablewidth{0pc}
\footnotesize
\tablecaption{Mean Properties of Cepheid Variables\label{tbl-8}}  
\tablehead{
\colhead{ID} & \colhead{Period} & \colhead{$\langle V \rangle$} & 
\colhead{$(\bv)_{\rm mag}$} & \colhead{$A_V$} & 
\colhead{$A_B$}   
          }
\startdata
V18 & 2.89 & 15.616 & 0.975 & 0.77 & 1.20 \\
V29 & 1.88 & \nodata & \nodata & \nodata & 1.05 \\ 
V36 & 3.10 & 15.558 & 0.850 & 1.05 & 1.45 \\ 
V37 & 10.0 & 14.707 & 1.173 & \nodata & \nodata \\ 
\enddata
\end{deluxetable}

\begin{deluxetable}{ccccccc}
\tablewidth{0pc}
\footnotesize
\tablecaption{Mean Properties of Binary Stars\label{tbl-9}}
\tablehead{
\colhead{ID} & \colhead{Period} & \colhead{$\langle V \rangle$}
& \colhead{$(\bv)_{\rm mag}$} & \colhead{$A_V$} & 
\colhead{$A_B$} & \colhead{Comments}
          }
\startdata
V14 & 2.16  & 16.179 & 0.511 & 1.30 & 1.30 & Detached \\ 
V38 & 0.412 & 18.270 & 1.092 & 0.46 & 0.56 & Contact \\ 
V39 & 0.537 & 17.996 & 0.751 & 0.28 & 0.30 & Contact \\ 
V40 & 0.311 & 18.806 & 1.106 & 0.39 & \nodata & Contact \\ 
V41 & 1.71  & 17.364 & 0.656 & 2.20 & 2.80 & Detached \\ 
V42 & 1.82  & 15.724 & 0.251 & 0.48 & 0.52 & Detached \\ 
V43 & 2.02  & 19.606 & 0.800 & 1.22 & 1.50 & Detached \\ 
V54 & 0.366 & 19.447 & 0.872 & 0.67 & 0.72 & Contact \\ 
V57 & 0.278 & 19.116 & 1.082 & 0.80 & 0.80 & Contact \\
\enddata
\end{deluxetable}

\begin{deluxetable}{cccccc}
\tablewidth{0pc}
\footnotesize
\tablecaption{Cluster properties\label{tbl-10}}
\tablehead{
\colhead{Cluster} & \colhead{Type} & \colhead{[Fe/H]}
& \colhead{$\langle P_{\rm ab} \rangle$} & \colhead{$\langle P_{\rm c} 
\rangle$} & \colhead{$N_{\rm c}/N_{\rm RR}$}
          }
\startdata
M3        & Oo~I  & $-1.6$ & 0.56 & 0.32 & 0.16 \\
M15       & Oo~II & $-2.2$ & 0.64 & 0.38 & 0.48 \\
NGC~6388  &   ?   & $-0.6$ & 0.71 & 0.36 & 0.71 \\ 
NGC~6441  &   ?   & $-0.5$ & 0.75 & 0.38 & 0.31 \\
\enddata
\end{deluxetable}

\begin{deluxetable}{ccccc}
\tablewidth{0pc}
\footnotesize
\tablecaption{RRc Parameters\label{tbl-11}}
\tablehead{
\colhead{ID} & \colhead{$M/M_{\odot}$} & 
\colhead{$\log\,(L/L_{\odot})$} & \colhead{$T_{\rm eff}$} & 
\colhead{$M_V$}
          }
\startdata 
V16 & 0.53 & 1.60 & 7583 & 0.87 \\ 
V23 & 0.46 & 1.66 & 7352 & 0.76 \\ 
V31 & 0.39 & 1.63 & 7424 & 0.76 \\ 
V34 & 0.54 & 1.58 & 7621 & 0.87 \\ 
Mean & 0.48$\pm$0.07 & 1.62$\pm$0.04 & 7495$\pm$128 & 0.82$\pm$0.06 \\
\enddata 
\end{deluxetable}

\begin{deluxetable}{cccccc}
\tablewidth{0pc}
\footnotesize
\tablecaption{RRab Parameters\label{tbl-12}}
\tablehead{ 
\colhead{ID} & \colhead{$M/M_{\odot}$} & 
\colhead{$\log\,(L/L_{\odot})$} & \colhead{$\log\,T_{\rm eff}$} & 
\colhead{$M_V$} & \colhead{[Fe/H]}  
          }
\startdata
V17 & 0.55 & 1.67 & 3.82 & 0.78 & --1.12 \\ 
V21 & 0.51 & 1.70 & 3.81 & 0.55 & --1.33 \\ 
V22 & 0.54 & 1.64 & 3.83 & 0.77 & --1.02 \\ 
V28 & 0.57 & 1.73 & 3.80 & 0.52 & --1.35 \\ 
Mean & 0.56$\pm$0.03 & 1.69$\pm$0.04 & 3.82$\pm$0.01 & 0.66$\pm$0.14 & --1.21$\pm$0.16 \\ 
\enddata
\end{deluxetable}

\begin{deluxetable}{ccccccc} 
\tablewidth{0pc} 
\footnotesize 
\tablecaption{Population II Cepheid Distance Estimates \label{tbl-13}} 
\tablehead{ 
\colhead{ID} & \colhead{$M_V$} & \colhead{$M_B$} & \colhead{$m_{0,V}$} & 
\colhead{$m_{0,B}$} & \colhead{$d_V$} & \colhead{$d_B$} 
          } 
\startdata 
V18 & --0.79 & --0.37 & 14.336 & 14.951 & 10.6 & 11.6 \\ 
V29 & --0.49 & --0.12 & \nodata & 14.395 & \nodata & 8.0 \\ 
V36 & --0.84 & --0.41 & 14.278 & 14.768 & 10.6 & 10.9 \\ 
V37 & --1.11 & --0.37 & 13.427 & 14.240 &  8.1 &  8.4 \\ 
\enddata 
\end{deluxetable}

\begin{deluxetable}{lcccc} 
\tablewidth{0pc}
\footnotesize
\tablecaption{Reddest Possible HB Morphology of an Oosterhoff Type~II Cluster 
\label{tbl-14}}
\tablehead{
  \colhead{$Z$} & 
  \colhead{$M_{\rm HB,ev}$} & 
  \colhead{$(B-R)/(B+V+R)_{\rm min}$} & 
  \colhead{$V/(B+V+R)_{\rm max}$} & 
  \colhead{$B/(B+V+R)_{\rm min}$}  
  }
\startdata
0.0005   & 0.6668   &       0.729     &       0.193     &    0.768 \\
0.001    & 0.6313   &       0.823     &       0.127     &    0.848 \\
0.002    & 0.6042   &       0.839     &       0.121     &    0.859 \\
0.006    & 0.5691   &       0.740     &       0.218     &    0.761 \\
\enddata
\end{deluxetable}


\begin{thebibliography}{100}

\bibitem[00]{00} Alcaino, G. 1981, \aaps, 44, 33 

\bibitem[1]{1} Alcock, C., et al.\ 1996, \aj, 111, 1146 

\bibitem[2]{2} Armandroff, T. E., \& Zinn, R. 1988, \aj, 96, 92

\bibitem[3]{3} Bingham, E. A., Cacciari, C., Dickens, R. J., \& 
Fusi Pecci, F. 1984, \mnras, 209, 765

\bibitem[4]{4} Blanco, V. 1992, \aj, 104, 734 

\bibitem[5]{5} Bono, G., Cassisi, S., Zoccali, M., \& Piotto, G. 2001, 
\apj, 546, L109 

\bibitem[6]{6} Borissova, J., Ivanov, V. D., \& Catelan, M. 2000, 
IBVS, 4919

\bibitem[7]{7} Brocato, E., Castellani, V., Scotti, G. A., Saviane, I., 
Piotto, G., \& Ferraro, F. R. 1998, \aap, 335, 929 

\bibitem[8]{8} Butler, D., Dickens, R. J., \& Epps, E. 1978, \apj, 225, 148

\bibitem[9]{9} Caloi, V. D'Antona, F., \& Mazzitelli, I. 1997, \aap, 320, 
823

\bibitem[0]{0} Caputo, F. 1981, \apss, 76, 329

\bibitem[a]{a} Carney, B. W. 1996, \pasp, 108, 900 

\bibitem[b]{b} Carney, B. W., Storm, J., \& Williams, C. 1993, \pasp, 105, 294 

\bibitem[c]{c} Carretta, E., Cacciari, C., Ferraro, F. R., Fusi Pecci, F., \& 
Tessicini, G. 1998, \mnras, 298, 1005 

\bibitem[d]{d} Castellani, V., \& Tornamb\`e, A. 1981, \aap, 96, 20 

\bibitem[e]{e} Catelan, M. 1992, \aap, 261, 457

\bibitem[f]{f} Catelan, M. 1993, \aaps, 98, 547

\bibitem[g]{g} Catelan, M., Bellazzini, M., Landsman, W. B., Ferraro, F. R., 
  Fusi Pecci, F., \& Galleti, S. 2001a, \aj, 122, 3171

\bibitem[h]{h} Catelan, M., de Freitas Pacheco, J. A., \& Horvath, J. E. 
  1996, \apj, 461, 231 

\bibitem[i]{i} Catelan, M., Ferraro, F. R., \& Rood, R. T. 2001b, \apj, 
560, 970 

\bibitem[j]{j} Clement, C. 2000, in The Impact of Large-Scale Surveys on 
Pulsating Star Research, ed.\ L. Szabados \& D. W. Kurtz (San Francisco: ASP), 
266 

\bibitem[k]{k} Clement, C. Dickens, R. J., \& Bingham, E. E. 1979, \aj, 84, 217 

\bibitem[l]{l} Clement, C., et al.\ 2001, \aj, 122, 2587  

\bibitem[m]{m} Clement, C., \& Rowe, J. 2000, \aj, 120, 2579 

\bibitem[n]{n} Clement, C., \& Rowe, J. 2001, \aj, 122, 1464 

\bibitem[o]{o} Clement, C., \& Shelton, I. 1997, \aj, 113, 1711  

\bibitem[p]{p} Clement, C., \& Shelton, I. 1999, \apj, 515, L85 

\bibitem[q]{q} Cohn, H. N., Lugger, P. M., Grindlay, J. E., \& Edmonds, P. D. 
2002, \apj, in press (astro-ph/0202059) 

\bibitem[r]{r} Dorman, B. 1992, \apjs, 81, 221

\bibitem[s]{s} Edmonds, P. D., Grindlay, J. E., Cohn, H., \& Lugger, P. 
2001, \apj, 547, 829 

\bibitem[t]{t} Fusi Pecci, F., Ferraro, F. R., Crocker, D. A., Rood, R. T., \& 
Buonanno, R. 1990, \aap, 238, 95 

\bibitem[u]{u} Gingold, R. A. 1976, \apj, 204, 116 

\bibitem[v]{v} Graham, J. A. 1982, \pasp, 94, 244  

\bibitem[w]{w} Harris, H. C. 1985, in Cepheids: Theory and Observations, 
ed.\ E. R. Madore (London: Cambridge University Press), 232 

\bibitem[x]{x} Harris, W. E. 1996, \aj, 112, 1487 

\bibitem[y]{y} Hazen, M. L., \& Hesser, B. H. 1986, \aj, 92, 1094 

\bibitem[z]{z} Iben, I., Jr., \& Rood, R. T. 1970, \apj, 161, 587 

\bibitem[aa]{aa} Jurcsik, J. 1998, \aap, 333, 571 

\bibitem[bb]{bb} Jurcsik, J., \& Kov\'{a}cs, G. 1996, \aap, 312, 111 

\bibitem[cc]{cc} Jurcsik, J., \& Kov\'{a}cs, G. 1999, New Astronomy Reviews, 
43, 463

\bibitem[dd]{dd} Kaluzny, J., Hilditch, R. W., Clement, C., \& Rucinski, S. M. 
1998, \mnras, 296, 347 

\bibitem[ee]{ee} Kaluzny, J., Kubiak, M., Szymanski, M., Udalski, A., 
Krzeminski, W., \& Mateo, M. 1997, \aaps, 125, 343 

\bibitem[ff]{ff} Kaluzny, J., Olech, A., Thompson, I., Pych, W., Krzeminski, W., 
\& Schwarzenberg-Cerny, A. 2000, \aaps, 143, 215 

\bibitem[gg]{gg} Kholopov, P. N. 1985, General Catalogue of 
Variable Stars (4th ed.; Moscow: Nauka) 

\bibitem[hh]{hh} Kov\'{a}cs, G. 1998, in A Half Century of Stellar Pulsation 
Interpretation: A Tribute to Arthur N. Cox, ed.\ P. A. Bradley \& J. A. 
Buzik (San Francisco: ASP), 52

\bibitem[ii]{ii} Kov\'{a}cs, G., \& Jurcsik, J. 1997, \aap, 322, 218 

\bibitem[jj]{jj} Landolt, A. U. 1992, \aj, 104, 340  

\bibitem[kk]{kk} Layden, A. C., Ritter, L. A., Welch, D. L., \& Webb, T. M. A. 
1999, \aj, 117, 1313 

\bibitem[ll]{ll} Layden, A. C., \& Sarajedini, A. 2000, \aj, 119, 1760 

\bibitem[mm]{mm} Lee, J.-W., \& Carney, B. W. 1999, \aj, 118, 1373 

\bibitem[nn]{nn} Lee, Y.-W. 1990, \apj, 363, 159 

\bibitem[oo]{oo} Lee, Y.-W., Demarque, P., \& Zinn, R. 1990, \apj, 350, 155

\bibitem[pp]{pp} Lloyd Evans, T., \& Menzies, J. W. 1973, in Variable Stars 
in Globular Clusters and in Related Systems, ed.\ J. D. Fernie 
(Boston: Reidel), 151 

\bibitem[qq]{qq} LLoyd Evans, T., \& Menzies, J. W. 1977, \mnras, 178, 163 

\bibitem[rr]{rr} McNamara, D. H., 1995, \aj, 109, 2134 

\bibitem[ss]{ss} Mengel, J. G. 1973, in Variable Stars in Globular Clusters and 
in Related Systems, ed.\ J. D. Fernei (Boston: Reidel), 214

\bibitem[tt]{tt} Mengel, J. G., \& Gross, P. G. 1976, \apss, 41, 407 

\bibitem[uu]{uu} Mohler, S. 2001, \pasp, 113, 1162 

\bibitem[vv]{vv} Nemec, J. M., Nemec, A. F. L., \& Lutz, T. E. 1994, \aj, 108, 
222 

\bibitem[ww]{ww} Olech, A., Kaluzny, J., Thompson, I. B., Pych, W., Krzeminski, W., 
\& Shwarzenberg-Czerny, A. 1999, \aj, 118, 442 

\bibitem[xx]{xx} Pancino, E., Ferraro, F. R., Bellazzini, M., Piotto, G., \& 
Zoccali, M. 2000, \apjl, 534, L83

\bibitem[yy]{yy} Petersen, J. O. 1986, \aap, 170, 59 

\bibitem[zz]{zz} Piotto, G., et al.\ 1997, in Advances in Stellar Evolution, ed.\ 
R. T. Rood \& A. Renzini (Cambridge: Cambridge University Press), 84 

\bibitem[a1]{a1} Piotto, G., et al.\ 1999, \aj, 118, 1727 

\bibitem[a2]{a2} Pritzl, B., Smith, H. A., Catelan, M., \& Sweigart, A. V. 
2000, \apj, 530, L41 

\bibitem[a3]{a4} Pritzl, B., Smith, H. A., Catelan, M., \& Sweigart, A. V. 
2001, 122, 2600 (Paper~I)

\bibitem[a5]{a5} Raimondo, G., Castellani, V., Cassisi, S., Brocato, E., \& 
Piotto, G. 2002, \apj, in press (astro-ph/0201123) 

\bibitem[a6]{a6} Ree, C. H., Yoon, S.-J., Rey, S.-C., Lee, Y.-W. 2002, in 
Omega Centauri: A Unique Window Into Astrophysics, ed.\ F. van~Leeuwen, 
G. Piotto, \& J. Hughes (Cambridge: Cambridge University Press), in press 
(astro-ph/0110689) 

\bibitem[a7]{a7} Reed, B. C., Hesser, J. E., Shawl, S. J. 1988, \pasp, 100, 545 

\bibitem[a8]{a8} Reimers, D. 1975, in Mem. Soc. R. Sci. Li\`ege 6 S\'er., 8, 369

\bibitem[a9]{a9} Renzini, A., \& Fusi Pecci, F. 1988, \araa, 26, 199

\bibitem[a0]{a0} Rey, S.-C., Lee, Y.-W., Joo, J.-M., Walker, A., \& Baird, S. 
2000, \aj, 119, 1824 

\bibitem[b1]{b1} Rich, R. M., et al.\ 1997, \apj, 484, L25

\bibitem[b2]{b2} Rich, R. M., Minniti, D., \& Liebert, J. 1993, \apj, 406, 489 

\bibitem[b3]{b3} Rood, R. T., \& Crocker, D. A. 1989, in IAU Colloq.\ 111, The 
Use of Pulsating Stars in Fundamental Problems of Astronomy, ed.\ E. G. 
Schmidt (Cambridge: Cambridge University Press), 103

\bibitem[b4]{b4} Salaris, M., Cassisi, S., \& Weiss, A. 2002, \pasp, in press 
  (astro-ph/0201387) 

\bibitem[b5]{b5} Sandage, A. 1990, \apj, 350, 603 

\bibitem[b6]{b6} Sandage, A., \& Wildey, R. 1967, \apj, 150, 469 

\bibitem[b7]{b7} Schlegel, D. J., Finkbeiner, D. P., \& Davis, M. 1998, 
\apj, 500, 525 

\bibitem[b8]{b8} Silbermann, N. A., \& Smith, H. A., 1995, \aj, 110, 704 

\bibitem[b9]{b9} Silbermann, N. A., Smith, H. A., Bolte, M., \& Hazen, M. L. 
1994, \aj, 107, 1764 

\bibitem[b0]{b0} Simon, N. R., \& Clement, C. 1993a, \apj, 410, 526 

\bibitem[c1]{c1} Simon, N. R., \& Clement, C. 1993b, in IAU Colloq.\ 139, New 
   Perspectives on Stellar Pulsation and Pulsating Variable Stars, ed.\ J. M. 
   Nemec \& J. M. Matthews (Cambridge: Cambridge University Press), 315 

\bibitem[c2]{c2} Simon, N. R., \& Teays, T. J. 1982, \apj, 261, 586  

\bibitem[c3]{c3} Smith, H. A. 1995, RR Lyrae Stars (Cambridge: Cambridge 
University Press) 

\bibitem[c4]{c4} Smith H. A., \& Wehlau, A. 1985, \apj, 298, 572 

\bibitem[c5]{c5} Stellingwerf, R. F., Gautschy, A., \& Dickens, R. J. 
1987, \apj, 313, L75 

\bibitem[c6]{c6} Sweigart, A. V. 1978, in The HR Diagram, ed.\ A. G. Davis
Philip \& D. S. Hayes (Dordrecht: Reidel), 333 

\bibitem[c7]{c7} Sweigart, A. V. 1987, \apjs, 65, 95

\bibitem[c8]{c8} Sweigart, A. V. 1999, in Spectrophotometric Dating of Stars
and Galaxies, ASP Conf.\ Ser.\ Vol.\ 192, ed.\ I. Hubeny, S. Heap, 
\& R. Cornett (San Francisco: ASP), 239

\bibitem[c9]{c9} Sweigart, A. V. 2002, in Highlights of Astronomy, Vol.\ 12, 
in press (astro-ph/0103133) 

\bibitem[c0]{c0} Sweigart, A. V., \& Catelan, M. 1998, \apj, 501, L63

\bibitem[d1]{d1} Sweigart, A. V. \& Gross, P. G. 1976, \apjs, 32, 367 

\bibitem[d2]{d2} Sweigart, A. V., Renzini, A., \& Tornamb\`e, A. 1987, \apj, 
312, 762 

\bibitem[d3]{d3} van Albada, T. S., \& Baker, N. 1971, \apj, 169, 311 

\bibitem[d4]{d4} van Albada, T. S., \& Baker, N. 1973, \apj, 185, 477 

\bibitem[d5]{d5} van den Bergh, S. 1967, \pasp, 79, 460 

\bibitem[d6]{d6} Walker, A. R. 1994, \aj, 108, 555 

\bibitem[d7]{d7} Walker, A. R. 2000, in The Impact of Large-Scale Surveys on 
Pulsating Star Research, ed.\ L. Szabados \& D. W. Kurtz (San Francisco: ASP), 
165

\bibitem[d8]{d8} Walker, A. R., \& Nemec, J. M. 1996, \aj, 112, 2026 

\bibitem[d9]{d9} Wallerstein, G. W. 1970, \apj, 160, 345 

\bibitem[d0]{d0} Yi, S., Lee, Y.-W., \& Demarque, P. 1993, \apj, 411, L25 

\bibitem[e1]{e1} Zinn, R. 1980, \apjs, 42, 19 

\bibitem[e2]{e2} Zinn, R., \& West, M. J. 1984, \apjs, 55, 45 

\bibitem[e3]{e3} Zoccali, M., Cassisi, S., Bono, G., Piotto, G., Rich, 
R. M., \& Djorgovski, S. G. 2000, \apj, 538, 289 

\bibitem[e4]{e4} Zoccali, M., Cassisi, S., Piotto, G., Bono, G., \& 
Salaris, M. 1999, \apj, 518, L49 

\end{thebibliography}
\end{document}